\documentclass[12pt]{article}
\usepackage{amsmath}
\usepackage{amssymb}
\usepackage{graphicx}
\usepackage[numbers,sort&compress]{natbib}
\usepackage[colorlinks,
            anchorcolor=blue,
            citecolor=green]
            {hyperref}
\title{Existence of global solutions to the massive Thirring model in the non-laboratory coordinates}

\author{Sucai Niu$^{a}$, 
Junyi Zhu$^{a}$\thanks{E-mail: jyzhu@zzu.edu.cn} and Xueru Wang$^{b}$\thanks{E-mail: xr\_wang@163.com}\\
\scriptsize{\sl $^{a}$ School of Mathematics and Statistics, Zhengzhou University, Zhengzhou, Henan 450001, China}\\
\scriptsize{\sl $^{b}$ School of Mathematics and Statistics, Xinyang Normal University, Xinyang, Henan 464000, China}
}

\date{}
\begin{document}
\newtheorem{theorem}{Theorem}
\newtheorem{definition}{Definition}
\newtheorem{lemma}{Lemma}
\newtheorem{corollary}{Corollary}
\newtheorem{proposition}{Proposition}
\newtheorem{remark}{Remark}
\hypersetup{CJKbookmarks=true}

\maketitle
\begin{abstract}
The massive Thirring model in the non-laboratory coordinates is considered by the Riemann-Hilbert approach. Existence of global solutions is shown for the cases of the associated Riemann-Hilbert problem without eigenvalues or resonances. The Lipschitz continuity of the map from the potential $v_0(x)\in H^2(\mathbb{R})\cap H^{1,1}(\mathbb{R})$ to the scattering data is given in the direct scattering transform. Two transform matrices are introduced to curb the convergence of the Volterra integral equations and the relevant estimates of the modified Jost functions. For small potential, the solvability of the Riemann-Hilbert problems without eigenvalues or resonances is discussed. The Lipschitz continuity of the map from the scattering data to the potential $v(x)$ is shown. The reconstructions for potential $u(x,t)$ and $v(x,t)$ are finished by considering the time dependence of the scattering data and by constructing the conservation laws obtain via the dressing method.


\end{abstract}
\newpage
 \tableofcontents

\section{Introduction}
The massive Thirring (MT) model has a rich physical background in two-dimensional quantum field theory \cite{ap3-91}, optical Bragg gratings \cite{pio33-203}, and diatomic chains with periodic couplings \cite{ap403-198}. The MT model is related to the Lax pair by the Coleman's correspondences \cite{prd14-472}. The MT model is integrable \cite{jl23-320}and can be used to study the inverse scattering \cite{lnc20-325,jpsj47-1327,jpsj65-722,tmp30-193,1801.00039,sam84-207,jpsj52-1084,P-S2019,2307.15323,2411.18140}, soliton solutions \cite{cmp112-423,jmp34-3039,jmp34-3054,prd14-472}, and construction of rogue waves \cite{pla379-1067}. Tau-function formulation for bright, dark soliton and breather solutions to the massive Thirring model was considered in \cite{sam150-35}. The Darboux transformation, B\"acklund transformation of the MT model and its connection with other integrable systems have been investigated in the literature\cite{lnc20-325,LJH1993,tmp99-617,umj31-362,tmp46-249,jpa46-035201,jpa48-235204}.
The orbital stability of Dirac solitons and spectral stability of the standing periodic wave in the integrable MT model in laboratory coordinates was discussed \cite{cpde41-227,jmp37-308,lmp104-21,sam154-e12789}.

The global well-posedness for derivative NLS equation with the inverse scattering transform method has been well studied in  \cite{imrn2018-5663,dpde14-271,cpde43-1151,jpam78-33}, and other integrable equations \cite{jde366-320,sam152-111,camb45-497,jde414-34}. In this paper, we extend the global well-posedness with the inverse scattering transform method to the Cauchy problem of the MT model in non-laboratory coordinates
\begin{equation}\label{mta1}
\begin{aligned}
iu_{x}(x,t)+v(x,t)-\vert v(x,t)\vert^{2}u(x,t)=0, \\
iv_t(x,t)+u(x,t)-\vert u(x,t)\vert^{2}v(x,t)=0,\\
u(x,0)=u_0(x)\in H^3(\mathbb{R})\cap H^{1,1}(\mathbb{R}),\\
 v(x,0)=v_0(x)\in H^2(\mathbb{R})\cap H^{1,1}(\mathbb{R}).
\end{aligned}
\end{equation}
Here we introduce the following notations
\[H^n(\mathbb{R})=\{f(x):\partial_x^jf\in L^2(\mathbb{R}), j=1,\cdots,n\},\]
\[H^{1,1}(\mathbb{R})=\{f(x):f,\partial_x f\in L^{2,1}(\mathbb{R})\},\]
\[\|f(x)\|_{L^{2,1}(\mathbb{R})}=\|(1+ x^2)^{1/2} f(x)\|_{L^2(\mathbb{R})}. \]
The MT model is equivalent to the compatibility condition of the following linear system
\begin{equation}\label{mta2}
\psi_x=W\psi, \quad W=-\frac{i}{2}(k^2\sigma_3+2kV-V^2\sigma_3),
\end{equation}
and
\begin{equation}\label{mta3}
 \psi_t=\tilde{W}\psi, \quad \tilde{W}=-\frac{i}{2}(k^{-2}\sigma_3+2k^{-1}U-U^3\sigma_3),
\end{equation}
with $\psi=\psi(x,t;k)$ is a vector eigenfunction and
\begin{equation}\label{mta4}
 V=\left(\begin{matrix}
0  & v  \\
-\overline{v} &0
 \end{matrix}\right), \quad
U=\left(\begin{matrix}
0 & u  \\
-\overline{u} &0
 \end{matrix}\right).
\end{equation}
Here the Pauli matrix $\sigma_3={\rm diag}(1,-1)$.

In general, the inverse scattering transform is utilized to discuss integrable equations, with spectral analysis typically confined to linear spectral problems. The time-dependent linear spectral problem under limit condition is only used to explore the time evolution of scattering data. It's important to note that for the MT equation in non-laboratory coordinates \eqref{mta1}, the potentials $v$ and $u$ are incorporated into the linear spectral problem and the time-dependent linear problem, respectively. Consequently, the standard inverse scattering transform method can only address aspects related to potential $v$, making it difficult to study the global existence of the MT model in non-laboratory coordinates.

In present paper, we extend the inverse scattering transform method to investigate the existence of global solutions for the MT equation in non-laboratory coordinates. Specifically, for the linear spectral problem, by introducing two transformation matrix functions, we complete the direct scattering transform, discuss the analyticity and asymptotic behavior of the related Jost functions, and derive prior estimates between the Jost functions and scattering data. We construct the associated Riemann-Hilbert problem, identify conditions for small potential $v$ without eigenvalues and resonance points, and investigate the solvability of the RH problem. Furthermore, we complete the inverse scattering transform related to the linear spectral problem and provide a reconstruction formula for potential $v$. For cases $x>0$ and $x<0$, we discuss prior estimates from potential $v$ to scattering data and their Lipschitz continuity.

To obtain a reconstruction formula for potential $u$, we discuss the time evolution of scattering data and employ the dressing method to derive infinite conservation laws for the MT equation \eqref{mta1}. The first two conservation laws are
\begin{equation}\label{mta5}
 (|v|^2)_t+(|u|^2)_x=0,
\end{equation}
and
\begin{equation}\label{mta6}
 i(v\bar{v}_x)_t+(u\bar{v})_x=0.
\end{equation}
This enables us to derive the reconstruction formula for the potential $u$ from the conservation law \eqref{mta5} and the second equation in \eqref{mta1}, as well as the reconstruction of $v$. We note that, for every $t\in[0,T]$, if $v,v_t\in H_x^2(\mathbb{R})\cap H_x^{1,1}(\mathbb{R})$, then $u\in H_x^3(\mathbb{R})\cap H_x^{1,1}(\mathbb{R})$, in view of the MT equation $iu_{x}+v-\vert v\vert^{2}u=0$ and
\[|u|\leq|v_t|+2|v|(\|v\|_{L_x^2}\|v_t\|_{L_x^2}).\]
 Furthermore, we discuss the Lipschitz continuity between potential $u$ and scattering data. Consequently, the Lipschitz continuity from potentials $v$ and $u$ to scattering data is established.

The paper is organized as follows. In section \ref{sec2}, two transformation matrices are introduced to define two new spectral problems, which are equivalent to two Volterra integral equations of the two vector Jost functions. The properties of analyticity and asymptotic behaviors of these Jost functions are discussed. Lipschitz continuity of the map from the potential $v$ to the scatter data is shown by virtue of the Lipschitz continuity of the map from the potential $v$ to the Jost vectors.
In section \ref{sec3}, the Riemann-Hilbert problem for original Jost functions to original linear spectral problem \eqref{mta2} is obtained, and the Riemann-Hilbert problem for the modified Jost functions are given. The two Riemann-Hilbert problems are defined by two sets of scattering coefficients $r(k)$ and $r_\pm(z)$, respectively. For small potential, the solvability of the Riemann-Hilbert problems with no eigenvalues or resonances are discussed.
In section \ref{sec4}, the inverse scattering transform is discussed. The properties of the scattering coefficients are discussed. The solutions of the Riemann-Hilbert problems are presented via the Cauchy project operators. The prior estimates for the scattering coefficients and the modified Jost vectors are obtained. Furthermore, the reconstruction formula for the potential $v$ is obtained, the prior estimates of $v$ about the scatter data for the cases of $x\in\mathbb{R}^+$ and $x\in\mathbb{R}^-$ are discussed. As results, Lipschitz continuity of the maps from scattering data to $v$ in the two cases  are shown.
In section \ref{sec6}, the time evolution of the scattering data is presented. Then the MT model is recovered and the infinite conservation laws are derived by virtue of dressing method. Moreover, The reconstruction for potential $u$ is obtained and the Lipschitz continuity of the maps from scattering data to $u$ is shown explicitly via the prior estimates of the time-dependent Jost functions are discussed. In section \ref{sec7}, a unique global solution in $H_x^2(\mathbb{R})(H_x^3(\mathbb{R}))\cap H_x^{1,1}(\mathbb{R})$ to the MT model in non-laboratory coordinates is shown according to the local solution and the uniformly prior estimates.

\setcounter{equation}{0}
\section{The direct scattering transform and Lipschitz continuity}\label{sec2}
Let $\psi$ be a solution of the linear spectral problem \eqref{mta2} and fix the time $t$. We introduce the two transformation matrices 
\begin{equation}\label{mtb1}
 T_1=\left(\begin{matrix}
 1&0\\
 2\bar{v}&2k
 \end{matrix}\right), \quad T_2=\left(\begin{matrix}
 2k&2v\\
 0&1
 \end{matrix}\right),
\end{equation}
and define two vector eigenfunctions $\psi_{1,2}=T_{1,2}\psi$, then we have
\begin{equation}\label{mtb2}
\psi_{1,x}=W_1\psi_1, \quad W_1=-\frac{i}{2}k^2\sigma_3+V_1, \quad
V_1=\frac{i}{2}\left(\begin{matrix}
 |v|^2&-v\\
 -4i\bar{v}_x&-|v|^2
 \end{matrix}\right),
\end{equation}
and
\begin{equation}\label{mtb3}
\psi_{2,x}=W_2\psi_2, \quad W_2=-\frac{i}{2}k^2\sigma_3+V_2, \quad
V_2=\frac{i}{2}\left(\begin{matrix}
 |v|^2&-4iv_x\\
 \bar{v}&-|v|^2
 \end{matrix}\right).
\end{equation}

For the linear system \eqref{mtb2} and \eqref{mtb3}, letting $z=k^2\in\mathbb{R}$ and introducing two Jost functions
\begin{equation}\label{mtb4}
 m_\pm(x;z)=\psi_1(x;z)\mathrm{e}^{\frac{i}{2}zx}, \quad n_\pm(x;z)=\psi_2(x;z)\mathrm{e}^{-\frac{i}{2}zx},
\end{equation}
with the boundary conditions at infinity
\begin{equation}\label{mtb5}
 m_\pm(x;z)\to e_1=\left(\begin{array}{c}
 1\\
 0
 \end{array}\right), \quad n_\pm(x;z)\to e_2=\left(\begin{array}{c}
 0\\
 1
 \end{array}\right), \quad x\to\pm\infty,
\end{equation}
we get that the Jost functions are the solutions of the following Volterra integral equations
\begin{equation}\label{mtb6}
 m_\pm(x;z)=e_1+\int_{\pm\infty}^x\left(\begin{matrix}
 1&0\\
 0&\mathrm{e}^{iz(x-y)}
 \end{matrix}\right)V_1(y)m_\pm(y;z)\mathrm{d}y,
\end{equation}
and
\begin{equation}\label{mtb7}
 n_\pm(x;z)=e_2+\int_{\pm\infty}^x\left(\begin{matrix}
 \mathrm{e}^{-iz(x-y)}&0\\
 0&1
 \end{matrix}\right)V_2(y)n_\pm(y;z)\mathrm{d}y.
\end{equation}

For every $z\in \mathbb{C}^\pm$, and for every vector $f(\cdot;z)\in L^\infty(\mathbb{R})$, we define the integral operators
\begin{equation}\label{mtb8}
{\cal{K}}_{\pm}f(x;z)=\int_{\pm\infty}^x\left(\begin{matrix}
 \mathrm{e}^{-iz(x-y)}&0\\
 0&1
 \end{matrix}\right)V_2(y)f(y;z)\mathrm{d}y,
\end{equation}
and for every $x_0\in \mathbb{R}$, we define a compact form $(j\in\mathbb{N})$
\begin{equation}\label{mtb8a}
\|v\|_{L^j(\pm\infty,x_0)'}=\left\{\begin{array}{cc}
\|v\|_{L^j(-\infty,x_0)}, & x\in(-\infty,x_0),\\
\|v\|_{L^j(x_0,+\infty)}, & x\in(x_0,+\infty).
 \end{array}\right.
\end{equation}
Then the eigenvalues of the matrix
\[\frac{1}{2}\left(\begin{matrix}
\|v\|^2_{L^2(\pm\infty,x_0)'}&\|v\|_{L^1(\pm\infty,x_0)'}\\
4\|v_x\|_{L^1(\pm\infty,x_0)'}&\|v\|^2_{L^2(\pm\infty,x_0)'}
\end{matrix}\right),\]
associated with the kernel of the left integral in \eqref{mtb8} are
\[\lambda_\pm=\frac{1}{2}\|v\|^2_{L^2(\pm\infty,x_0)'}\pm\sqrt{\|v\|_{L^1(\pm\infty,x_0)'}\|v_x\|_{L^1(\pm\infty,x_0)'}}.\]
If
\begin{equation}\label{mtb8b}
\lambda_+=\frac{1}{2}\|v\|^2_{L^2(\pm\infty,x_0)'}+\sqrt{\|v\|_{L^1(\pm\infty,x_0)'}\|v_x\|_{L^1(\pm\infty,x_0)'}}<1,
\end{equation}
the operators ${\cal{K}_\pm}$ defined in \eqref{mtb8} are contraction, in view of the fact that the eigenvalues $\lambda_\pm$ are located in the unit circle.

\begin{lemma}\label{lem1}
Let $v\in L^2(\mathbb{R})\cap L^1(\mathbb{R}), v_x\in L^1(\mathbb{R})$, and the condition \eqref{mtb8b} is valid. For every $z\in\mathbb{R}$, there exist unique
solutions $m_\pm(\cdot;z)\in L^\infty(\mathbb{R}), n_\pm(\cdot;z)\in L^\infty(\mathbb{R})$ to the integral equations \eqref{mtb6} and \eqref{mtb7}. For every $x\in\mathbb{R}$, $m_\mp(x;\cdot)$ and $n_\pm(x;\cdot)$ are continued
analytically in $\mathbb{C}^\pm$.
\end{lemma}
{\bf Proof}~~  We only prove the statement for the Jost functions $n_\pm(x;z)$. The integral equation \eqref{mtb7} can be written as $n_\pm=e_2+{\cal{K}}_{\pm}n_\pm$. By the Banach fixed point theorem, for every $x_0\in \mathbb{R}$ and for every $z\in \mathbb{C}^\pm$, there exist unique solutions $n_\pm(\cdot;z)\in L^\infty(\pm\infty,x_0)'$.

To prove the analyticity of $n_\pm(x;\cdot)\in \mathbb{C}^\pm$ for every $x\in \mathbb{R}$,  we define the vector form $\|f(\cdot;z)\|_{L^\infty(\pm\infty,x_0)'}$ and the matrix form $\|A\|_{L^j(\pm\infty,x_0)'}$ as
\[\|f(\cdot;z)\|_{L^\infty(\pm\infty,x_0)'}=\sum\limits_{j=1}^2\|f_j(\cdot;z)\|_{L^\infty(\pm\infty,x_0)'},\]
and
\[\|A(\cdot;z)\|_{L^j(\pm\infty,x_0)'}=\sum\limits_{l,m=1}^2\|A_{lm}(\cdot;z)\|_{L^j(\pm\infty,x_0)'}, \quad j\in\mathbb{N},\]
then, for every $z\in \mathbb{C}^\pm$, we find that
\begin{equation}\label{mtb9}
 \|{\cal{K}}_{\pm}f(\cdot;z)\|_{L^\infty(\pm\infty,x_0)'}\leq\|V_2\|_{L^1(\pm\infty,x_0)'}\|f(\cdot;z)\|_{L^\infty(\pm\infty,x_0)'},
\end{equation}
with
\[\|V_2\|_{L^1(\pm\infty,x_0)'}=\|v\|^2_{L^2(\pm\infty,x_0)'}+\frac{1}{2}\|v\|_{L^1(\pm\infty,x_0)'}+2\|v_x\|_{L^1(\pm\infty,x_0)'}.\]

Now we define the Neumann series $\sum\limits_{l=0}^\infty w_\pm^{<l>}(x;z)$, where
\[w_\pm^{<0>}(x;z)=e_2, \quad w_\pm^{<l+1>}(x;z)={\cal{K}}_{\pm}w_\pm^{<l>}(x;z).\]
For every $z\in \mathbb{C}^\pm$, from \eqref{mtb9}, we have
\[\|w_\pm^{<l>}(x;z)\|_{L_x^\infty} \leq \|{\cal{K}}_{\pm}^le_2\|_{L_x^\infty}
\leq \frac{(\|V_2\|_{L^1(\pm\infty,x_0)'})^l}{l!}.\]
If $v\in L^2(\mathbb{R})\cap L^1(\mathbb{R}), v_x\in L^1(\mathbb{R})$, the Neumann series $\sum\limits_{l=0}^\infty w_{\pm}^{<l>}(x;z)$ converges absolutely and uniformly for every $x\in\mathbb{R}$ and for every $z\in\mathbb{C}^\pm$.
Thus, the summation of the Neumann series give the Jost solutions
\[n_\pm(x;z)=\sum\limits_{l=0}^\infty w_{\pm}^{<l>}(x;z),\]
which are analytic in $z\in\mathbb{C}^\pm$ for every $x\in\mathbb{R}$, and
\begin{equation}\label{mtb9a}
\|n_\pm(x;z)\|_{L_x^\infty}\leq \mathrm{e}^{\|V_2\|_{L^1(\mathbb{R})}}.
\end{equation}
This finishes the proof for $n_\pm(x;z)$. The proof for other Jost functions is analogous. \qquad $\Box$

\begin{lemma}\label{lem2}
Let $v\in L^2(\mathbb{R})\cap L^1(\mathbb{R}), v_x\in L^1(\mathbb{R})$. For every $x\in\mathbb{R}$, the Jost functions $m_\pm(x;z)$ and $n_\pm(x;z)$ admit the following asymptotic behaviors along a contour in the domains of their
 analyticity
\begin{equation}\label{mtb10}
\lim\limits_{z\to\infty}m_\pm(x;z)=\mathrm{e}^{i\nu_\pm(x)}e_1, \quad
\lim\limits_{z\to\infty}n_\pm(x;z)=\mathrm{e}^{-i\nu_\pm(x)}e_2.
\end{equation}
Moreover, If $v\in C^1(\mathbb{R})$, then
\begin{equation}\label{mtb12}
\lim\limits_{z\to\infty}z[m_\pm(x;z)-\mathrm{e}^{i\nu_\pm(x)}e_1]=\mathrm{e}^{i\nu_\pm(x)}[\bar\mu_\pm(x)e_1+2i\bar{v}_x(x)e_2], \quad z\in \mathbb{C}^\mp,
\end{equation}
\begin{equation}\label{mtb13}
\lim\limits_{z\to\infty}z[n_\pm(x;z)-\mathrm{e}^{-i\nu_\pm(x)}e_2]=\mathrm{e}^{-i\nu_\pm(x)}[-2i{v}_x(x)e_1+\mu_\pm(x)e_2], \quad z\in\mathbb{C}^\pm.
\end{equation}
Here
\begin{equation}\label{mtb14}
\nu_\pm(x)=\frac{1}{2}\int_{\pm\infty}^x|v(y)|^2\mathrm{d}y, \quad \mu_\pm(x)=\int_{\pm\infty}^x \bar{v}(y)v_y(y)\mathrm{d}y.
\end{equation}
\end{lemma}
{\bf Proof}~~ We prove the statement for the Jost function $n_\pm(x;z)$ only. Let $n_\pm=(n_\pm^{(1)},n_\pm^{(2)})^T$, and rewrite the Volterra integel equation \eqref{mtb7},
\begin{equation}\label{mtb15}
n_\pm^{(1)}(x;z)=\frac{i}{2}\int_{\pm\infty}^x \mathrm{e}^{-iz(x-y)}\big[|v(y)|^2n_\pm^{(1)}(y;z)-4iv_y(y)n_\pm^{(2)}(y;z)\big]\mathrm{d}y,
\end{equation}
\begin{equation}\label{mtb16}
n_\pm^{(2)}(x;z)=1+\frac{i}{2}\int_{\pm\infty}^x \big[\bar{v}(y)n_\pm^{(1)}(y;z)-|v(y)|^2n_\pm^{(2)}(y;z)\big]\mathrm{d}y.
\end{equation}
Considering the bounded condition \eqref{mtb9a} 
for every $v\in L^2(\mathbb{R})\cap L^1(\mathbb{R}), v_x\in L^1(\mathbb{R})$. The integrand of the integral equation \eqref{mtb15} is bounded for every $z\in\mathbb{C}^\pm$ and tend to zero for every $y\in(\pm\infty,x)$ and $z\to\infty$ in $\mathbb{C}^\pm$. By Lebesgue's dominated convergence theorem, we have
$$\lim\limits_{z\to\infty}n_\pm^{(1)}(x;z)=0.$$
Thus, from \eqref{mtb16}, $\lim\limits_{z\to\infty}n_\pm^{(2)}(x;z)$ satisfy the inhomogeneous integral equation
\[\lim\limits_{z\to\infty}n_\pm^{(2)}(x;z)=1-\frac{i}{2}\int_{\pm\infty}^x|v(y)|^2\lim\limits_{z\to\infty}n_\pm^{(2)}(y;z)\mathrm{d}y,\]
which implies the unique solution $\lim\limits_{z\to\infty}n_\pm^{(2)}(x;z)=\mathrm{e}^{-i\nu_\pm(x)}$. This proves the limit for $n_\pm(x;z)$.

To prove the limit \eqref{mtb13}, we add the condition $v\in C^1(\mathbb{R})$. For every $x\in\mathbb{R}$ and every small $\delta>0$, we rewrite the integral equation \eqref{mtb15} as
\[\begin{aligned}
n_\pm^{(1)}(x;z)&=\int_{\pm\infty}^{x\pm\delta}\mathrm{e}^{-iz(x-y)}Y(y;z)\mathrm{d}y
+Y(x;z)\int_{x\pm\delta}^x\mathrm{e}^{-iz(x-y)}\mathrm{d}y\\
&\quad+\int_{x\pm\delta}^x\mathrm{e}^{-iz(x-y)}[Y(y;z)-Y(x;z)]\mathrm{d}y\\
&=A_{1,\pm}+A_{2,\pm}+A_{3,\pm},
\end{aligned}\]
where
\[Y(x;z)=\frac{i}{2}\big[|v(x)|^2n_\pm^{(1)}(x;z)-4iv_x(x)n_\pm^{(2)}(x;z)\big].\]

For every $y\in (\pm\infty,x\pm\delta)$ and every $z\in\mathbb{C}^\pm$,
\[|\mathrm{e}^{-iz(x-y)}|=\mathrm{e}^{{\rm Im}z(x-y)}\leq\mathrm{e}^{\mp\delta{\rm Im}z}. \]
Since for every $z\in\mathbb{ C}^\pm$, $n_\pm\in L_x^\infty(\mathbb{R})$ and $v\in L^2(\mathbb{R})\cap L^1(\mathbb{R}), v_x\in L^1(\mathbb{R})$, then $Y(\cdot;z)\in L^1(\mathbb{R})$ and
\[|A_{1,\pm}|\leq\mathrm{e}^{\mp\delta{\rm Im}z}\|Y(\cdot;z)\|_{L^1(\mathbb{R})}, \quad z\in\mathbb{C}^\pm.\]

Under the added condition $v\in C^1(\mathbb{R})$, for every $z\in\mathbb{C}^\pm$, the function $Y(\cdot;z)\in C[x,x\pm\delta]$ and is bounded. Thus,
\[|A_{3,\pm}|\leq\frac{2}{|{\rm Im}z|}\|Y(\cdot;z)-Y(x;z)\|_{L^\infty[x,x\pm\delta]}, \quad z\in\mathbb{C}^\pm.\]
In addition,
\[A_{2,\pm}=\frac{1}{iz}(1-\mathrm{e}^{\pm iz\delta})\psi(x;z), \quad z\in\mathbb{C}^\pm.\]

If taking $\delta=(\pm{\rm Im}z)^{-1/2}, z\in\mathbb{C}^\pm$, then $\delta\to0$ as $|{\rm Im}z|\to\infty$ and
\[\lim\limits_{z\to\infty \atop {\rm Im}z\gtrless0}|\mathrm{e}^{\pm iz\delta}|
=\lim\limits_{z\to\infty \atop {\rm Im}z\gtrless0}\mathrm{e}^{-(\pm{\rm Im}z)^{1/2}}=0.\]
Furthermore,
\[|zA_{1,\pm}|\to0, \quad |zA_{3,\pm}|\to0, \quad z\to\infty, z\in\mathbb{C}^\pm.\]

As a result,
\[\lim\limits_{z\to\infty}zn_\pm^{(1)}(x;z)=\lim\limits_{z\to\infty}(-i)\psi(x;z)=-2iv_x\mathrm{e}^{-i\nu_\pm(x)}, \quad z\in\mathbb{C}^\pm. \]

Differentiating equation \eqref{mtb16} with respect to $x$, we have
\[n_{\pm,x}^{(2)}(x;z)+\frac{i}{2}|v|^2n_\pm^{(2)}(x;z)=\frac{i}{2}\bar{v}(x)n_\pm^{(1)}(x;z),\]
which equivalents to another integral equation
\[n_\pm^{(2)}(x;z)=\mathrm{e}^{-i\nu_\pm(x)}+\mathrm{e}^{-i\nu_\pm(x)}
\frac{i}{2}\int_{\pm\infty}^x\mathrm{e}^{i\nu_\pm(y)}\bar{v}(y)n_\pm^{(1)}(y;z)\mathrm{d}y,\]
in view of \eqref{mtb14}. Thus
\[\begin{aligned}
\lim\limits_{z\to\infty}z[n_\pm^{(2)}(x;z)-\mathrm{e}^{-i\nu_\pm(x)}]
&=\mathrm{e}^{-i\nu_\pm(x)}
\frac{i}{2}\int_{\pm\infty}^x\mathrm{e}^{i\nu_\pm(y)}\bar{v}(y)\lim\limits_{z\to\infty}zn_\pm^{(1)}(y;z)\mathrm{d}y\\
&=\mu_\pm(x)\mathrm{e}^{-i\nu_\pm(x)}.
\end{aligned}\]
The limit \eqref{mtb13} is prove. The statement for the Jost function $m_\pm(x;z)$ can be proved similarly. \qquad$\square$

\begin{proposition}\label{pro1}
If $w\in {H}^{1}(\mathbb{R})$, then
\begin{equation}
\label{eq26}
\sup\limits_{x\in \mathbb{R}}\left \|\int^{x}_{\pm\infty}\mathrm{e}^{\mathrm -iz(x-y)}w(y)\mathrm{d}y\right  \|_{{L}^{2}_{z}(\mathbb{R})}\leq\sqrt{2\pi}\parallel w\parallel_{{L}^{2}(\mathbb{R})}.
\end{equation}
\begin{equation}
\label{eq27}
\sup\limits_{x\in \mathbb{R}} \left\|w(x)-{i}z\int^{x}_{\pm\infty}\mathrm{e}^{-iz(x-y)}w(y)\mathrm{d}y\right\|_{{L}^{2}_{z}(\mathbb{R})}\leq\sqrt{2\pi}\| \partial_{x}w\|_{{L}^{2}(\mathbb{\mathbb{R}})}.
\end{equation}
In addition, if $w\in {L}^{2,1}(\mathbb{R})$, then, for every $x_{0}\in \mathbb{R}^{\pm}$,
\begin{equation}
\label{eq27a}
\sup\limits_{x\in (\pm\infty,x_0)}\left\|\langle x\rangle\int^{x}_{\pm\infty}\mathrm{e}^{-iz(x-y)}w(y)\mathrm{d}y\right\|_{{L}^{2}_{z}(\mathbb{R})}\leq\sqrt{2\pi}\| w\|_{{L}^{2,1}(\pm\infty,x_0)'}.
\end{equation}
where $\langle x\rangle=(1+x^{2})^{\frac{1}{2}}$ and $(\pm\infty,x_0)'$ is same as the notation defined in \eqref{mtb8a}.

The bounded statements about $\mathrm{e}^{iz(x-y)}$ for the case $-\infty$ have been given in \cite{imrn2018-5663}.
\end{proposition}
{\bf Proof.}~~ For every $x\in\mathbb{R}$ and every $z\in\mathbb{R}$, we write
\[f_\pm(x,z)=\int_{\pm\infty}^x\mathrm{e}^{-iz(x-y)}w(y)\mathrm{d}y
=\int_{\pm\infty}^0\mathrm{e}^{izy_1}w(y_1+x)\mathrm{d}y_1.\]
then
\[\begin{aligned}
\|f_\pm(x;z)\|_{L_z^2}^2&=2\pi\int_{\pm\infty}^0\mathrm{d}y_1\int_{\pm\infty}^0\mathrm{d}y_2\delta(y_1-y_2)w(y_1+x)\overline{w(y_2+x)}\\
&=\mp2\pi\int_{\pm\infty}^0|w(y+x)|^2\mathrm{d}y=\mp2\pi\int_{\pm\infty}^x|w(y)|^2\mathrm{d}y\\
&\leq2\pi\|w\|_{L^2}.
\end{aligned}\]
Thus the bound \eqref{eq26} is proved if $w\in L^2$.

For $y\leq x\leq0$ or $y\geq x\geq0$, then $(1+y^2)^{-1}\leq(1+x^2)^{-1}$, and we further have
\[|f_\pm(x;z)\|_{L_z^2}^2\leq\mp\frac{2\pi}{1+x^2}\int_{\pm\infty}^x(1+y^2)|w(y)|^2\mathrm{d}y\leq\frac{2\pi}{1+x^2}\|w\|^2_{L^{2,1}},\]
which proves the bound \eqref{eq27a} if $w\in H^1$.

Next, we prove the bound \eqref{eq27}. The condition $w\in H^1$ ensure that the Fourier transform is exit, and $w\to0$ as $|x|\to\infty$. In addition, we have $\|w\|_{L^\infty}\leq\frac{1}{\sqrt{2}}\|w\|_{H^1}$, or $w\in L^\infty$. Since
\[w(x)-iz\int_{\pm\infty}^x|w(y)|^2\mathrm{d}y=\int_{\pm\infty}^x\mathrm{e}^{-iz(x-y)}\partial_yw(y)\mathrm{d}y,\]
then we find, by a similar way, that
\[\left\|w(x)-iz\int^{x}_{\pm\infty}\mathrm{e}^{-iz(x-y)}w(y)\mathrm{d}y\right\|=\mp2\pi\int_{\pm\infty}^x|\partial_yw(y)|^2\mathrm{d}y,\]
which gives the bound \eqref{eq27}. \qquad $\square$

\begin{lemma}\label{lem3}
If $v\in {H}^{1,1}(\mathbb{R})$, then for every $x\in \mathbb{R}^{\pm}$, one gets
\begin{equation}
\label{eq28}
m_{\pm}(x;\cdot)-\mathrm{e}^{i\nu_\pm(x)}e_{1}\in {H}^{1}(\mathbb{R})   ,   \quad n_{\pm}(x;\cdot)-\mathrm{e}^{-i\nu_{\pm}(x)}e_{2}\in {H}^{1}(\mathbb{R}),
\end{equation}
Furthermore, if $v\in {H}^{1}(\mathbb{R})\cap{H}^{2}(\mathbb{R})$, then for every $x\in \mathbb{R}^{\pm}$
\begin{equation}
\label{eq29}
z[m_{\pm}(x;\cdot)-\mathrm{e}^{i\nu_\pm(x)}e_{1}]-\mathrm{e}^{i\nu_\pm(x)}[\bar\mu_\pm(x)e_1+2i\bar{v}_x(x)e_2]\in {L}^{2}_{z}(\mathbb{R}),
\end{equation}
\begin{equation}
\label{eq30}
z[n_{\pm}(x;\cdot)-\mathrm{e}^{-i\nu_{\pm}(x)}e_{2}]-\mathrm{e}^{-i\nu_\pm(x)}[-2i{v}_x(x)e_1+\mu_\pm(x)e_2]\in {L}^{2}_{z}(\mathbb{R}).
\end{equation}
\end{lemma}
{\bf Proof}~~ As an example, we only prove the statement for $n_\pm(x;z)$, and the proof for other Jost
 functions is analogous. To this end, we rewrite the integral equation \eqref{mtb7} as
\[n_\pm=e_2+{\cal{K}}_\pm n_\pm,\]
where the operator ${\cal{K}}_\pm$ is define in \eqref{mtb8}. It is verified that
\begin{equation}\label{mtb25}
(I-{\cal{K}}_\pm)[n_\pm(x;z)-\mathrm{e}^{-i\nu_\pm(x)}e_2]=h_\pm(x)e_1,
\end{equation}
where
\begin{equation}\label{mtb26}
 h_\pm(x)=2\int_{\pm\infty}^x \mathrm{e}^{-iz(x-y)}v_y(y)\mathrm{e}^{-i\nu_\pm(y)}\mathrm{d}y.
\end{equation}
If $v\in H^1(\mathbb{R})$, then from the estimate \eqref{eq26}, we know that
\begin{equation}\label{mtb27}
\sup\limits_{x\in\mathbb{R}}\|h_\pm(x;z)\|_{L_z^2(\mathbb{R})}\leq2\sqrt{2\pi}\|v_x\|_{L^2(\mathbb{R})}.
\end{equation}
and furthermore, for every $x_0\in\mathbb{R}^\pm$,
\begin{equation}\label{mtb27a}
\sup\limits_{x\in(\pm\infty,x_0)}\|\langle x\rangle h_\pm(x;z)\|_{L_z^2(\mathbb{R})}\leq2\sqrt{2\pi}\|v\|_{H^{1,1}(\mathbb{R})}.
\end{equation}
In addition, the proof of the Lemma \ref{lem1} implies that $(I-{\cal{K}}_\pm)^{-1}$ is exist on the space $L_x^\infty(\mathbb{R},L_z^2(\mathbb{R}))$, and is bounded
\begin{equation}\label{mtb28}
 \|(I-{\cal{K}}_\pm)^{-1}\|_{L_x^\infty L_z^2(\mathbb{R})\to L_x^\infty L_z^2(\mathbb{R})}\leq \mathrm{e}^{\|V_2\|_{L^1(\mathbb{R})}}.
\end{equation}
Hence, from \eqref{mtb27}, \eqref{mtb28} and \eqref{mtb25}, we find that
\begin{equation}\label{mtb30}
\sup\limits_{x\in\mathbb{R}}\|n_\pm(x;z)-\mathrm{e}^{-i\nu_\pm(x)}e_2\|_{L_z^2(\mathbb{R})}\leq2\sqrt{2\pi}\|v\|_{H^{1}(\mathbb{R})}\mathrm{e}^{\|V_2\|_{L^1(\mathbb{R})}},
\end{equation}
and for every $x_0\in\mathbb{R}^\pm$,
\begin{equation}\label{mtb31}
\sup\limits_{x\in(\pm\infty,x_0)}\big\|\langle x\rangle [n_\pm(x;z)-\mathrm{e}^{-i\nu_\pm(x)}e_2]\big\|_{L_z^2(\mathbb{R})}\leq2\sqrt{2\pi}\mathrm{e}^{\|V_2\|_{L^1(\mathbb{R})}}\|v\|_{H^{1,1}(\mathbb{R})},
\end{equation}
in terms of the estimate \eqref{mtb27a}.
It is noted that in $\|V_2\|_{L^1}$, $\|v\|_{L^2}\leq\|v\|_{L^{2,1}}$ and $\|v\|_{L^1}\leq\sqrt{\pi}\|v\|_{L^{2,1}}$.

To finish the proof the statement for $n_\pm$, one needs to consider bounded for $\partial_zn_\pm(x,z)$.
To this end, we introduce a vector
\begin{equation}\label{mtb32}
q_\pm(x;z)=\partial_zn_\pm(x;z)+\left(\begin{matrix}
ix&0\\
0&0
\end{matrix}\right)n_\pm(x;z),
\end{equation}
and consider
\begin{equation}\label{mtb33}
(I-{\cal{K}}_\pm)q_\pm(x;z)=[h_\pm^{(1)}(x;z)+h_\pm^{(2)}(x;z)]e_1+h_\pm^{(3)}(x;z)e_2,
\end{equation}
where
\begin{equation}\label{mtb34}
 \begin{aligned}
h_\pm^{(1)}(x;z)&=2i\int_{\pm\infty}^x\mathrm{e}^{-iz(x-y)}yv_y(y)[n_\pm^{(2)}(y;z)-\mathrm{e}^{-i\nu_\pm(y)}]\mathrm{d}y,\\
h_\pm^{(2)}(x;z)&=2i\int_{\pm\infty}^x\mathrm{e}^{-iz(x-y)}yv_y(y)\mathrm{e}^{-i\nu_\pm(y)}\mathrm{d}y,\\
h_\pm^{(3)}(x;z)&=\frac{1}{2}\int_{\pm\infty}^xy\bar{v}(y)n_\pm^{(1)}(y;z)\mathrm{d}y.
 \end{aligned}
\end{equation}
Now, for every $x_0\in\mathbb{R}^\pm$ and every $z\in \mathbb{C}^\pm$, we get that
$$
\begin{aligned}
\sup\limits_{x\in(\pm\infty,x_0)}\big\|h_\pm^{(1)}(x;z)\big\|_{L_z^2(\mathbb{R})}&=2\|v_x\|_{L^1(\mathbb{R})}
\sup\limits_{x\in(\pm\infty,x_0)}\big\|\langle x\rangle[n_\pm^{(2)}(x;z)-e^{-i\nu_\pm(x)}]\big\|_{L_z^2(\mathbb{R})} ,\\
\sup\limits_{x\in(\pm\infty,x_0)}\big\|h_\pm^{(3)}(x;z)\big\|_{L_z^2(\mathbb{R})}&=\frac{1}{2}\|v_x\|_{L^1(\mathbb{R})}
\sup\limits_{x\in(\pm\infty,x_0)}\|\langle x\rangle n_\pm^{(1)}(x;z)\|_{L_z^2(\mathbb{R})} ,\\
\sup\limits_{x\in(\pm\infty,x_0)}\big\|h_\pm^{(2)}(x;z)\big\|_{L_z^2(\mathbb{R})}&=2\sqrt{2\pi}\|v_x\|_{L^{2,1}(\mathbb{R})}  .
\end{aligned}$$
By virtue of the estimate \eqref{mtb31}, we know that $\sup\limits_{x\in(\pm\infty,x_0)}\big\|h_\pm^{(j)}(x;z)\big\|_{L_z^2(\mathbb{R})}, (j=1,2,3)$ are bounded. Furthermore, from equation \eqref{mtb33} and the boundedness of the operator $(1-{\cal{K}}_\pm)^{-1}$, we prove that, for every $x_0\in\mathbb{R}^\pm$, $q_\pm(x;z)\in L_x^\infty((\pm\infty,x_0),L_z^2(\mathbb{R}))$. Since $xn_\pm(x;z)$ is bounded in $L_x^\infty((\pm\infty,x_0),L_z^2(\mathbb{R}))$ according to the estimate \eqref{mtb31}, then from \eqref{mtb32}, we get that $\partial_zn_\pm(x;z)\in L_x^\infty((\pm\infty,x_0),L_z^2(\mathbb{R}))$. This prove the statement for $n_\pm(x;z)$ in \eqref{eq28}.

Next, we prove the estimate \eqref{eq30}. On use of \eqref{mtb25}, we find that
\begin{equation}\label{mtb35}
(1-{\cal{K}}_\pm)\big[z(n_\pm(x;a)-\mathrm{e}^{-i\nu_\pm(x)}e_2)-\mathrm{e}^{-i\nu_\pm(x)}(-2iv_xe_1+\mu_\pm(x)e_2)\big]
=g_\pm(x;z)e_1,
\end{equation}
where
\[\begin{aligned}
g_\pm(x;z)=&2i[v_x\mathrm{e}^{-i\nu_\pm(x)}-iz\int_{\pm\infty}^x\mathrm{e}^{-iz(x-y)}v_y(y)\mathrm{e}^{-i\nu_\pm(y)}\mathrm{d}y]\\
&+\int_{\pm\infty}^x\mathrm{e}^{-iz(x-y)}v_y(y)\mathrm{e}^{-i\nu_\pm(y)}(|v(y)|^2+2\mu_\pm(y))\mathrm{d}y.
\end{aligned}\]
Here the identity
\[n_\pm^{(2)}(x;z)=1+\frac{i}{2}\int_\pm^x\big(\bar{v}(y)n_\pm^{(1)}(y;z)-|v(y)|^2n_\pm^{(2)}(y;z)\big)\mathrm{d}y,\]
is used.

Note that $v\in H^1(\mathbb{R})$ implies that $v\in L^\infty(\mathbb{R})$, then
\[|\mu_\pm|\leq\int_{\pm\infty}^x|v||v_y|\mathrm{d}y\leq\|v\|_{L^2(\mathbb{R})}\|v_x\|_{L^2(\mathbb{R})},\]
\[|v|^2=\int_{\pm\infty}^xv_y\bar{v}+v\bar{v}_y\mathrm{d}y\leq2\|v\|_{L^2(\mathbb{R})}\|v_x\|_{L^2(\mathbb{R})}.\]
From the estimate \eqref{eq27}, we have
\[\begin{aligned}
&\sup\limits_{x\in\mathbb{R}}\bigg\|\int_{\pm\infty}^x\mathrm{e}^{-iz(x-y)}v_y(y)\mathrm{e}^{-i\nu_\pm(y)}(|v(y)|^2+2\mu_\pm(y))\mathrm{d}y\bigg\|_{L_z^2(\mathbb{R})}\\
&\leq\sqrt{2\pi}\||v(y)|^2+2\mu_\pm(y)\|_{L^\infty(\mathbb{R})}\|v_{x}\|_{L^2(\mathbb{R})}\\
&\leq4\sqrt{2\pi}\|v\|_{L^2(\mathbb{R})}\|v_x\|_{L^2(\mathbb{R})}^2,
\end{aligned}\]
and
\[\begin{aligned}
&\sup\limits_{x\in\mathbb{R}}\bigg\|v_x\mathrm{e}^{-i\nu_\pm(x)}-iz\int_{\pm\infty}^x\mathrm{e}^{-iz(x-y)}v_y(y)\mathrm{e}^{-i\nu_\pm(y)}\mathrm{d}y\bigg\|_{L_z^2(\mathbb{R})}\\
&\leq\sqrt{2\pi}\big(\|v_{xx}\|_{L^2(\mathbb{R})}+\|v\|_{L^2(\mathbb{R})}\|v_x\|_{L^2(\mathbb{R})}^2\big).
\end{aligned}\]
By using of the triangle inequality, we get
\[\sup\limits_{x\in\mathbb{R}}\|g_\pm(x;z)\|_{L_z^2(\mathbb{R})}\leq\sqrt{2\pi}\big[\|v_{xx}\|_{L^2(\mathbb{R})}+5\|v\|_{L^2(\mathbb{R})}\|v_x\|_{L^2(\mathbb{R})}^2\big].\]
As a result, using the condition together with the bound of the operator $(1-{\cal{K}}_\pm)^{-1}$, we prove the estimate \eqref{eq30}. \qquad$\square$

\begin{lemma}\label{lem4}
The maps
\begin{equation}\label{mtb36}
 H^{1,1}(\mathbb{R})\ni v\mapsto[m_\pm(x;z)-\mathrm{e}^{i\nu_\pm(x)}e_1,n_\pm(x;z)-\mathrm{e}^{-i\nu_\pm(x)}e_2]\in L^\infty(\mathbb{R}^\pm;H_z^1(\mathbb{R})),
\end{equation}
and
\begin{equation}\label{mtb36a}
 H^{1}(\mathbb{R})\cap H^2\ni v\mapsto[\hat{m}_\pm(x;z),\hat{n}_\pm(x;z)]\in L^\infty(\mathbb{R}^\pm;L_z^2(\mathbb{R})),
\end{equation}
are Lipschitz continuous, where
\[\begin{aligned}
\hat{m}_\pm(x;z)=z[m_\pm(x;z)-\mathrm{e}^{i\nu_\pm(x)}e_1]-\mathrm{e}^{i\nu_\pm(x)}[\bar\mu_\pm(x)e_1+2i\bar{v}_x(x)e_2],\\
\hat{n}_\pm(x;z)=z[n_\pm(x;z)-\mathrm{e}^{-i\nu_\pm(x)}e_2]-\mathrm{e}^{-i\nu_\pm(x)}[-2i{v}_x(x)e_1+\mu_\pm(x)e_2].
\end{aligned}\]
\end{lemma}
{\bf Proof}~~ Let $v,\tilde{v}\in H^{1,1}(\mathbb{R})$ satisfy the condition $\|v\|_{H^{1,1}},\|\tilde{v}\|_{H^{1,1}}\leq B$ for some $B>0$. Let the associated Jost functions are $(m_\pm,n_\pm)$ and $(\tilde{m}_\pm,\tilde{n}_\pm)$, respectively.
From \eqref{mtb25}, we consider
\begin{equation}\label{mtb37}
\begin{aligned}
& (n_\pm-\mathrm{e}^{-i\nu_\pm}e_2)-(\tilde{n}_\pm-\mathrm{e}^{-i\tilde\nu_\pm}e_2)\\
&=(I-{\cal{K}}_\pm)^{-1}(h_\pm-\tilde{h}_\pm)e_1+(I-{\cal{K}}_\pm)^{-1}({\cal{K}}_\pm-{\cal{\tilde{K}}}_\pm)(I-{\cal{\tilde{K}}}_\pm)^{-1}\tilde{h}_\pm e_1.
\end{aligned}
\end{equation}
Note that
\[\begin{aligned}
h_\pm-\tilde{h}_\pm=&2\int_{\pm\infty}^x\mathrm{e}^{-iz(x-y)}\mathrm{e}^{-i\nu_\pm(y)}(v_y-\tilde{v}_y)\mathrm{d}y\\
&+2\int_{\pm\infty}^x\mathrm{e}^{-iz(x-y)}\tilde{v}_y(\mathrm{e}^{-i\nu_\pm(y)}-\mathrm{e}^{-i\tilde\nu_\pm(y)})\mathrm{d}y.
\end{aligned}\]
On using the inequality $|\mathrm{e}^z-1|\leq \mathrm{e}^{|z|}-1\leq|z|\mathrm{e}^{|z|}$ obtained by Taylor series, we have
\[\begin{aligned}
|\mathrm{e}^{-i\nu_\pm(y)}-\mathrm{e}^{-i\tilde\nu_\pm(y)}|&=|\mathrm{e}^{-i(\nu_\pm(y)-\tilde\nu_\pm(y))}-1|\\
&\leq|\nu_\pm-\tilde\nu_\pm|\mathrm{e}^{|\nu_\pm-\tilde\nu_\pm|}=\frac{1}{2}\big|\int_{\pm\infty}^x|v|^2-|\tilde{v}|^2\mathrm{d}y\big|\mathrm{e}^{\frac{1}{2}(\|v\|_{L^2}+\|\tilde{v}\|_{L^2})}\\
&\leq\frac{1}{2} \mathrm{e}^{\frac{1}{2}(\|v\|_{L^2}+\|\tilde{v}\|_{L^2})}\bigg(\big|\int_{\pm\infty}^xv(\bar{v}-\tilde{\bar{v}})\mathrm{d}y\big|+\big|\int_{\pm\infty}^x({v}-\tilde{v})\tilde{\bar{v}}\mathrm{d}y\big|\bigg),
\end{aligned}\]
which implies that
\begin{equation}\label{mtb38}
\|\mathrm{e}^{-i\nu_\pm(x)}-\mathrm{e}^{-i\tilde\nu_\pm(x)}\|_{L^\infty}\leq \frac{1}{2} \mathrm{e}^{\frac{1}{2}(\|v\|_{L^2}+\|\tilde{v}\|_{L^2})}(\|v\|_{L^2}+\|\tilde{v}\|_{L^2})\|v-\tilde{v}\|_{L^2}.
\end{equation}

By virtue of the bound \eqref{eq26}, we have
\[\begin{aligned}
\sup\limits_{x\in\mathbb{R}}\|h_\pm-\tilde{h}_\pm\|_{L_z^2(\mathbb{R})}&\leq\sqrt{2\pi}\big[2\|v_x-\tilde{v}_x\|_{L^2}+\mathrm{e}^{\|v\|_{L^2}+\|\tilde{v}\|_{L^2}}\|\tilde{v}_x\|_{L^2}(\|v\|_{L^2}+\|\tilde{v}\|_{L^2})\|v-\tilde{v}\|_{L^2}\big]\\
&\leq C(B)\|v-\tilde{v}\|_{H^1(\mathbb{R})},
\end{aligned}\]
where $C(B)$ is a $B$-dependent positive constant.

In addition, for every $f\in L^\infty$, we find that
\[({\cal{K}}_\pm-{\cal{\tilde{K}}}_\pm)f=\frac{i}{2}\int_{\pm\infty}^x\left(\begin{matrix}
\mathrm{e}^{-iz(x-y)}&0\\
0&1
\end{matrix}\right)\left(\begin{matrix}
|v|^2-|\tilde{v}|^2&-4i(v_y-\tilde{v}_y)\\
\bar{v}-\tilde{\bar{v}}&|\tilde{v}|^2-|v|^2
\end{matrix}\right)f(y;z)\mathrm{d}y,\]
which implies that
\[\sup\limits_{x\in\mathbb{R}}\|({\cal{K}}_\pm-{\cal{\tilde{K}}}_\pm)f\|_{L_z^2(\mathbb{R})}\leq C_1(B)\|v-\tilde{v}\|_{H^1(\mathbb{R})}\|f\|_{L^\infty(\mathbb{R})},\]
where $C_1(B)$ is a $B$-dependent positive constant.

Thus, on use of the estimate \eqref{mtb27} and the bound of the operator $(I-{\cal{K}}_\pm)^{-1}$, we can get that
\begin{equation}\label{mtb39}
\sup\limits_{x\in\mathbb{R}}\|(n_\pm-\mathrm{e}^{-i\nu_\pm}e_2)-(\tilde{n}_\pm-\mathrm{e}^{-i\tilde\nu_\pm}e_2)\|_{L_z^2(\mathbb{R})}\leq C(B)\|v-\tilde{v}\|_{H^1(\mathbb{R})}.
\end{equation}
Furthermore, for every $x_0\in \mathbb{R}^\pm$,
\begin{equation}\label{mtb40}
\sup\limits_{x\in(\pm\infty,x_0)}\|\langle x\rangle[(n_\pm-\mathrm{e}^{-i\nu_\pm}e_2)-(\tilde{n}_\pm-\mathrm{e}^{-i\tilde\nu_\pm}e_2)]\|_{L_z^2(\mathbb{R})}\leq C(B)\|v-\tilde{v}\|_{H^{1,1}(\mathbb{R})}.
\end{equation}

To finish the proof of \eqref{mtb36}, we need to estimate $\partial_zn_\pm-\partial_z\tilde{n}_\pm$. It is noted that
\[\begin{aligned}
\partial_zn_\pm-\partial_z\tilde{n}_\pm=q_\pm-\tilde{q}_\pm-\left(\begin{matrix}
ix&0\\
0&0
\end{matrix}\right)\big[(n_\pm-e^{-i\nu_\pm}e_2)-(\tilde{n}_\pm-e^{-i\tilde\nu_\pm}e_2)\big].
\end{aligned}\]
in terms of \eqref{mtb32}. Now from \eqref{mtb33}, we have
\[\begin{aligned}
q_\pm-\tilde{q}_\pm=&(I-{\cal{K}}_\pm)^{-1}[(h_\pm^{(1)}-\tilde{h}_\pm^{(1)})e_1+(h_\pm^{(2)}-\tilde{h}_\pm^{(2)})e_1+(h_\pm^{(3)}-\tilde{h}_\pm^{(3)})e_2]\\
&+(I-{\cal{K}}_\pm)^{-1}({\cal{K}}_\pm-{\cal{\tilde{K}}}_\pm)(I-{\cal{\tilde{K}}}_\pm)^{-1}[\tilde{h}_\pm^{(1)})e_1+\tilde{h}_\pm^{(2)})e_1+\tilde{h}_\pm^{(3)})e_2].
\end{aligned}\]
For every $x_0\in\mathbb{R}^\pm$, by using \eqref{mtb40} and \eqref{mtb31}, we can get
\[\begin{aligned}
\sup\limits_{x\in(\pm\infty,x_0)}\|h_\pm^{(1)}-\tilde{h}_\pm^{(1)}\|_{L_z^2(\mathbb{R})}&\leq2\sqrt{2\pi}\|v_x-\tilde{v}_x\|_{L^1(\mathbb{R})}\sup\limits_{x\in(\pm\infty,x_0)}\|\langle x\rangle(n_\pm^{(2)}-e^{-i\nu_\pm})\|_{L_z^2(\mathbb{R})}\\
&+2\sqrt{2\pi}\|\tilde{v}_x\|_{L^1(\mathbb{R})}\sup\limits_{x\in(\pm\infty,x_0)}\|\langle x\rangle[(n_\pm^{(2)}-\mathrm{e}^{-i\nu_\pm})-(\tilde{n}_\pm^{(2)}-\mathrm{e}^{-i\tilde\nu_\pm})]\|_{L_z^2(\mathbb{R})}\\
&\leq C_3(B)\|v-\tilde{v}\|_{H^{1,1}(\mathbb{R})}.
\end{aligned}\]
Similarly
\[\sup\limits_{x\in(\pm\infty,x_0)}\|h_\pm^{(2)}-\tilde{h}_\pm^{(2)}\|_{L_z^2(\mathbb{R})}\leq C_4(B)\|v-\tilde{v}\|_{H^{1,1}(\mathbb{R})},\]
\[\sup\limits_{x\in(\pm\infty,x_0)}\|h_\pm^{(3)}-\tilde{h}_\pm^{(3)}\|_{L_z^2(\mathbb{R})}\leq C_5(B)\|v-\tilde{v}\|_{H^{1,1}(\mathbb{R})}.\]
Thus, for every $x_0\in\mathbb{R}^\pm$
\[\sup\limits_{x\in(\pm\infty,x_0)}\|q_\pm-\tilde{q}_\pm\|_{L_z^2(\mathbb{R})}\leq C_6(B)\|v-\tilde{v}\|_{H^{1,1}(\mathbb{R})},\]
which, together with the estimate \eqref{mtb40}, implies that
\begin{equation}\label{mtb41}
 \sup\limits_{x\in(\pm\infty,x_0)}\|\partial_zn_\pm-\partial_z\tilde{n}_\pm\|_{L_z^2(\mathbb{R})}\leq C_7(B)\|v-\tilde{v}\|_{H^{1,1}(\mathbb{R})}.
\end{equation}

As a result, we obtain that
\begin{equation}\label{mtb42}
 \sup\limits_{x\in(\pm\infty,x_0)}\|(n_\pm-\mathrm{e}^{-i\nu_+}e_2)-(\tilde{n}_\pm-\mathrm{e}^{-i\tilde\nu_+}e_2)\|_{H_z^1(\mathbb{R})}\leq C_8(B)\|v-\tilde{v}\|_{H^{1,1}(\mathbb{R})}.
\end{equation}
Here $C_j(B), j=2,\cdots,8$ are $B$-dependent positive constants.

A similar procedure applied to $m_\pm(x;z)$ and $\tilde{m}_\pm(x;z)$ gives rise to a Lipshitz condition as \eqref{mtb42}. So the Lipshitz continous \eqref{mtb36} is proved. The Lipshitz continous \eqref{mtb37} can be proved similarly. \qquad $\square$

Next, We consider the Jsot functions $\varphi(x;k)$ and $\phi_\pm(x;k)$ of the spectral problem \eqref{mta2}, and give the following definition by using the Jost function $m_\pm(x;z)$ and $n_\pm(x;z)$
\begin{equation}\label{mtb43}
\varphi_\pm(x;k)=T_1^{-1}(x;k)m_\pm(x;z), \quad
\phi_\pm(x;k)=T_2^{-1}(x;k)n_\pm(x;z),
\end{equation}
where $T_j, 1=1,2$ are given in \eqref{mtb1}. Then they admit the boundary condition
\begin{equation}\label{mtb44}
 \varphi_\pm(x;k)\to e_1, \quad \phi_\pm(x;k)\to e_2, \quad x\to\pm\infty.
\end{equation}
It is readily verified that the Jost functions $\varphi_-(x;\cdot), \phi_+(x;\cdot)$ are analytic in the first and third quadrant of the $k$ plane (that is ${\rm Im} k^2>0$). In addition, the Jost functions $\varphi_+(x;\cdot), \phi_-(x;\cdot)$  are analytic in the second and fourth quadrant of the $k$ plane (where ${\rm Im} k^2>0$).

For the one-order linear equation \eqref{mta2}, two set of Jost functions $\varphi_-,\phi_-$ and $\varphi_+,\phi_+$ are connected by the so-called scattering coefficients, for every $x\in \mathbb{R}$ and every $k^2\in\mathbb{R}\setminus\{0\}$,
\begin{equation}\label{mtb45}
\begin{aligned}
\varphi_-(x;k)=a(k)\varphi_+(x;k)+b(k)\mathrm{e}^{ik^2x}\phi_+(x;k),\\
\phi_-(x;k)=c(k)\mathrm{e}^{-ik^2x}\varphi_+(x;k)+d(k)\phi_+(x;k).
\end{aligned}
\end{equation}
For two columns $\varphi,\phi$, we define the Wronsky determinant $W[\varphi,\phi]=\det(\varphi,\phi)$. If $\varphi,\phi$ are vector solution of the linear spectral problemt \eqref{mta2}, then their Wronsky determinant is independent of the variable $x$. Hence, for the Jost functions $\varphi_-,\phi_-$ and $\varphi_+,\phi_+$, we have
\begin{equation}\label{mtb46}
 W[\varphi_\pm(x;k),\phi_\pm(x;k)]=1,
\end{equation}
in view of the boundary condition \eqref{mtb44}.

It is noted that the two sets of Jost functions $\varphi_\pm(x;k)$ and $\phi_\pm(x;k)$ admit the following symmetry condition
\begin{equation}\label{mtb47}
\phi_\pm(x;k)=\sigma_1\sigma_3\overline{\varphi_\pm(x;-\bar{k})},
\end{equation}
where $\sigma_1=\left(\begin{matrix}
0&1\\
1&0
\end{matrix}\right)$ and $\sigma_3$ are Pauli matrices. From \eqref{mtb45},\eqref{mtb46} and \eqref{mtb47}, we find that
\begin{equation}\label{mtb48}
a(k)=W[\varphi_-(x;k)\mathrm{e}^{-\frac{i}{2}k^2x},\phi_+(x;k)\mathrm{e}^{\frac{i}{2}k^2x}]=W[\varphi_-(0;k),\phi_+(0;k)],
\end{equation}
\begin{equation}\label{mtb48a}
b(k)=W[\varphi_+(x;k)\mathrm{e}^{-\frac{i}{2}k^2x},\varphi_-(x;k)\mathrm{e}^{-\frac{i}{2}k^2x}]=W[\varphi_+(0;k),\varphi_-(0;k)],
\end{equation}
and
\[d(k)=\overline{a(-\bar{k})}, \quad c(k)=-\overline{b(-\bar{k})},\]
\begin{equation}\label{mtb49}
 a(k)\overline{a(-\bar{k})}+b(k)\overline{b(-\bar{k})}=1.
\end{equation}

\begin{lemma}\label{lem5}
If $v\in H^{1,1}(\mathbb{R})$, then the function $a(k)$ is continued analytically in $\mathbb{C}^+$ with respect to $z=k^2$. In addition,
\begin{equation}\label{mtb50}
a(k)-e^{i\nu}, \quad kb(k), \quad k^{-1}b(k)\in H_z^1(\mathbb{R}),
\end{equation}
where $\nu=\frac{1}{2}\int_{-\infty}^\infty|v(y)|^2\mathrm{d}y=\nu_-(x)-\nu_+(x)$. Moreover, if $v\in H^2(\mathbb{R})\cap H^{1,1}(\mathbb{R})$, then
\begin{equation}\label{mtb51}
kb(k), \quad k^{-1}b(k)\in L_z^{2,1}(\mathbb{R}).
\end{equation}
\end{lemma}
{\bf Proof}~~ The transformation \eqref{mtb43} and the Volterra integral equation \eqref{mtb6} and \eqref{mtb7} imply that
\begin{equation}
\label{eq44}
\varphi_{\pm}(x;k)=e_{1}-\frac{i}{2}\int^{x}_{\pm\infty}\left(\begin{matrix}
1   &  0\\
0   &   \mathrm{e}^{ik^{2}(x-y)}
\end{matrix}\right)[2kV(y)-V^2(y)\sigma_3]\varphi_{\pm}(y;k)\mathrm{d}y,
\end{equation}
\begin{equation}
\label{eq45}
\phi_{\pm}(x;k)=e_{2}-\frac{i}{2}\int^{x}_{\pm\infty}\left(\begin{matrix}
 \mathrm{e}^{-ik^{2}(x-y)} &  0\\
0   &   1
\end{matrix}\right)[2kV(y)-V^2(y)\sigma_3]\phi_{\pm}(y;k)\mathrm{d}y.
\end{equation}
Substituting them into \eqref{mtb45}, we can get that
\begin{equation}\label{mtb52}
a(k)=1-\frac{i}{2}\int_{-\infty}^\infty|v(x)|^2\varphi_-^{(1)}(x;k)+2kv(x)\varphi_-^{(2)}(x;k)\mathrm{d}x,
\end{equation}
\begin{equation}\label{mtb52a}
b(k)=\frac{i}{2}\int_{-\infty}^\infty\mathrm{e}^{-ik^2x}[2k\bar{v}(x)\varphi_-^{(1)}(x;k)+|v(x)|^2\varphi_-^{(2)}(x;k)]\mathrm{d}x,
\end{equation}
where $\varphi_\pm=(\varphi_\pm^{(1)},\varphi_\pm^{(1)})^T$ and $\phi_\pm=(\phi_\pm^{(1)},\phi_\pm^{(1)})^T$.
Note that equation \eqref{mtb43} gives
\begin{equation}\label{mtb44a}
\begin{aligned}
\varphi^{(1)}_{\pm}(x;k)=m^{(1)}_{\pm}(x;z),\quad \varphi^{(2)}_{\pm}(x;k)=\frac{1}{2k}[-2\overline{v}(x)m^{(1)}_{\pm}(x;z)+m^{(2)}_{\pm}(x;z)], \\
\phi^{(1)}_{\pm}(x;k)=\frac{1}{2k}[n^{(1)}_{\pm}(x;z)-2v(x)n^{(2)}_{\pm}(x;z)], \quad\phi^{(2)}_{\pm}(x;k)=n^{(2)}_{\pm}(x;z),
\end{aligned}
\end{equation}
which imply that $\varphi^{(1)}_{\pm}(x;k)$ and $\phi^{(2)}_{\pm}(x;k)$ are even in $k$, $\varphi^{(2)}_{\pm}(x;k)$  and $\phi^{(1)}_{\pm}(x;k)$ are odd in $k$. Substituting \eqref{mtb44a} into \eqref{mtb52} and considering the asymptotic behaviors \eqref{mtb10} of the Jost functions $m_-(x;k)$, we have
\begin{equation}\label{mtb57}
\lim\limits_{k\to\infty}a(k)=1+\mathrm{e}^{i\nu_-(x)}|_{x=-\infty}^\infty=\mathrm{e}^{i\nu}.
\end{equation}

Now, we prove $a(k)-\mathrm{e}^{i\nu}\in H_z^1(\mathbb{R})$. To this end, by virtue of \eqref{mtb48} and \eqref{mtb44a}, we consider
\begin{equation}\label{mtb58}
\begin{aligned}
a(k)-\mathrm{e}^{i\nu}&=m_-^{(1)}(0;z)n_+^{(2)}(0;z)-\mathrm{e}^{i\nu_-(0)}\mathrm{e}^{-i\nu_+(0)}-\varphi_-^{(2)}(0;k)\phi_+^{(1)}(0;k)\\
&=[m_-^{(1)}(0;z)-\mathrm{e}^{i\nu_-(0)}][n_+^{(2)}(0;z)-\mathrm{e}^{-i\nu_+(0)}]+\mathrm{e}^{i\nu_-(0)}[n_+^{(2)}(0;z)-\mathrm{e}^{-i\nu_+(0)}]\\
&\quad+\mathrm{e}^{-i\nu_+(0)}[m_-^{(1)}(0;z)-\mathrm{e}^{i\nu_-(0)}]-k^{-1}\varphi_-^{(2)}(0;k)v(0)\mathrm{e}^{-i\nu_+(0)}\\
&\quad+k^{-1}\varphi_-^{(2)}(0;k)[k\phi_+^{(1)}(0;k)+v(0)\mathrm{e}^{-i\nu_+(0)}].
\end{aligned}
\end{equation}
From the representation of $\phi_+^{(1)}$ in \eqref{mtb44a}, we know that, for every $x\in\mathbb{R}^+$,
\begin{equation}\label{mtb59}
 2[k\phi_+^{(1)}(x;k)+v(x)\mathrm{e}^{-i\nu_+(x)}]
=n^{(1)}_{+}(x;z)-2v(x)[n^{(2)}_{+}(x;z)-\mathrm{e}^{-i\nu_+(x)}]\in H_z^1(\mathbb{R}),
\end{equation}
in view of the property of $n_\pm(x;z)$ in \eqref{eq28}. In addition, from the Volterra integral equation of $\varphi_-$ and \eqref{mtb44a}, we get that
\begin{equation}\label{mtb60}
\begin{aligned}
&k^{-1}\varphi_-^{(2)}(x;k)-\frac{i}{2}\int_{-\infty}^x\mathrm{e}^{ik^2(x-y)}|v(y)|^2k^{-1}\varphi_-^{(2)}(y;k)\mathrm{d}y\\
&=i\int_{-\infty}^x\mathrm{e}^{iz(x-y)} \bar{v}(y)[m_-^{(1)}(y,z)-\mathrm{e}^{i\nu_+(y)}]\mathrm{d}y
+i\int_{-\infty}^x\mathrm{e}^{iz(x-y)} \bar{v}(y)\mathrm{e}^{i\nu_+(y)}\mathrm{d}y,
\end{aligned}
\end{equation}
If we introduce an operator $\mathcal{T}$ for every function $f\in L^\infty(\mathbb{R})$
\[{\mathcal{T}}f=\frac{i}{2}\int_{-\infty}^x\mathrm{e}^{ik^2(x-y)}|v(y)|^2f(y)\mathrm{d}y,\]
then $(I-{\mathcal{T}})^{-1}$ is exist and bounded for $(\|v\|_{L^2})/2<1$. Thus
\eqref{mtb60} can be rewritten as
\[\begin{aligned}
k^{-1}\varphi_-^{(2)}(x;k)&=(I-\mathcal{T})^{-1}i\int_{-\infty}^x\mathrm{e}^{iz(x-y)} \bar{v}(y)\mathrm{e}^{i\nu_+(y)}\mathrm{d}y\\
&\quad+(I-\mathcal{T})^{-1}\big[i\int_{-\infty}^x\mathrm{e}^{iz(x-y)} \bar{v}(y)[m_-^{(1)}(y,z)-\mathrm{e}^{i\nu_+(y)}]\mathrm{d}y
\big],
\end{aligned}\]
which implies that $k^{-1}\varphi_-^{(2)}(x;k)\in H_z^1(\mathbb{R})$ by a similar discussion as in Lemma \ref{lem3}.
Again using the statements \eqref{eq28} in Lemma \ref{lem3} together with this fact and \eqref{mtb59}, we prove, from
\eqref{mtb58}, that $a(k)-\mathrm{e}^{i\nu}\in H_z^1(\mathbb{R})$.

Moreover, come to consider $kb(k)$ and $k^{-1}b(k)$. we note, from \eqref{mtb48a}, \eqref{mtb44a} and \eqref{eq28}, that
for $v\in H^{1,1}(\mathbb{R})$,
\[\begin{aligned}
2kb(k)=&[m_+^{(1)}(0;z)-\mathrm{e}^{i\nu_+(0)}]m_-^{(2)}(0;z)-m_+^{(2)}(0;z)[m_-^{(1)}(0;z)-\mathrm{e}^{i\nu_-(0)}]\\
&+\mathrm{e}^{i\nu_+(0)}m_-^{(2)}(0;z)-\mathrm{e}^{i\nu_-(0)}m_+^{(2)}(0;z)\in H_z^1(\mathbb{R}),
\end{aligned}\]
and
\[k^{-1}b(k)=m_+^{(1)}(0;z)k^{-1}\varphi_-^{(2)}(0;k)-m_-^{(1)}(0;z)k^{-1}\varphi_+^{(2)}(0;k)\in H_z^1(\mathbb{R}).\]
This finish the proof of \eqref{mtb50}.

To prove the statement \eqref{mtb51}, we rewrite $2kzb(k)$ as
\[\begin{aligned}
2zkb(k)
&=m_+^{(1)}(0;z)[zm_-^{(2)}(0;z)-2i\bar{v}_x(0)\mathrm{e}^{i\nu_-(0)}]\\
&\quad+2i\bar{v}_x(0)\mathrm{e}^{i\nu_-(0)}[m_+^{(1)}(0;z)-\mathrm{e}^{i\nu_+(0)}]\\
&\quad-m_-^{(1)}(0;z)[zm_+^{(2)}(0;z)-2i\bar{v}_x(0)\mathrm{e}^{i\nu_+(0)}]\\
&\quad-2i\bar{v}_x(0)\mathrm{e}^{i\nu_+(0)}[m_-^{(1)}(0;z)-\mathrm{e}^{i\nu_-(0)}]\in L_z^2(\mathbb{R}).
\end{aligned}\]
which implies that, for every $v\in H^2\cap H^{1,1}$, $kb(k)\in L_z^{2,1}(\mathbb{R})$. $k^{-1}b(k)\in L_z^{2,1}(\mathbb{R})$ can be proved similarly. The Lemma \ref{lem5} is proved. \qquad $\square$

\begin{lemma}\label{lem6}
The map
\begin{equation}\label{mtb61}
H^{1,1}(\mathbb{R})\ni v\mapsto a(k)-\mathrm{e}^{i\nu}, \quad kb(k), \quad k^{-1}b(k)\in H_z^1(\mathbb{\mathbb{R}}),
\end{equation}
and
\begin{equation}\label{mtb62}
H^2(\mathbb{R})\cap H^{1,1}(\mathbb{R})\ni v\mapsto kb(k), \quad k^{-1}b(k)\in L_z^{2,1}(\mathbb{R}),
\end{equation}
are Lipshitz continuous.
\end{lemma}
{\bf Proof}~~ Let $a(k), b(k)$ and $\tilde{a}(k), \tilde{b}(k)$ are the associated scattering coefficients with $v$ and $\tilde{v}$. Using the results in the proof of the Lemma \ref{lem5} and Lemma \ref{lem4}, we can get
\begin{equation}
\label{eq66}
\begin{aligned}
&\|a(k)-\mathrm{e}^{i\nu}-\tilde{a}(k)+\mathrm{e}^{i\tilde\nu}\|_{H^{1}_{z}(\mathbb{R})}\\
&+\| kb(k)-k\tilde{b}(k)\|_{H^{1}_{z}(\mathbb{R})}+\| k^{-1}b(k)-k^{-1}\tilde{b}(k)\|_{H^{1}_{z}(\mathbb{R})}\\
&\leq C(B)\| v-\tilde{v}\|_{H^{1.1}}.
\end{aligned}
\end{equation}
In addition, we have
\begin{equation}
\label{eq66a}
\| kb(k)-k\tilde{b}(k)\|_{L^{2,1}_{z}(\mathbb{R})}+\| k^{-1}b(k)-k^{-1}\tilde{b}(k)\|_{L^{2,1}_{z}(\mathbb{R})}\leq C(B)\| v-\tilde{v}\|_{H^{1,1}(\mathbb{R})\cap H^{2}(\mathbb{R})}.
\end{equation}
Here $C(B)$ is $B$-dependent positive constant. \qquad $\square$

\begin{lemma}\label{lem7}
The scattering coefficients $a(k)$ and $b(k)$ are even and odd function in $k\in \mathbb{R}\cup i\mathbb{R}$, respectively.
Moreover,
\begin{equation}
\label{eq67}
\begin{aligned}
&| a(k)|^{2}-| b(k)|^{2}=1 \quad k\in \mathbb{R},\\
&|a(k)|^{2}+| b(k)|^{2}=1 \quad k\in i\mathbb{R}.
\end{aligned}
\end{equation}
\end{lemma}
{\bf Proof}~~ Since, $\varphi^{(1)}_{\pm}(x;k)$ and $\phi^{(2)}_{\pm}(x;k)$ are even in $k$, $\varphi^{(2)}_{\pm}(x;k)$  and $\phi^{(1)}_{\pm}(x;k)$ are odd in $k$, equations \eqref{mtb48} and \eqref{mtb48a} imply that $a(k)$ is a even function and $b(k)$ is odd. From \eqref{mtb49}, we can get the second equation in \eqref{eq67} for $k\in i\mathbb{R}$. For $k\in \mathbb{R}$, equation \eqref{mtb49} takes the form $a(k)\overline{a(-{k})}+b(k)\overline{b(-{k})}=1$, which implies the first equation in \eqref{eq67}.  \qquad $\square$

\setcounter{equation}{0}
\section{Riemann-Hilbert problems}\label{sec3}
We rewrite integral equation \eqref{mtb6} as a compact form $m_\pm=e_1+\tilde{{\cal{K}}}_\pm m_\pm$, where
\[\tilde{{\cal{K}}}_\pm f(x)=\int_{\pm\infty}^x\left(\begin{matrix}
 1&0\\
 0&\mathrm{e}^{iz(x-y)}
 \end{matrix}\right)V_1(y)f(y)\mathrm{d}y,\]
for every $f\in L^\infty(\mathbb{R})$, where $V_1$ is given in \eqref{mtb2}. It is verified that $I-\tilde{{\cal{K}}}$ is inverse and for every $z\in \mathbb{C}^{\mp}$,
\[\|(I-\tilde{{\cal{K}}}_\pm)^{-1}\|_{L_x^\infty L_z^2\to L_x^\infty L_z^2}\leq \mathrm{e}^{\|V_1\|_{L^1}},\]
if $v\in L^2(\mathbb{R})\cap L^2(\mathbb{R}), v_x\in L^1(\mathbb{R})$.
In addition, we write the equations \eqref{mtb52} and \eqref{mtb44a} as
\[\begin{aligned}
a(k)-1&=\frac{i}{2}\int_{-\infty}^\infty\big[|v(x)|^2m_-^{(1)}(x;z)-v(x)m_-^{(2)}(x;z)\big]\mathrm{d}x\\
&:=\tilde{\cal{T}}m_-(x;z)=\tilde{\cal{T}}(I-\tilde{{\cal{K}}}_-)^{-1}e_1.
\end{aligned}\]
Thus for every $v\in L^2(\mathbb{R})\cap L^1(\mathbb{R}), v_x\in L^1(\mathbb{R})$, we have
\[|a(k)-1|\leq\frac{1}{2}(\|v\|_{L^1}+\|v\|_{L^2}^2)\mathrm{e}^{\|V_1\|_{L^1}}.\]
The continuity of $a(k)$ implies that
\begin{equation}\label{mtc1}
|a(k)|\geq 1-\frac{1}{2}(\|v\|_{L^1}+\|v\|_{L^2}^2)\mathrm{e}^{\|V_1\|_{L^1}}>0,
\end{equation}
for small enough $v$. Hence, for small enough $v$, $a(k)$ has no zero in $k$( or $z$ in view of that $a(k)$ is even) space.

According to the analyticity  of the scattering coefficient $a(k)$ and the Jost functions $\varphi_\pm(x;k)$ and $\phi_\pm(x;k)$, we define two sectionally analytic functions
\begin{equation}\label{mtc2}
\Psi_+(x;k)=\bigg(\frac{\varphi_-(x;k)}{a(k)},\phi_+(x;k)\bigg), \quad
\Psi_-(x;k)=\bigg(\varphi_+(x;k),\frac{\phi_-(x;k)}{\overline{a(-k)}}\bigg),
\end{equation}
where $\Psi_+(x;k)$ analytic in the first and third quadrant of the $k$ plane (that is ${\rm Im} k^2>0$) and $\Psi_-(x;k)$ are analytic in the second and fourth quadrant of the $k$ plane (where ${\rm Im} k^2>0$). For every $x\in\mathbb{R}$, there exists a condition
\begin{equation}\label{mtc3}
 \Psi_+(x;k)-\Psi_-(x;k)=\Psi_-(x;k)S(x;k), \quad k\in\mathbb{R}\cup i\mathbb{R},
\end{equation}
where
\begin{equation}
\label{mtc4}
S(x;k)=\left(\begin{matrix}
-\mid r(k)\mid^{2} &  -\overline{r(k)}\mathrm{e}^{- ik^{2}x}\\
r(k)\mathrm{e}^{ ik^{2}x}  &  0
\end{matrix}\right) ,\quad k\in \mathbb{R}.
\end{equation}
\begin{equation}
\label{mtc5}
S(x;k)=\left(\begin{matrix}
\mid r(k)\mid^{2} &  \overline{r(k)}\mathrm{e}^{- ik^{2}x}\\
r(k)\mathrm{e}^{ ik^{2}x}  &  0
\end{matrix}\right) ,\quad k\in  i\mathbb{R}.
\end{equation}
with the reflection coefficient
\begin{equation}\label{mtc6}
 r(k)=\frac{b(k)}{a(k)},
\end{equation}
which is well-defined in view of the condition \eqref{mtc1}.
From \eqref{mtb52},\eqref{mtb52a} and \eqref{mtb44a}, we know that the function $a(k)$ is even and $b(k)$ is odd. So, we have $r(-k)=-r(k)$ and $r(0)=0$.
In addition, from \eqref{mtb10}, \eqref{mtb57} and \eqref{mtb44a},  we get the normalization condition
\begin{equation}\label{mtc7}
\Psi_\pm(x;k)\to\mathrm{e}^{i\nu_+(x)\sigma_3}, \quad k\to\infty,
\end{equation}
in view of the fact $\nu_-(x)-\nu_+(x)=\nu$. Thus, the sectionally analytic functions \eqref{mtc2}, the jump condition \eqref{mtc3} and the normalization condition \eqref{mtc7} construct a Riemann-Hilbert problem.
It is noted that the matrix $S$ is Hermitian for $k\in i\mathbb{R}$. In this case, the Riemann-Hilbert problem \eqref{mtc3},\eqref{mtc5} and \eqref{mtc7} exist a unique solution. However, the matrix $S$ is not a Hermitian for $k\in\mathbb{R}$. But from \eqref{eq67}, we find that
\begin{equation}\label{mtc8}
 1-|r(k)|^2=\frac{1}{|a(k)|^2}\geq C_0^2>0, \quad k\in\mathbb{R},
\end{equation}
where $C_0=\sup\limits_{k\in\mathbb{R}}|a(k)|$. The condition \eqref{mtc8} implies that the Riemann-Hilbert problem \eqref{mtc3},\eqref{mtc4} and \eqref{mtc7} also exist a unique solution.

Now we come to reconstruct the Riemann-Hilbert problem on the Jost function $m_\pm(x;z)$ and $n_\pm(x;z)$. To this end, we define two new reflection coefficients
\begin{equation}\label{mtc9}
 r_+(z)=-\frac{b(k)}{2ka(k)}, \quad r_-(z)=\frac{2kb(k)}{a(k)},
\end{equation}
which admit
\begin{equation}\label{mtc10}
r_-(z)=-4zr_+(z),
\end{equation}
and
\begin{equation}\label{mtc11}
\begin{aligned}
\overline{r_+(z)}r_-(z)=-|r(k)|^2, \quad z\in\mathbb{R}^+,k\in\mathbb{R},\\
\overline{r_+(z)}r_-(z)=|r(k)|^2, \quad z\in\mathbb{R}^-,k\in i\mathbb{R}.
\end{aligned}
\end{equation}

\begin{lemma}\label{lem8}
Let $a(k)$ satisfies the constraint condition \eqref{mtc1}. If $v\in H^{1,1}(\mathbb{R})$, then $r_\pm(z)\in H^1(\mathbb{R})$,
and if  $v\in H^2(\mathbb{R})\cap H^{1,1}(\mathbb{R})$, then $r_\pm(z)\in L^{2,1}(\mathbb{R})$. Moreover, the map
\begin{equation}\label{mtc12}
 H^2(\mathbb{R})\cap H^{1,1}(\mathbb{R})\ni v\mapsto(r_-,r_+)\in H^{1,1}(\mathbb{R})\cap L^{2,1}(\mathbb{R}),
\end{equation}
is Lipschitz continuous.
\end{lemma}
{\bf Proof}~~ The first statement on $r_\pm(z)$ can be proved by using Lemma \ref{lem5}. Let $r_\pm$ and $\tilde{r}_\pm$ be associated with $v$ and $\tilde{v}$, then the following representations
\begin{equation}\label{mtc13}
\begin{aligned}
r_{-}-\tilde{r}_{-}=\frac{2k(b-\tilde{b})}{a}+\frac{2k\tilde{b}}{a\tilde{a}}[\tilde{a}-\mathrm{e}^{i\tilde\nu}-a+\mathrm{e}^{i\nu}]
+\frac{2k\tilde{b}}{a\tilde{a}}[\mathrm{e}^{i\tilde\nu}-\mathrm{e}^{i\nu}],\\
r_{+}-\tilde{r}_{+}=\frac{(b-\tilde{b})}{2ka}+\frac{\tilde{b}}{2ka\tilde{a}}[\tilde{a}-\mathrm{e}^{i\tilde\nu}-a+\mathrm{e}^{i\nu}]
+\frac{\tilde{b}}{2ka\tilde{a}}[\mathrm{e}^{i\tilde\nu}-\mathrm{e}^{i\nu}],
\end{aligned}
\end{equation}
implies the Lipschitz continuity of the map \eqref{mtc12} in terms of the Lemma \ref{lem6}. \quad$\square$

By virtue of the transformation \eqref{mtb43}, we define two new sectionally analytic functions
\begin{equation}\label{mtc14}
\begin{aligned}
\Phi_+(x;z)=\mathrm{e}^{-i\nu_+(x)\sigma_3}\left(\frac{m_-(x;z)}{a(z)},p_+(x;z)\right), \\ \Phi_-(x;z)=\mathrm{e}^{-i\nu_+(x)\sigma_3}\left(m_+(x;z),\frac{p_-(x;z)}{\overline{a(z)}}\right),
\end{aligned}
\end{equation}
where
\begin{equation}\label{mtc15}
 p_\pm(x;z)=\frac{1}{2k}T_1(x;k)T_2^{-1}(x;k)n_\pm(x;z).
\end{equation}
Note that $a(-k)=a(k)$, we write it as $a(z), z=k^2$. Then the jump condition \eqref{mtc3} gives to
\begin{equation}\label{mtc16}
\Phi_+(x;z)-\Phi_-(x;z)=\Phi_-(x;z)R(x;z), z\in \mathbb{R},
\end{equation}
where
\begin{equation}\label{mtc16a}
R(x;z)=\left(
\begin{matrix}
\overline{r_+(z)}r_-(z)&\overline{r_+(z)}\mathrm{e}^{-izx}\\
r_-(z)\mathrm{e}^{izx}&0
\end{matrix}\right).
\end{equation}

Since equation \eqref{mtc15} implies that, for every $x\in\mathbb{R}$,
\begin{equation}\label{mtc17}
\lim\limits_{z\to\infty}p_\pm(x,;z)=\mathrm{e}^{-i\nu_\pm(x)}e_2,
\end{equation}
then using it and the asymptotic behaviors \eqref{mtb10}, \eqref{mtb57}, normalization condition for $\Phi_\pm(x;z)$ will be
\begin{equation}\label{mtc18}
 \Phi_\pm(x;z)\to I, \quad z\to\infty.
\end{equation}

For the Riemann-Hilbert problem \eqref{mtc14},\eqref{mtc16} and \eqref{mtc18}, the matrix $R(x;z)$ is not Hermitian. So it is difficult to construct its unique solution. Here, for every $k\in \mathbb{C}\setminus\{0\}$, we introduce two matrices
\begin{equation}\label{mtc19}
\tau_1(k)=\left(\begin{matrix}
1&0\\
0&2k
\end{matrix}\right), \quad \tau_2(k)=\left(\begin{matrix}
(2k)^{-1}&0\\
0&1
\end{matrix}\right).
\end{equation}
It is verified that
\[\tau_1^{-1}(k)R(x;z)\tau_1(k)=\tau_2^{-1}(k)R(x;z)\tau_2(k)=S(x;k).\]

The Riemann-Hilbert problem for $\Phi_\pm(x;z)$ can be rewritten as
\begin{equation}\label{mtc20}
 \begin{aligned}
\Xi_{+1,2}(x;k)-\Xi_{-1,2}(x;k)&=\Xi_{-1,2}(x;k)S(x;k)+F_{1,2}(x;k), \quad k\in\mathbb{R}\cap i\mathbb{R},\\
\Xi_{\pm1,2}(x;k)&\to0, \quad k\to\infty,
 \end{aligned}
\end{equation}
where
\begin{equation}\label{mtc21}
 \Xi_{\pm1,2}(x;k)=(\Phi_\pm(x;z)-I)\tau_{1,2}(k), \quad F_{1,2}(x;k)=\tau_{1,2}(k)S(x;k).
\end{equation}
Here, the function $\Xi_{+1,2}(x;\cdot)$ is analytic in the first and third quadrant of the $k$ plane, and $\Xi_{-1,2}(x;\cdot)$ is analytic in the second and fourth quadrant of the $k$ plane.

\setcounter{equation}{0}
\section{The inverse scattering transform and Lipschitz continuity}\label{sec4}
\subsection{Solvability of the Riemann-Hilbert problem}
Given the scattering coefficients $r_\pm\in H^1(\mathbb{R})\cap L^{2,1}(\mathbb{R})$ admitting constraint condition \eqref{mtc10}, we will discuss the solvability of the Riemann-Hilbert problem \eqref{mtc20}.

\begin{proposition}\label{pro2}
If $r_\pm(z)\in H^1(\mathbb{R})\cap L^{2,1}(\mathbb{R})$, then $r(k)\in L_z^{2,1}(\mathbb{R})\cap L_z^\infty(\mathbb{R})$.
In addition, if $r_-(z)\in H^1(\mathbb{R})\cap L^{2,1}(\mathbb{R})$, then $\|kr_-(z)\|_{L_z^\infty}\leq\|r_-\|_{H^1\cap L^{2,1}(\mathbb{R})}$
\end{proposition}
{\bf Proof}.~~From \eqref{mtc11}, we know that $|r(k)|^2=-{\rm sgn }z\overline{r_+(z)}r_-(z)$. So, by virtue of the Schwartz inequality, we find that if $r_\pm(z)\in L^{2,1}(\mathbb{R})$, then $r(k)\in L^{2,1}(\mathbb{R})$.

Since $\|f\|_{L^\infty}\leq\frac{1}{\sqrt{2}}\|f\|_{H^1}$, then $r_\pm\in L_z^\infty$ for $r_\pm\in H^1(\mathbb{R})$ and furthermore $r(k)\in L_z^\infty(\mathbb{R})$.

To prove the second statement, we find that
\[\begin{aligned}
|zr_-(z)|^2&\leq\int_0^z|r_-^2(z)+2zr_-(z)\partial_zr_{-}(z)|\mathrm{d}z\\
&\leq\|r_-\|_{L^2}^2+2\|r_-\|_{L^{2,1}}\|\partial_zr_-\|_{L_z^{2}}\leq\|r_-\|^2_{H^1\cap L^{2,1}(\mathbb{R})}.
\end{aligned}\]
The Proposition is proved.  \qquad $\square$

\begin{proposition}\label{pro3}
If $r_\pm(z)\in H^1(\mathbb{R})\cap L^{2,1}(\mathbb{R})$ and satisfying the condition \eqref{mtc10}, then $z^{-2}kr_-,k^{-1}r_+\in L_z^2(\mathbb{R})\cap L_z^\infty(\mathbb{R})$.
\end{proposition}
{\bf Proof}.~~ Given a small $\epsilon>0$, let $\Omega=\mathbb{R}\setminus(-\epsilon,\epsilon)$ and $\chi$ be the eigenfunction of the interval $\Omega$. From \eqref{mtc10}, that is $r_-(z)=-4zr_+(z)$,
and \eqref{mtc6},\eqref{mtc9}, we know that $k^{-1}r_+(z)=-z^{-2}kr_-(z)/4=-z^{-1}r(k)/2$, then
\[\begin{aligned}
\|k^{-1}r_+(z)\|_{L^2(\mathbb{R})}\leq\frac{1}{2}\|kr_-(z)\|_{L^\infty}\|z^{-2}\|_{L^2(\epsilon,+\infty)}+\frac{1}{2}\|z^{-1}\|_{L^{1/2}(-\epsilon,\epsilon)}^{1/5}\|r(k)\|_{L^8(-\epsilon,\epsilon)}^{4/5}.
\end{aligned}\]
Note that $\|r(k)\|_{L^8(-\epsilon,\epsilon)}\leq c\|r(k)\|_{L^2(-\epsilon,\epsilon)}\leq c\|r(k)\|_{L^{2,1}(-\epsilon,\epsilon)}$, we also get $k^{-1}r_+\in L_z^2\cap L_z^\infty(\mathbb{R})$. \qquad $\square$

Now, we introduce the Cauchy operator and the project operator. For every $f\in L^p(R), 1\leq p<\infty$, the Cauchy operator ${\mathcal{C}}_z$ is
\begin{equation}
\label{mtd1}
{\mathcal{C}}_z(f)=\frac{1}{2\pi i}\int_{\mathbb{R}}\frac{f(s)}{s-z}\mathrm{d}s,\quad z\in \mathbb{C}\setminus \mathbb{R}.
\end{equation}
The function $\mathcal{C}_z(f)(z)$ is analytic off the real line. The projection operators ${\mathcal{P}}^{\pm}$ are given explicitly
\begin{equation}\label{mtd2}
{\mathcal{P}}^{\pm}(f)(z)=\lim\limits_{\varepsilon\to 0}\int_{\mathbb{R}}\frac{f(s)}{s-(z\pm\varepsilon i)}\mathrm{d}s,\quad z\in \mathbb{R}.
\end{equation}

\begin{lemma}\cite{imrn2018-5663}\label{lem9}
For every $r(k)\in L_z^2(\mathbb{R})\cap L_z^\infty(\mathbb{R})$ satisfying the constraint \eqref{mtc8}, and for every $F(k)\in L_z^2(\mathbb{R})$, there exist a unique solution $G(k)\in L_z^2(\mathbb{R})$ to the linear equation
\begin{equation}\label{mtd3}
 (I-{\mathcal{P}}^{-}_{s})G(k)=F(k),\quad k\in \mathbb{R}\cup i\mathbb{R},
\end{equation}
where ${\mathcal{P}}^{-}_{s}G={\mathcal{P}}^{-}(GS)$. In addition, the operator $(I-{\mathcal{P}}^{-}_{s})$ is inverse and for every $f(z)\in L^2(R)$ ,
\begin{equation} \label{mtd4}
 \|(I-{\mathcal{P}}^{-}_{s})^{-1}f\|_{{L}^{2}_{z}(\mathbb{R})}\leq C\| f\|_{{L}^{2}_{z}(\mathbb{R})},
\end{equation}
where $C$ is a constant depending on $\| r(k)\|_{L^{\infty}_{z}(\mathbb{R})}$.
\end{lemma}

From the propositions \ref{pro2} and \ref{pro3}, we find that, for every $x\in\mathbb{R}$, in Riemann-Hilbert problem given in \eqref{mtc20}, $S(x;k)\in L_z^1(\mathbb{R})\cap L_z^\infty(\mathbb{R})$ and $F(x;k)\in L_z^2(\mathbb{R})$.

\begin{corollary}\label{cor1}
Let $r_\pm(z)\in H^1(\mathbb{R})\cap L^{2,1}(\mathbb{R})$ satisfying the constraint \eqref{mtc10} and \eqref{mtc8}. For every $x\in\mathbb{R}$,
the Riemann-Hilbert problem given in \eqref{mtc20} exist a unique solution
\[\Xi_{\pm1,2}(x;k)=\Phi_{\pm}(x;z)\tau_{1,2}(k)-\tau_{1,2}(k).\]
\end{corollary}
{\bf Proof}.~~ From \eqref{mtd3}, we find the unique solution $\Xi_{-1,2}(x;k)$ to the equation
\begin{equation}\label{mtd5}
\Xi_{-1,2}(x;k)={\mathcal{P}}^{-}(\Xi_{-1,2}(x;k)S(x;k)+F_{1,2}(x;k))(z),\quad z\in\mathbb{R}.
\end{equation}
With $\Xi_{-1,2}(x;k)$ in hand, we define $\Xi_{+1,2}(x;k)$ by
\begin{equation}\label{mtd6}
\Xi_{+1,2}(x;k)={\mathcal{P}}^{+}(\Xi_{-1,2}(x;k)S(x;k)+F_{1,2}(x;k))(z),\quad z\in\mathbb{R}.
\end{equation}
Then the unique solution of the Riemann-Hilbert problem can be obtained by analytic continuation via the Cauchy operator
\begin{equation}
\label{mtd7}
\Xi_{\pm1,2}(x;k)={\mathcal{C}}_z\big(\Xi_{-1,2}(x;k)S(x;k)+F_{1,2}(x;k)\big),\quad z\in \mathbb{C}^{\pm}.
\end{equation}
The corollary is proved. \qquad $\square$

In the following, we will give the estimates on solutions to the Riemann-Hilbert problem given in \eqref{mtc16}.
We define the column vectors of the matrices $\Phi_\pm$ as
\begin{equation}\label{mtd8}
\Phi_\pm(x;z)=\big(\xi_\pm(x;z), \eta_\pm(x;z)\big).
\end{equation}
Then from \eqref{mtc21}
\begin{equation}\label{mtd9}
\Xi_{\pm1}(x;k)=\Phi_{\pm}(x;z)\tau_{1}(k)-\tau_{1}(k)=\big(\xi_{\pm}-e_{1},2k(\eta_{\pm}-e_{2})\big),
\end{equation}
and
\begin{equation}\label{mtd10}
\Xi_{\pm2}(x;k)=\Phi_{\pm}(x;z)\tau_{2}(k)-\tau_{2}(k)=\big((2k)^{-1}(\xi_{\pm}-e_{1}),\eta_{\pm}-e_{2}\big).
\end{equation}
Since
\[F_{1,2}(x;k)=\tau_{1,2}(k)S(x;k)=R(x;z)\tau_{1,2}(k),\]
then
\begin{equation}\label{mtd11}
\Xi_{\pm1,2}(x;k)S(x;k)+F_{1,2}(x;k)=\Phi_{\pm}(x;z)R(x;z)\tau_{1,2}(k).
\end{equation}
Substituting it into \eqref{mtd5},\eqref{mtd6} and taking the first column for the solution $\Xi_{\pm1}(x;k)$, we have
\begin{equation}\label{mtd12}
\xi_{\pm}-e_{1}={\mathcal{P}}^{\pm}((\Phi_{-}R)^{[1]}(x;\cdot))(z),
\end{equation}
whereas taking the second column for the solution $\Xi_{\pm2}(x;k)$, we have
\begin{equation}\label{mtd13}
\eta_{\pm}-e_{2}={\mathcal{P}}^{\pm}((\Phi_{-}R)^{[2]}(x;\cdot))(z).
\end{equation}
In addition, we have
\begin{equation}\label{mtd12a}
\begin{aligned}
(2k)^{-1}(\xi_{\pm}-e_{1})&={\mathcal{P}}^{\pm}((2k)^{-1}(\Phi_{-}R)^{[1]}(x;\cdot))(z),\\
2k(\eta_{\pm}-e_{2})&={\mathcal{P}}^{\pm}((2k)(\Phi_{-}R)^{[2]}(x;\cdot))(z).
\end{aligned}
\end{equation}

Collecting equations \eqref{mtd12} and \eqref{mtd13}, we have
\begin{equation}\label{mtd14}
\Phi_{\pm}(x;z)=I+{\mathcal{P}}^{\pm}(\Phi_{-}(x;\cdot)R(x;\cdot))(z),\quad z\in \mathbb{R},
\end{equation}
and the solution of the Riemann-Hilbert problem \eqref{mtc16}
\begin{equation}\label{mtd15}
\Phi_{\pm}(x;z)=I+{\mathcal{C}}(\Phi_{-}(x;\cdot)R(x;\cdot))(z),\quad z\in \mathbb{C}^\pm.
\end{equation}

\begin{lemma}\label{lem10}
Let $r_\pm(z)\in H^1(\mathbb{R})\cap L^{2,1}(\mathbb{R})$ satisfying the constraint \eqref{mtc10} and \eqref{mtc8}.
There is a positive constant $C$ that only depends on $\|r_\pm\|_{L^\infty}$ such that the unique solution to the
integral equations \eqref{mtd14} enjoys the estimate for every $x\in \mathbb{R}$
\begin{equation}\label{mtd16}
\|\Phi_{\pm}(x;\cdot)-I\|_{{L}^{2}(\mathbb{R})}\leq C(\| r_{+}\|_{{L}^{2}(\mathbb{R})}+\| r_{-}\|_{{L}^{2}(\mathbb{R})}).
\end{equation}
\end{lemma}
{\bf Proof}.~~ From Proposition \ref{pro2}, we know that if $r_{\pm}(z)\in {H}^{1}(\mathbb{R})\cap {L}^{2.1}(\mathbb{R})$, then $r(k)\in {L}^{2}_{z}(\mathbb{R})\cap {L}^{\infty}_{z}(\mathbb{R})$. There is a positive constant $C$ that only depends on $\|r_\pm\|_{L^\infty}$ such that
\begin{equation}\label{mtd17}
\| R(x;z)\tau_{1,2}(k)\| _{{L}^{2}_{z}(\mathbb{R})}\leq C(\| r_{+}\|_{{L}^{2}(\mathbb{R})}+\|r_{-}\|_{{L}^{2}(\mathbb{R})}).
\end{equation}

From \eqref{mtd12} and \eqref{mtd9}, we know that, for every $x\in\mathbb{R}$
\[\|(\Phi_{\pm}(x;z)-I)^{[1]}\|_{{L_z}^{2}(\mathbb{R})}=\|{\mathcal{P}}^{\pm}(\Phi_{-}(x;k)R(x;k))^{[1]}(z)\|_{L_z^2}=\|\Xi_{\pm1}^{[1]}(x;k)\|_{L_z^2}.\]
In addition, from the first column of equation \eqref{mtd5}, we have
\[\Xi_{-1}^{[1]}(x;k)=(I-{\mathcal{P}}_s^-)^{-1}{\mathcal{P}}^-F_1^{[1]}=(I-{\mathcal{P}}_s^-)^{-1}{\mathcal{P}}^-(R(x;k)\tau_1(k))^{[1]}.\]
Then the bounded condition \eqref{mtd4} implies that
\[\|\Xi_{-1}^{[1]}(x;k)\|_{L_z^2}\leq C_1\|{\mathcal{P}}^-(R(x;k)\tau_1(k))^{[1]}\|_{L_z^2}
\leq C_1\|(R(x;k)\tau_1(k))^{[1]}\|_{L_z^2},\]
where $C_1$ is a positive constant dependent on $\|r\|_{L^\infty}$. Here we have used the important inequality $\|{\mathcal{P}}^\pm f\|_{L_z^2}\leq\|f\|_{L_z^2}$ for every $f\in L^2(\mathbb{R})$\cite{cpam51-0697}.
In addition, from \eqref{mtd6} and \eqref{mtd11}, we have
\[\|\Xi_{+1}^{[1]}(x;k)\|_{L_z^2}\leq \|{\mathcal{P}}^+(R(x;k)\tau_1(k))^{[1]}\|_{L_z^2}
\leq \|(R(x;k)\tau_1(k))^{[1]}\|_{L_z^2}.\]
Thus, using \eqref{mtd17}, we get
\[\|(\Phi_{\pm}(x;z)-I)^{[1]}\|_{{L_z}^{2}(\mathbb{R})}\leq C(\| r_{+}\|_{{L}^{2}(\mathbb{R})}+\| r_{-}\|_{{L}^{2}(\mathbb{R})}),\]
where the positive constant $C$ only depends on $\|r_\pm\|_{L^\infty}$.

The estimate for $(\Phi_{\pm}(x;z)-I)^{[2]}$ can be obtained from \eqref{mtd13}, \eqref{mtd5} by a similar way. The Lemma is proved. \qquad$\square$

From \eqref{mtc14} and \eqref{mtc16}, we find that the column $(\Phi_-R)^{[1]}$ in \eqref{mtd12} can be rewritten as
\[(\Phi_-R)^{[1]}(x;z)=\mathrm{e}^{-i\nu_+(x)\sigma_3}\left(m_+(x;z)\overline{r_+(z)}\mathrm{e}^{-izx}+\frac{p_-(x;z)}{\overline{a(z)}}\right)r_-(z)\mathrm{e}^{izx},\]
which further reduces to
\[(\Phi_-R)^{[1]}(x;z)=\mathrm{e}^{-i\nu_+(x)\sigma_3}p_+(x;z)r_-(z)\mathrm{e}^{izx}=r_-(z)\mathrm{e}^{izx}\Phi_+^{[2]}(x;z),\]
in view of the jump condition \eqref{mtc16}. Similarly, we have
\[(\Phi_-R)^{[2]}(x;z)=\overline{r_+(z)}\mathrm{e}^{-izx}\Phi_-^{[1]}(x;z).\]

\begin{figure}[h]
\setlength{\unitlength}{0.1in}
\begin{picture}(20,10)
\put(9,8){\line(1,0){15}}
\put(13,8){\vector(-1,0){1}}
\put(21,8){\vector(1,0){1}}
\put(22.5,6.5){\makebox(2, 1)[l]{$\mathrm{Re}k$}}
\put(17,2){\line(0,1){12}}
\put(17,11){\vector(0,-1){1}}
\put(17,5){\vector(0,1){1}}
\put(17.5,13){\makebox(2, 1)[l]{$\mathrm{Im}k$}}
\put(16,0){\makebox(2, 1)[l]{(left)}}
\put(36,8){\line(1,0){15}}
\put(40,8){\vector(1,0){1}}
\put(48,8){\vector(1,0){1}}
\put(49.5,6.5){\makebox(2, 1)[l]{$\mathrm{Re}k$}}
\put(44,2){\line(0,1){12}}
\put(44,11){\vector(0,-1){1}}
\put(44,5){\vector(0,-1){1}}
\put(44.5,13){\makebox(2, 1)[l]{$\mathrm{Im}k$}}
\put(43,0){\makebox(2, 1)[l]{(right)}}
\end{picture}
\caption{If $f(k)$ is an even function, the orientation of the integral contour is defined in left;
if $f(k)$ is an odd function, the orientation of the integral contour is defined in right.}
\label{fig1}
\end{figure}
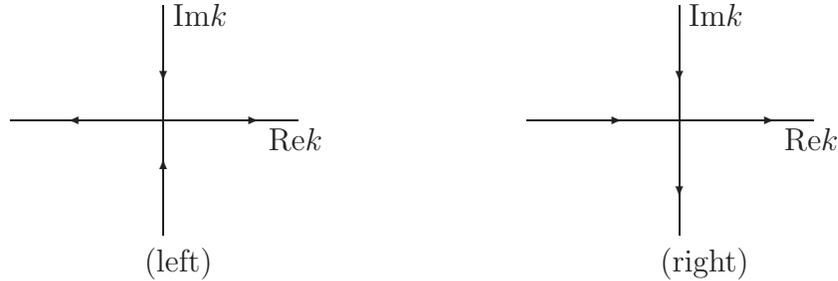

Hence, using the obtained columns $(\Phi_-R)^{[1]}$ and $(\Phi_-R)^{[2]}$, we get, from \eqref{mtd12}, \eqref{mtd13} and \eqref{mtd8}, that
\begin{equation}\label{mtd18}
\begin{aligned}
\xi_-(x;z)-e_1={\cal{P}}^-\left(r_-(z)\mathrm{e}^{izx}\eta_+(x;z)\right),\\
\eta_+(x;z)-e_2={\cal{P}}^+\left(\overline{r_+(z)}\mathrm{e}^{-izx}\xi_-(x;z)\right).
\end{aligned}
\end{equation}
In addition, from \eqref{mtd12a}, we have
\begin{equation}\label{mtd18a}
\begin{aligned}
(2k)^{-1}(\xi_-(x;z)-e_1)={\cal{P}}^-\left((2k)^{-1}r_-(z)\mathrm{e}^{izx}\eta_+(x;z)\right),\\
(2k)(\eta_+(x;z)-e_2)={\cal{P}}^+\left((2k)\overline{r_+(z)}\mathrm{e}^{-izx}\xi_-(x;z)\right).
\end{aligned}
\end{equation}
To ensure the consistency for the two sets of expressions in \eqref{mtd18} and \eqref{mtd18a}, we consider the projection
${\cal{P}}^\pm(kf(k)), k\in\mathbb{R}\cap i\mathbb{R}$, where the orientation of the integral contour is defined in Fig \ref{fig1}.

\vspace{1cm}

From \eqref{mtd18}, we define a new matrix
\begin{equation}\label{mtd19}
 M(x;z)=(\xi_-(x;z)-e_1,\eta_+(x;z)-e_2),
\end{equation}
then the system \eqref{mtd18} can be rewritten as
\begin{equation}
\label{b64}
M(x;z)-{\mathcal{P}}^{+}(M(x;z)R_{+}(x;z))-{\mathcal{P}}^{-}(M(x;z)R_{-}(x;z))=F(x;z),
\end{equation}
where
\begin{equation}
\label{b65}
R_{+}(x;z)=\left(\begin{matrix}
0 &  \overline{r_{+}(z)}\mathrm{e}^{-izx}\\
0   &  0
\end{matrix}\right),\quad R_{-}(x;z)=\left(\begin{matrix}
0 &  0\\
r_{-}(z)\mathrm{e}^{izx}   &  0
\end{matrix}\right),
\end{equation}
\begin{equation}
\label{b66}
F(x;z)=\big(e_{2}{\mathcal{P}}^{-}(r_{-}(z)\mathrm{e}^{izx}),e_{1}{\mathcal{P}}^{+}(\overline{r_{+}(z)}\mathrm{e}^{-izx})\big).
\end{equation}

Since ${\mathcal{P}}^{+}-{\mathcal{P}}^{-}=I$ and
\[R_{+}(x;z)+R_{-}(x;z)=\big(I-R_{+}(x;z)\big)R(x;z),\]
then equation \eqref{b64} can be rewritten as
\begin{equation}\label{mtd23}
G(x;z)-{\mathcal{P}}^{-}(G(x;z)R(x;z))=F(x;z),
\end{equation}
where
\begin{equation}\label{mtd24}
\begin{aligned}
G(x;z)&=M(x;z)\big(I-R_+(x;z)\big)\\
&=\left(\begin{matrix}
\xi_-^{(1)}-1&\eta_+^{(1)}-\overline{r_+}\mathrm{e}^{-izx}(\xi_-^{(1)}-1)\\
\xi_-^{(2)}&\eta_+^{(2)}-1-\overline{r_+}\mathrm{e}^{-izx}\xi_-^{(2)}
\end{matrix}\right).
\end{aligned}
\end{equation}

\begin{proposition}\label{pro5}
For every $x_0\in\mathbb{R}^\pm$, and every $r_\pm(z)\in H^1(\mathbb{R})$, we have
\begin{equation}
\label{b57}
\sup\limits_{x\in (x_{0},\pm\infty)}\| \langle x\rangle{\mathcal{P}}^{\pm}(z^{-j}\overline{r_{+}(z)}\mathrm{e}^{-izx})\|_{{L}^{2}_{z}(\mathbb{R})}\leq \| z^{-j}r_{+}\|_{{H}^{1}(\mathbb{R})}, \quad j=0,1,
\end{equation}
\begin{equation}
\label{b58}
\sup\limits_{x\in (x_{0},\pm\infty)}\| \langle x\rangle{\mathcal{P}}^{\pm}(r_{-}(z)\mathrm{e}^{izx})\|_{{L}^{2}_{z}(\mathbb{R})}\leq \| r_{-}\|_{{H}^{1}(\mathbb{R})}.
\end{equation}
where $\langle x\rangle=(1+x^{2})^{\frac{1}{2}}$. In addition
\begin{equation}
\label{b59}
\sup\limits_{x\in \mathbb{R}}\| {\mathcal{P}}^{\pm}(z^{-j}\overline{r_{+}(z)}\mathrm{e}^{-izx})\|_{{L}^{\infty}_{z}(\mathbb{R})}\leq \frac{1}{\sqrt{2}}\| z^{-j}r_{+}\|_{{H}^{1}(\mathbb{R})}, \quad j=0,1,
\end{equation}
\begin{equation}
\label{b60}
\sup\limits_{x\in \mathbb{R}}\| {\mathcal{P}}^{\pm}(r_{-}(z)\mathrm{e}^{izx})\|_{{L}^{\infty}_{z}(\mathbb{R})}\leq \frac{1}{\sqrt{2}}\| r_{-}\|_{{H}^{1}(\mathbb{R})}.
\end{equation}
Moreover, if $r_{\pm}(z)\in {L}^{2,1}(\mathbb{R})$, then
\begin{equation}
\label{b61}
\sup\limits_{x\in \mathbb{R}}\| {\mathcal{P}}^{\pm}(z\overline{r_{+}(z)}\mathrm{e}^{-izx}\|_{{L}^{2}_{z}(\mathbb{R})}\leq \| zr_{+}(z)\|_{{L}^{2}(\mathbb{R})},
\end{equation}
\begin{equation}
\label{b62}
\sup\limits_{x\in \mathbb{R}}\| {\mathcal{P}}^{\pm}(zr_{-}(z)\mathrm{e}^{izx})\|_{{L}^{2}_{z}(\mathbb{R})}\leq \| zr_{-}(z)\|_{{L}^{2}(\mathbb{R})}.
\end{equation}
\end{proposition}
{\bf Proof}.~~For a function $\rho(z)\in L^2(\mathbb{R})$, the Fourier transform and its its inverse denoted by
\[{\cal{F}}[\rho](y)=\int_{\mathbb{R}}\rho(z)\mathrm{e}^{-izy}dz, \]
and the Parseval equality is $\|\rho(z)\|_{L^2(\mathbb{R})}=\frac{1}{\sqrt{2\pi}}\|{\cal{F}}[\rho](y)\|_{L^2(\mathbb{R})}$.

For the bounds with the exponential $\mathrm{e}^{-izx}$, we consider the project
\begin{equation}\label{b63a}
\begin{aligned}
{\cal{P}}^\pm(\rho(z)\mathrm{e}^{-izx})&=\frac{1}{2\pi i}\lim\limits_{\epsilon\to0^+}\int_{\mathbb{R}}\frac{\rho(s)\mathrm{e}^{-isx}}{s-(z\pm i\epsilon)}\mathrm{d}s\\
&=\frac{1}{2\pi}\int_{\mathbb{R}}\mathrm{d}y{\cal{F}}[\rho](y)\frac{1}{2\pi i}\lim\limits_{\epsilon\to0^+}\int_{\mathbb{R}}\frac{\mathrm{e}^{is(y-x)}}{s-(z\pm i\epsilon)}\mathrm{d}s\\
&=\frac{-1}{2\pi}\int_{\pm\infty}^x{\cal{F}}[\rho](y)\mathrm{e}^{-iz(x-y)}\mathrm{d}y,
\end{aligned}
\end{equation}
where the following equation obtained by residue theorem is used
\[\frac{1}{2\pi i}\lim\limits_{\epsilon\to0^+}\int_{\mathbb{R}}\frac{\mathrm{e}^{is(y-x)}}{s-(z\pm i\epsilon)}\mathrm{d}s
=\pm\chi(\pm(y-x))\mathrm{e}^{-iz(x-y)},\]
where $\chi(x)$ is the Heaviside function.

The bound \eqref{b61} can be proved by choosing $\rho(z)=z\overline{r_{+}(z)}$, and by using Proposition \ref{pro2} and the bound \eqref{eq26} in Proposition \ref{pro1}, as well as the Parseval equality.

The bound \eqref{b57} can be proved by choosing $\rho(z)=z^{-j}\overline{r_{+}(z)}, (j=0,1)$, and by using Proposition \ref{pro2}, \ref{pro3} and the bound \eqref{eq27a}. To prove \eqref{b59}, we consider \eqref{b63a} with $\rho(z)=z^{-j}\overline{r_{+}(z)}, (j=0,1)$,
\[\begin{aligned}
\sup\limits_{x\in \mathbb{R}}\| {\mathcal{P}}^{\pm}(z^{-j}\overline{r_{+}(z)}\mathrm{e}^{-izx})\|_{{L}^{\infty}_{z}(\mathbb{R})}
&\leq\frac{1}{2\pi}\|{\cal{F}}[z^{-j}\overline{r_{+}(z)}](y)\|_{L^1(\mathbb{R})}\\
&\leq\frac{\sqrt{\pi}}{2\pi}\|{\cal{F}}[z^{-j}\overline{r_{+}(z)}](y)\|_{L^{2,1}(\mathbb{R})}\\
&\leq\frac{1}{\sqrt{2}}\|z^{-j}{r_{+}(z)}\|_{H^1(\mathbb{R})}.
\end{aligned}\]

For the bounds with the exponential $\mathrm{e}^{izx}$ in \eqref{b58},\eqref{b60} and \eqref{b62} can be proved similarly. \qquad $\square$

\begin{lemma}\label{lem11}
For every $x_0\in\mathbb{R}^+$ and every $r_\pm(z)\in H^1(\mathbb{R})$, the unique solution of the integral equations in \eqref{mtd18} admits the following estimate
\begin{equation}
\label{b50}
\sup\limits_{x\in (x_{0},+\infty)}\| \langle x\rangle\xi^{(2)}_{-}(x;z)\|_{{L}^{2}_{z}(\mathbb{R})}\leq C\| r_{-}\|_{{H}^{1}(\mathbb{R})},
\end{equation}
\begin{equation}
\label{b51}
\sup\limits_{x\in (x_{0},+\infty)}\| \langle x\rangle\eta^{(1)}_{+}(x;z)\|_{{L}^{2}_{z}(\mathbb{R})}\leq C\| r_{+}\|_{{H}^{1}(\mathbb{R})},
\end{equation}
where $C$ is a positive constant depending on $\| r_{\pm}\|_{L^{\infty}(\mathbb{R})}$. In addition, for every $r_{\pm}\in {H}^{1}(\mathbb{R})\cap {{L}}^{2,1}(\mathbb{R})$, we have
\begin{equation}
\label{b52}
\sup\limits_{x\in \mathbb{R}}\left\| \partial_{x}\xi^{(2)}_{-}(x;z)\right\|_{{L}^{2}_{z}(\mathbb{R})}\leq C\left(\| r_{+}\|_{{H}^{1}(\mathbb{R})\cap {L}^{2,1}(\mathbb{R})}+\| r_{-}\|_{{H}^{1}(\mathbb{R})\cap {L}^{2,1}(\mathbb{R})}\right),
\end{equation}
\begin{equation}
\label{b53}
\sup\limits_{x\in \mathbb{R}}\| \partial_{x}\eta^{(1)}_{+}(x;z)\|_{{L}^{2}_{z}(\mathbb{R})}\leq C\left(\| r_{+}\|_{{H}^{1}(\mathbb{R})\cap L^{2,1}}+\| r_{-}\|_{{H}^{1}(\mathbb{R})\cap {L}^{2,1}(\mathbb{R})}\right).
\end{equation}
\end{lemma}
{\bf Proof}~~ Note that $R(x;z)\tau_{1,2}(k)=\tau_{1,2}(k)S(x;k)$, equation \eqref{mtd23} can be rewritten as
\begin{equation}\label{mtd35}
 G_{1,2}(x;k)-{\cal{P}}_s^-(G_{1,2}(x;k))=F(x;z)\tau_{1,2}(k),
\end{equation}
where $G_{1,2}(x;k)=G(x;z)\tau_{1,2}(k)$ and the operator ${\cal{P}}_s^-$ is defined in Lemma \ref{lem9}.
Since the second row of $F(x;z)$ and $F(x;z)\tau_1(k)$ share the same vector $({\mathcal{P}}^{-}(r_{-}(z)\mathrm{e}^{izx}),0)$,
then for the case of $G_1(x;k)$ in \eqref{mtd35}, considering its second row and using the bound \eqref{mtd4}, we have
\begin{equation}
\label{b54}
\| \xi^{(2)}_{-}(x;z)\|_{{L}^{2}_{z}(\mathbb{R})}\leq C\|{\mathcal{P}}^{-}(r_{-}(z)\mathrm{e}^{izx})\|_{{L}^{2}_{z}(\mathbb{R})}.
\end{equation}
and
\begin{equation}
\label{b55}
\| 2k(\eta^{(2)}_{+}-1-\overline{r_{+}(z)}\mathrm{e}^{-izx}\xi^{(2)}_{-}\|_{{L}^{2}_{z}(\mathbb{R})}\leq C\|{\mathcal{P}}^{-}(r_{-}(z)\mathrm{e}^{izx}))\|_{{L}^{2}_{z}(\mathbb{R})},
\end{equation}
where $C$ is a positive constant depending on $\| r_{\pm}\|_{L^{\infty}(\mathbb{R})}$. Equations \eqref{b54} and \eqref{b58}
prove the bound \eqref{b50}.

From the definition \eqref{mtc9} and Proposition \ref{pro2}, we know that  $\|2k\overline{r_{+}(z)}\|_{L_z^2(\mathbb{R})}=\|r(k)\|_{L_z^2(\mathbb{R})}$.
Thus, using \eqref{b54} and \eqref{b55}
\begin{equation}\label{mtd38}
\begin{aligned}
&\| 2k(\eta^{(2)}_{+}-1)\|_{{L}^{2}_{z}(\mathbb{R})}\\
&\leq\| 2k(\eta^{(2)}_{+}-1-\overline{r_{+}(z)}\mathrm{e}^{-izx}\xi^{(2)}_{-})\|_{{L}^{2}_{z}(\mathbb{R})}
+\| 2k\overline{r_{+}(z)}\mathrm{e}^{-izx}\xi^{(2)}_{-}\|_{{L}^{2}_{z}(\mathbb{R})}\\
&\leq C_1\|{\mathcal{P}}^{-}(r_{-}(z)\mathrm{e}^{izx})\|_{{L}^{2}_{z}(\mathbb{R})},
\end{aligned}
\end{equation}
where $C_1$ is a positive constant depending on $\| r_{\pm}\|_{L^{\infty}(\mathbb{R})}$.

Similarly, the first row of $F(x;z)$ and $F(x;z)\tau_2(k)$ share the same vector $(0,{\mathcal{P}}^{+}(\overline{r_{+}(z)}\mathrm{e}^{izx}))$,
then for the case of $G_2(x;k)$ in \eqref{mtd35}, considering its first row and using the bound \eqref{mtd4}, we have
\begin{equation}
\label{b71}
\| (2k)^{-1}(\xi^{(1)}_{-}-1)\|_{{L}^{2}_{z}(\mathbb{R})}\leq C\|{\mathcal{P}}^{+}(\overline{r_{+}(z)}\mathrm{e}^{-izx})\|_{{L}^{2}_{z}(\mathbb{R})},
\end{equation}
\begin{equation}
\label{b72}
\| \eta^{(1)}_{+}(x;z)-\overline{r}_{+}(z)\mathrm{e}^{-izx}(\xi^{(1)}_{-}-1)\|_{{L}^{2}_{z}(\mathbb{R})}\leq C\|{\mathcal{P}}^{+}(\overline{r_{+}(z)}\mathrm{e}^{-izx})\|_{{L}^{2}_{z}(\mathbb{R})}.
\end{equation}
which imply that
\begin{equation}
\label{b73}
\begin{aligned}
\| \eta^{(1)}_{+}(x;z)\|_{L^{2}_{z}}&\leq\| \eta^{(1)}_{+}(x;z)-\overline{r_{+}(z)}\mathrm{e}^{-izx}(\xi^{(1)}_{-}-1)+(2k)\overline{r_{+}(z)}\mathrm{e}^{-izx}(2k)^{-1}(\xi_{-}^{(1)}-1)\|_{{L}^{2}_{z}(\mathbb{R})}\\
&\leq C_2\|{\mathcal{P}}^{+}(\overline{r_{+}(z)}\mathrm{e}^{-izx}\|_{{L}^{2}_{z}(\mathbb{R})},
\end{aligned}
\end{equation}
in terms of $|2k\overline{r_{+}(z)}|=|r(k)|$, where $C_2$ are a positive constant depending on $\| r_{\pm}\|_{L^{\infty}(\mathbb{R})}$. The bound \eqref{b51} can be proved by virtue of estimate \eqref{b57}.

To prove the bounds \eqref{b53} and \eqref{b53}, we consider the derivative of equation \eqref{b64} with respect to $x$ and let
\[\begin{aligned}
G'(x;z)&=\partial_xM(x;z)(I-R_+(x;z))\\
&=\left(\begin{matrix}
\xi_{-,x}^{(1)}&\eta_{+,x}^{(1)}-\overline{r_+}\mathrm{e}^{-izx}\xi_{-,x}^{(1)}\\
\xi_{-,x}^{(2)}&\eta_{+,x}^{(2)}-\overline{r_+}\mathrm{e}^{-izx}\xi_{-,x}^{(2)}
\end{matrix}\right).
\end{aligned}\]
Then we have
\begin{equation}\label{b73aa}
G'(x;z)-{\mathcal{P}}^{-}(G'(x;z)R(x;z))={F}'(x;z),
\end{equation}
where
\begin{equation}
\label{b73a}
\begin{aligned}
{F}'&=F_{x}+{\mathcal{P}}^{+}(MR_{+,x})+{\mathcal{P}}^{-}(MR_{-,x})\\
&=i\left| {{\begin{array}{*{20}c}
e_{2}{\mathcal{P}}^{-}(zr_{-}(z)\mathrm{e}^{izx})  &   e_{1}{\mathcal{P}}^{+}(-z\overline{r_{+}(z)}\mathrm{e}^{-izx})\\
\end{array} }} \right|\\
&+i\left(\begin{matrix}
\mathcal{P}^{^{-}}(zr_{-}(z)\mathrm{e}^{izx}\eta^{(1)}_{+}) & \mathcal{P}^{^{+}}(-z\overline{r_{+}(z)}\mathrm{e}^{-izx}(\xi^{(1)}_{-}-1))\\
  \mathcal{P}^{^{-}}(zr_{-}(z)\mathrm{e}^{izx}(\eta^{(2)}_{+}-1))&  \mathcal{P}^{^{+}}(-z\overline{r_{+}(z)}\mathrm{e}^{-izx}\xi^{(1)}_{-})
\end{matrix}\right).
\end{aligned}
\end{equation}
In Proposition \ref{pro2}, we know that $kr_-(z)\in L_z^\infty$.  Then the second row vector of ${F}'\tau_1(k)$ and the first row vector of ${F}'\tau_2(k)$ belong to $L_z^2(\mathbb{R})$. The bounds \eqref{b53} and \eqref{b53} can be proved in a similar way. \qquad $\square$

\subsection{Reconstruction for the potential \texorpdfstring{$v$}{}}\label{sec5}
Now we will discuss the potential reconstruction.
From \eqref{mtb12}, we have
\begin{equation}\label{mte22}
\overline{v_x(x)}\mathrm{e}^{i\nu_\pm(x)}=-\frac{i}{2}\lim\limits_{z\to\infty}zm_\pm^{(2)}(x;z).
\end{equation}
The transformation \eqref{mtc15} and the limit \eqref{mtb10} imply that
\begin{equation}\label{mte23}
v(x)\mathrm{e}^{-i\nu_\pm(x)}=-2\lim\limits_{z\to\infty}zp_\pm^{(1)}(x;z).
\end{equation}

Moreover, via the solution \eqref{mtd15}, we have
\begin{equation}\label{mte24}
\lim\limits_{z\to\infty}z(\eta_\pm(x;z)-e_2)=-\frac{1}{2i\pi}\int_{\mathbb{R}}\overline{r_+(z)}\mathrm{e}^{-izx}\xi_-(x;z)\mathrm{d}z,
\end{equation}
and
\begin{equation}\label{mte25}
\lim\limits_{z\to\infty}z(\xi_\pm(x;z)-e_1)
=-\frac{1}{2i\pi}\int_{\mathbb{R}}r_-(z)\mathrm{e}^{izx}(\eta_-(x;z)+\xi_-(x;z)\overline{r_+(z)}\mathrm{e}^{-izx})\mathrm{d}z.
\end{equation}

Firstly, we consider the case of $x\in\mathbb{R}^+$. Using the second column of the jump condition \eqref{mtc16}, equation \eqref{mte25} can be rewritten as
\begin{equation}\label{mte26}
\lim\limits_{z\to\infty}z(\xi_-(x;z)-e_1)=-\frac{1}{2i\pi}\int_{\mathbb{R}}r_-(z)\mathrm{e}^{izx}\eta_+(x;z)\mathrm{d}z.
\end{equation}
By virtue of the definitions \eqref{mtc14} and \eqref{mtd8}, we find that
\[\mathrm{e}^{-i\nu_+(x)\sigma_3}p_+(x;z)=\eta_+(x;z), \quad \mathrm{e}^{-i\nu_+(x)\sigma_3}m_+(x;z)=\xi_-(x;z).\]
Thus the reconstruction for $v$ can be given
\begin{equation}\label{mte27}
\overline{v_x(x)}\mathrm{e}^{2i\nu_+(x)}=\frac{1}{4\pi}\int_{\mathbb{R}}r_-(z)\mathrm{e}^{izx}\eta_+^{(2)}(x;z)\mathrm{d}z,
\end{equation}
and
\begin{equation}\label{mte28}
v(x)\mathrm{e}^{-2i\nu_+(x)}=\frac{1}{i\pi}\int_{\mathbb{R}}\overline{r_+(z)}\mathrm{e}^{-izx}\xi_-^{(1)}(x;z)\mathrm{d}z.
\end{equation}
From the integral equations \eqref{mte27} and \eqref{mte28}, we can obtain the following estimate.
\begin{lemma}\label{lem12}
If $r_{\pm}(z)\in {H}^{1}(\mathbb{R})\cap {L}^{2,1}(\mathbb{R})$ and admit the condition \eqref{mtc8}, then $v\in{H}^{2}(\mathbb{R}^+)\cap {H}^{1,1}(\mathbb{R}^+)$. In addition, we have
\begin{equation}
\label{b74}
\| v\|_{{H}^{2}(\mathbb{R}^{+})\cap {H}^{1,1}(\mathbb{R}^{+})}\leq C(\| r_{+}\|_{{H}^{1}(\mathbb{R})\cap {L}^{2,1}(\mathbb{R})}+\| r_{-}\|_{{H}^{1}(\mathbb{R})\cap {L}^{2,1}(\mathbb{R})}),
\end{equation}
where $C$ is a positive constant depending on $\| r_{\pm}\|_{{H}^{1}(\mathbb{R})\cap {L}^{2,1}(\mathbb{R})}$.
\end{lemma}
{\bf Proof}~~ We rewrite the construction formula \eqref{mte28} as
\begin{equation}\label{mte29}
\begin{aligned}
v(x)\mathrm{e}^{-2i\nu_+(x)}&=\frac{1}{\pi i}\int_{\mathbb{R}}\overline{r_{+}(z)}\mathrm{e}^{-izx}dz\\
&+\frac{1}{\pi i}\int_{\mathbb{R}}\overline{r_{+}(z)}\mathrm{e}^{-izx}(\xi_{-}^{(1)}(x;z)-1)\mathrm{d}z.
\end{aligned}
\end{equation}
If $\overline{r_+(z)}\in H^1(R)$, then the integral in the first term on the left hand side is the Fourier transform denoted by ${\cal{F}}[\overline{r_+(z)}]$, then ${\cal{F}}[\overline{r_+(z)}]\in L_x^{2,1}(\mathbb{R})$. To control the second term, letting it as $I(x)$ and using the equation \eqref{mtd18}, we have
\begin{equation}\label{mte30}
\begin{aligned}
I(x)&=\int_{-\infty}^{+\infty}\overline{r_{+}(z)}\mathrm{e}^{-izx}
\frac{1}{2i\pi}\int_{-\infty}^\infty\frac{r_-(s)\mathrm{e}^{isx}\eta_+^{(1)}(x;s)}{s-(z-i0)}\mathrm{d}s\mathrm{d}z\\
&=-\int_{-\infty}^{+\infty}r_-(s)\mathrm{e}^{isx}\eta_+^{(1)}(x;s)
\frac{1}{2i\pi}\int_{-\infty}^\infty\frac{\overline{r_{+}(z)}\mathrm{e}^{-izx}}{z-(s+i0)}\mathrm{d}z\mathrm{d}s\\
&=-\int_{-\infty}^{+\infty}r_-(z)\mathrm{e}^{izx}\eta_+^{(1)}(x;z){\cal{P}}^+\big(\overline{r_{+}(z)}\mathrm{e}^{-izx}\big)\mathrm{d}z.
\end{aligned}
\end{equation}

We introduce two functions
\[f=\frac{|\eta_+^{(1)}|}{(\int_{\mathbb{R}}\langle x\rangle|\eta_+^{(1)}|^2\mathrm{d}z)^{1/2}}, \quad
g=\frac{|{\cal{P}}^+(\overline{r_+(z)}\mathrm{e}^{-izx})|}{(\int_{\mathbb{R}}\langle x\rangle|{\cal{P}}^+(\overline{r_+(z)}\mathrm{e}^{-izx})|^2\mathrm{d}z)^{1/2}},\]
then the inequality $fg\leq\frac{1}{2}(f^2+g^2)$ implies that $\int_{\mathbb{R}}\langle x\rangle fg\mathrm{d}z\leq1$, that is the weighted h\"older inequality
\[\int_{\mathbb{R}}\langle x\rangle|\eta_+^{(1)}||{\cal{P}}^+(\overline{r_+(z)}\mathrm{e}^{-izx})|\mathrm{d}z\leq
\|\langle x\rangle \eta_+^{(1)} \|_{L_z^2}\|\langle x\rangle{\cal{P}}^+(\overline{r_+(z)}\mathrm{e}^{-izx})\|_{L_z^2}.\]
By the bounds \eqref{b57},\eqref{b51} and the weighted h\"older inequality, for every $x\in\mathbb{R}^+$, we have
\[\begin{aligned}
\sup\limits_{x\in (x_{0},+\infty)}|\langle x\rangle I(x)|&\leq\| r_{-}(z)\|_{{L}^{\infty}(\mathbb{R})}\sup\limits_{x\in (x_{0},+\infty)}\|\langle x\rangle{\mathcal{P}}^{+}(\overline{r_{+}(z)}\mathrm{e}^{-izx})\|_{{L}^{2}_{z}(\mathbb{R})}\\
&\quad\times\sup\limits_{x\in (x_{0},+\infty)}\|\langle x\rangle\eta^{(1)}_{+}(x;z)\|_{{L}^{2}_{z}(\mathbb{R})}\\
&\leq C\| r_{+}(z)\|^2_{{H}^{1}(\mathbb{R})},
\end{aligned}\]
where the positive constant $C$ only depend on $\|r_\pm(z)\|_{L^\infty}$.
Then from \eqref{mte29} and using the triangle inequality, we have
\begin{equation}
\label{b74ab}
\| v\|_{{L}^{2,1}(\mathbb{R}^{+})}\leq C(1+\| r_{+}\|_{{H}^{1}(\mathbb{R})})\| r_{+}\|_{{H}^{1}(\mathbb{R})}.
\end{equation}

For the reconstruction formula \eqref{mte27}, we also have
\begin{equation}
\label{b74a}
\begin{aligned}
\overline{v_x(x)}\mathrm{e}^{2i\nu_+(x)}&=\frac{1}{4\pi}\int_{\mathbb{R}}r_{-}(z)\mathrm{e}^{izx}\mathrm{d}z\\
&+\frac{1}{4\pi}\int_{\mathbb{R}}r_{-}(z)\mathrm{e}^{izx}(\eta^{(2)}_{+}-1)\mathrm{d}z.
\end{aligned}
\end{equation}
A similar discussion gives
\begin{equation}
\label{b74b}
\| {v}_{x}\|_{{L}^{2,1}(\mathbb{R}^+)}\leq C(1+\| r_{-}\|_{{H}^{1}}(\mathbb{R}))\| r_{-}\|_{{H}^{1}(\mathbb{R})},
\end{equation}
Thus the estimates \eqref{b74} and \eqref{b74b} imply
\begin{equation}
\label{b74c}
\|v\|_{{H}^{1,1}(\mathbb{R}^+)}\leq C(\|r_{+}\|_{{H}^{1}}+\| r_{-}\|_{{H}^{1}}).
\end{equation}

To prove the statement on $v\in H^2(\mathbb{R}^+)$, we consider
\begin{equation}
\label{b75}
\begin{aligned}
I'(x)&=-i\int_{\mathbb{R}}z\overline{r_{+}(z)}\mathrm{e}^{-izx}(\xi^{(1)}_{-}-1)\mathrm{d}z+\int_{\mathbb{R}}\overline{r_{+}(z)}\mathrm{e}^{-izx}\partial_{x}\xi^{(1)}_{-}\mathrm{d}z\\
&=i\int_{\mathbb{R}}r_{-}(z)\mathrm{e}^{izx}\eta^{(1)}_{+}(x;z){\mathcal{P}}^{+}(z\overline{r_{+}(z)}\mathrm{e}^{-izx})\mathrm{d}z\\
&-i\int_{\mathbb{R}}zr_{-}(z)\mathrm{e}^{izx}\eta^{(1)}_{+}(x;z){\mathcal{P}}^{+}(\overline{r_{+}(z)}\mathrm{e}^{-izx})\mathrm{d}z\\
&-\int_{\mathbb{R}}r_{-}(z)\mathrm{e}^{izx}\partial_{x}\eta^{(1)}_{+}(x;z){\mathcal{P}}^{+}(\overline{r_{+}(z)}\mathrm{e}^{-izx})\mathrm{d}z.
\end{aligned}
\end{equation}
For every $x\in\mathbb{R}^+$, using the bounds \eqref{b51},\eqref{b53}, \eqref{b57} and \eqref{b61}, we have
\[\begin{aligned}
&\sup\limits_{x\in (x_{0},+\infty)}| \langle x\rangle I'(x)|\\
&\leq\| r_{-}\|_{{L}^{\infty}(\mathbb{R})}\sup\limits_{x\in(x_{0},+\infty)}\| \langle x\rangle \eta^{(1)}_{+}(x;z)\|_{{L}^{2}_{z}(\mathbb{R})}\sup\limits_{x\in(x_{0},+\infty)}\| {\mathcal{P}}^{+}( zr_{+}(z)\mathrm{e}^{-izx})\|_{{L}^{2}_{z}(\mathbb{R})}\\
&+\| zr_{-}(z)\|_{{L}^{2}_{z}(\mathbb{R})}\sup\limits_{x\in(x_{0},+\infty)}\| \langle x\rangle\eta^{(1)}_{+}(x;z)\|_{{L}^{2}_{z}(\mathbb{R})}\sup\limits_{x\in(x_{0},+\infty)}\| {\mathcal{P}}^{+}(\overline{r_{+}(z)}e^{-izx})\|_{{L}^{\infty}_{z}(\mathbb{R})}\\
&+\| r_{-}\|_{{L}^{\infty}(\mathbb{R})}\sup\limits_{x\in(x_{0},+\infty)}\|\partial_{x}\eta^{(1)}_{+}(x;z)\|_{{L}^{2}_{z}(\mathbb{R})}\sup\limits_{x\in(x_{0},+\infty)}\| \langle x\rangle{\mathcal{P}}^{+}(\overline{r_{+}(z)}\mathrm{e}^{-izx})\|_{{L}^{2}_{z}(\mathbb{R})}\\
&\leq C\| r_{-}\|_{{H}^{1}(\mathbb{R})\cap {L}^{2,1}(\mathbb{R})}\| r_{+}\|_{{H}^{1}\cap L^{2,1}(\mathbb{R})}(\| r_{+}\|_{{H}^{1}(\mathbb{R})\cap {L}^{2,1}(\mathbb{R})}+\| r_{-}\|_{{H}^{1}\cap {L}^{2,1}(\mathbb{R})}).
\end{aligned}\]
This bound on $\sup\limits_{x\in (x_{0},+\infty)}| \langle x\rangle I'(x)|$
is sufficient to control $I'(x)$ in $L^2(\mathbb{R}^+)$ norm and hence the derivative of \eqref{mte29}
 in $x$. Using the same analysis for the derivative of \eqref{b74a} in $x$ yields similar estimates.
\begin{lemma}\label{lem13}
The map
\begin{equation}
\label{b76}
{H}^{1}(\mathbb{R})\cap {L}^{2,1}(\mathbb{R})\ni(r_{-},r_{+})\longmapsto v\in {H}^{2}(\mathbb{R}^+)\cap {H}^{1,1}(\mathbb{R}^+),
\end{equation}
is Lipschitz continuous.
\end{lemma}
{\bf Proof}~~ For convenience, we let $f(x)=v(x)\mathrm{e}^{-2i\nu_+(x)}$. Let $v, \tilde{v}\in {H}^{2}(\mathbb{R}^+)\cap {H}^{1,1}(\mathbb{R}^+)$, and the associated reflection coefficients are $r_\pm(z)$ and $\tilde{r}_\pm(z)$.
From \eqref{mte29}, we find that
\[\begin{aligned}
f(x)-\tilde{f}(x)=&\frac{1}{\pi i}\int_{\mathbb{R}}\big(\overline{r_{+}(z)}-\overline{\tilde{r}_{+}(z)}\big)\mathrm{e}^{-izx}dz\\
&+\frac{1}{\pi i}\int_{\mathbb{R}}\big(\overline{r_{+}(z)}-\overline{\tilde{r}_{+}(z)}\big)\mathrm{e}^{-izx}(\xi_{-}^{(1)}(x;z)-1)\mathrm{d}z\\
&+\frac{1}{\pi i}\int_{\mathbb{R}}\overline{\tilde{r}_{+}(z)}\mathrm{e}^{-izx}\big(\xi_{-}^{(1)}(x;z)-\tilde\xi_{-}^{(1)}(x;z)\big)\mathrm{d}z.
\end{aligned}\]
Repeating the same estimates in Lemma \ref{lem12}, we can get that
\[\|f-\tilde{f}\|_{L^{2,1}(\mathbb{R}^+)}\leq C(\|r_{+}-\tilde{r}_{+}\|_{{H}^{1}(\mathbb{R})}).\]
where $C$ is a positive constant depending on $\| r_{\pm}\|_{{H}^{1}(\mathbb{R})\cap {L}^{2,1}(\mathbb{R})}$ and $\| \tilde{r}_{\pm}\|_{{H}^{1}(\mathbb{R})\cap {L}^{2,1}(\mathbb{R})}$.
In addition, if let $g(x)=\overline{v_x(x)}\mathrm{e}^{2i\nu_+(x)}$, and repeating the same estimates in Lemma \ref{lem12}, we can get that
\[\|g-\tilde{g}\|_{L^{2,1}(\mathbb{R}^+)}\leq C(\|r_{-}-\tilde{r}_{-}\|_{{H}^{1}(\mathbb{R})}).\]

Note that
\[|v-\tilde{v}|\leq|f-\tilde{f}|+|\tilde{v}||\mathrm{e}^{i\nu_+}-\mathrm{e}^{i\tilde\nu_+}|,\]
and
\[\|\mathrm{e}^{i\nu_+}-\mathrm{e}^{i\tilde\nu_+}\|_{L^\infty}\leq \frac{1}{2} \mathrm{e}^{\|v\|_{L^2}+\|\tilde{v}\|_{L^2}}(\|v\|_{L^2}+\|\tilde{v}\|_{L^2})\|v-\tilde{v}\|_{L^2},\]
in view of the estimate \eqref{mtb38} in Lemma \ref{lem4}. For small potential such that
\[\frac{1}{2} \mathrm{e}^{\|v\|_{L^{2,1}}+\|\tilde{v}\|_{L^{2,1}}}(\|v\|_{L^{2,1}}+\|\tilde{v}\|_{L^{2,1}})\|\tilde{v}\|_{L^{2,1}}<1,\]
we can get
\[\|v-\tilde{v}\|_{L^{2,1}(\mathbb{R}^+)}\leq C\|r_{+}-\tilde{r}_{+}\|_{{H}^{1}(\mathbb{R})}.\]
where $C$ is a positive constant depending on $\| r_{\pm}\|_{{H}^{1}(\mathbb{R})\cap {L}^{2,1}(\mathbb{R})}$ and $\| \tilde{r}_{\pm}\|_{{H}^{1}(\mathbb{R})\cap {L}^{2,1}(\mathbb{R})}$.

Similarly, from
\[|v_x-\tilde{v}_x|\leq|g-\tilde{g}|+|\tilde{v}_x||\mathrm{e}^{i\nu_+}-\mathrm{e}^{i\tilde\nu_+}|,\]
we can obtain
\[\|v_x-\tilde{v}_x\|_{L^{2,1}(\mathbb{R}^+)}\leq C\|r_{-}-\tilde{r}_{-}\|_{{H}^{1}(\mathbb{R})}.\]
Thus, we have
\[\|v-\tilde{v}\|_{H^{1,1}(\mathbb{R}^+)}\leq C[\|r_{+}-\tilde{r}_{+}\|_{{H}^{1}(\mathbb{R})}+\|r_{-}-\tilde{r}_{-}\|_{{H}^{1}(\mathbb{R})}],\]
where $C$ is a positive constant depending on $\| r_{\pm}\|_{{H}^{1}(\mathbb{R})\cap {L}^{2,1}(\mathbb{R})}$ and $\| \tilde{r}_{\pm}\|_{{H}^{1}(\mathbb{R})\cap {L}^{2,1}(\mathbb{R})}$.

The estimate for $\|v-\tilde{v}\|_{H^{2}(\mathbb{R}^+)}$ can be obtained repeating the same procedure. The Lemma is proved. \qquad $\square$

Secondly, we consider the case $x\in\mathbb{R}^-$. Recall that $\lim\limits_{z\to\infty}a(z)=\mathrm{e}^{i\nu}$ and $\nu=\nu_--\nu_+$. From \eqref{mtb12} in Lemma \ref{lem2} and the transformation \eqref{mtc15}, as well as the definitions \eqref{mtd8} and \eqref{mtc14}, we have
\begin{equation}\label{mte38}
\overline{v}_{x}\mathrm{e}^{2i\nu_+(x)}=\frac{1}{2i}\lim\limits_{ z\to\infty}z\xi^{(2)}_{+}(x;z),
\end{equation}
\begin{equation}\label{mte38a}
v\mathrm{e}^{-2i\nu_+(x)}=-2\lim\limits_{ z\to\infty}z\eta^{(1)}_{-}(x;z).
\end{equation}
If we still use \eqref{mte24} and \eqref{mte25} to obtain the integral representations which will take the same form as \eqref{mte27} and \eqref{mte28}. Unfortunately, for every $x\in\mathbb{R}^-$, $\mathrm{e}^{\pm izx}$ in the representations will be difficult to control in $z\in \mathbb{C}^\pm$. So the estimate for $x\in\mathbb{R}^-$ cannot proceed. To overcome it, we introduce a scalar Riemann-Hilbert problem \cite{aom137-295}
\begin{equation}\label{mte39}
\begin{aligned}
&\delta_{+}(z)-\delta_{-}(z)=\overline{r_{+}(z)}r_{-}(z), \quad z\in \mathbb{R},\\
&\delta_{\pm}(z)\to1, \quad   z \to\infty.
\end{aligned}
\end{equation}
For every $r_{\pm}\in {H}^{1}\cap {L}^{2,1}(\mathbb{R})$ satisfying the constraint condition \eqref{mtc8}, there exists unique sectionally analytic function 
\begin{equation}\label{mte40}
\delta(z)=\mathrm{e}^{{\mathcal{C}}\log(1+\overline{r_{+}}r_{-})},\quad z\in \mathbb{C}^{\pm},
\end{equation}
for the scalar Riemann-Hilbert problem. The limits $\delta_\pm(z)$ will be
\begin{equation}\label{mte41}
\delta_{\pm}(z)=\mathrm{e}^{{\mathcal{P}}^{\pm}\log(1+\overline{r_{+}}r_{-})}, \quad z\in \mathbb{R},
\end{equation}
as $z\in \mathbb{C}^\pm$ approaches to a point on the real axis by any non-tangential contour in $\mathbb{C}^\pm$.

\begin{proposition}\label{pro5a}
For every $r_{\pm}(z)\in {H}^{1}\cap {L}^{2,1}(\mathbb{R})$ satisfying the constraint condition \eqref{mtc8}, $\delta_+(z)\delta_-(z)r_\pm(z)\in {H}^{1}(\mathbb{R})\cap {L}^{2,1}(\mathbb{R})$.
\end{proposition}
{\bf Proof}~~ For the project operators ${\cal{P}}_\pm$ defined in \eqref{mtd2}, we know ${\cal{P}}_++{\cal{P}}_-=-i{\cal{H}}$,
where ${\cal{H}}$ is the Hilbert operator. From \eqref{mte41}, for every $z\in \mathbb{R}$, we can get
\begin{equation}\label{mte42}
|\delta_+(z)\delta_-(z)|=\big|\mathrm{e}^{-i{\cal{H}}\log(1+\overline{r_+}r_-)}\big|=1,
\end{equation}
in terms of ${\cal{H}}\log(1+\overline{r_+}r_-)$ is real according to \eqref{mtc11}, if $\log(1+\overline{r_+}r_-)\in L^2(\mathbb{R})$.
So if $r_\pm\in L^{2,1}(\mathbb{R})$, then $\delta_+(z)\delta_-(z)r_\pm(z)\in {L}^{2,1}(\mathbb{R})$.

To prove $\delta_+(z)\delta_-(z)r_\pm(z)\in {H}^{1}(\mathbb{R})$, we have
\[\partial_z(\delta_-\delta_+r_\pm)=(\delta_-\delta_+r_\pm)\big(-i\partial_z{\cal{H}}\log(1+\overline{r_+}r_-)\big)+\delta_-\delta_+\partial_z(r_\pm).\]
Since $\|{\cal{H}}f\|_{L^2(\mathbb{R})}=\|f\|_{L^2(\mathbb{R})}$ and $\partial_z({\cal{H}}f)=-{\cal{H}}(\partial_zf)$, for every $f\in L^2$, then
$\|\partial_z({\cal{H}}f)\|_{L^2}=\|\partial_zf\|_{L^2}$. Thus, $\delta_+(z)\delta_-(z)r_\pm(z)\in {H}^{1}(\mathbb{R})$ is proved for every $r_{\pm}(z)\in {H}^{1}(\mathbb{R})\cap {L}^{2,1}(\mathbb{R})$. \qquad $\square$

Now, we can define modified reflection coefficients
\begin{equation}\label{mte43}
r_{\pm,\delta}(z)=\overline{\delta_+(z)}\overline{\delta_-(z)}r_\pm(z),
\end{equation}
for every $z\in\mathbb{R}$, and define new jump matrix
\begin{equation}\label{mte44}
R_{\delta}(x;z)=\left(\begin{matrix}
0&\overline{r_{+,\delta}(z)}\mathrm{e}^{-izx}\\
r_{-,\delta}(z)\mathrm{e}^{izx}&\overline{r_{+,\delta}(z)}r_{-,\delta}(z)
\end{matrix}\right).
\end{equation}
In addition, if we define new sectionall analytic functions
\begin{equation}\label{mte45}
 \Phi_{\pm,\delta}(x;z)=\Phi_\pm(x;z){\rm diag}(\delta_\pm^{-1}(z),\delta_\pm(z)),
\end{equation}
then the Riemann-Hilbert problem \eqref{mtc16} can be rewritten as
\begin{equation}\label{mte46}
\Phi_{+,\delta}(x;z)-\Phi_{-,\delta}(x;z)=\Phi_{-,\delta}(x;z)R_\delta(x;z), \quad z\in \mathbb{R},
\end{equation}
and normalization condition will be
\begin{equation}\label{mte47}
\Phi_{\pm,\delta}(x;z)\to I, \quad z\to\infty,
\end{equation}
in view of the condition in \eqref{mte39}.  The solution to the Riemann-Hilbert problem \eqref{mte46} and \eqref{mte47} takes the following form
\begin{equation}\label{mte48}
\Phi_{\pm,\delta}(x;z)=I+{\cal{C}}\big(\Phi_{-,\delta}(x;\cdot)R_\delta(x;\cdot)\big)(z), \quad z\in\mathbb{C}^\pm.
\end{equation}

If let the columns of the function $\Phi_{\pm,\delta}(x;z)$ be
\begin{equation}\label{mte49}
 \Phi_{\pm,\delta}(x;z)=\big(\xi_{\pm,\delta}(x;z),\eta_{\pm,\delta}(x;z)\big),
\end{equation}
then the first column of the jump condition \eqref{mte46} gives to
\[\xi_{+\delta}(x;z)=\xi_{-,\delta}(x;z)+\eta_{-,\delta}(x;z)r_{-,\delta}(z)\mathrm{e}^{-izx}.\]
Moreover, from \eqref{mte48}, we have
\begin{equation}\label{mte50}
 \lim\limits_{z\to\infty}z(\xi_{\pm,\delta}-e_1)=-\frac{1}{2\pi i}\int_{\mathbb{R}}r_{-,\delta}(z)\mathrm{e}^{izx}\eta_{-,\delta}(x;z)\mathrm{d}z,
\end{equation}
and
\begin{equation}\label{mte51}
 \lim\limits_{z\to\infty}z(\eta_{\pm,\delta}-e_2)=-\frac{1}{2\pi i}\int_{\mathbb{R}}\overline{r_{+,\delta}(z)}\mathrm{e}^{-izx}\xi_{+,\delta}(x;z)\mathrm{d}z.
\end{equation}
It is remarked that the exponentials $\mathrm{e}^{\pm izx}$ is bound in $z\in\mathbb{C}^\mp$, for every $x\in\mathbb{R}^-$.
So we can use them to get the reconstruction formula. Since $\delta_\pm(z)\to 1, z\to\infty$, then from \eqref{mte38},\eqref{mte45}, as well as \eqref{mte50} and \eqref{mte51}, and for every $x\in \mathbb{R}^-$, we have
\begin{equation}\label{mte52}
\begin{aligned}
v\mathrm{e}^{-2i\nu_+(x)}&=-2\lim\limits_{z\to\infty}z\eta_-^{(1)}(x;z)=-2\lim\limits_{z\to\infty}z\eta_{-,\delta}^{(1)}(x;z)\\
&=\frac{1}{\pi i}\int_{\mathbb{R}}\overline{r_{+,\delta}(z)}\mathrm{e}^{-izx}\xi_{+,\delta}^{(1)}(x;z)\mathrm{d}z,
\end{aligned}
\end{equation}
\begin{equation}\label{mte53}
\begin{aligned}
\bar{v}_x\mathrm{e}^{2i\nu_+(x)}&=-\frac{i}{2}\lim\limits_{z\to\infty}z\xi_-^{(2)}(x;z)=-\frac{i}{2}\lim\limits_{z\to\infty}z\xi_{-,\delta}^{(2)}(x;z)\\
&=\frac{1}{4\pi }\int_{\mathbb{R}}{r_{-,\delta}(z)}\mathrm{e}^{izx}\eta_{-,\delta}^{(2)}(x;z)\mathrm{d}z.
\end{aligned}
\end{equation}

For the solution \eqref{mte48}, as $z\in \mathbb{C}^\pm$ approaches to a point on the real axis by any non-tangential contour in $\mathbb{C}^\pm$, we can get
\begin{equation}\label{mte54}
\xi_{+,\delta}(x;z)-e_1={\cal{P}}^+\big(r_{-,\delta}(z)e^{izx}\eta_{-,\delta}(x;z)\big),
\end{equation}
\begin{equation}\label{mte55}
\eta_{-,\delta}(x;z)-e_2={\cal{P}}^-\big(\overline{r_{+,\delta}(z)}e^{-izx}\xi_{-,\delta}(x;z)\big),
\end{equation}
in terms of the jump condition \eqref{mte46}. Collect them, we have the following contract form
\begin{equation}\label{mte56}
G_\delta(x;z)-{\cal{P}}^-\big(G_\delta(x;z)R_\delta(x;z)\big)=F_\delta(x;z),
\end{equation}
where
\[G_\delta(x;z)=\big(\xi_{+,\delta}(x;z)-e_1, \eta_{-,\delta}(x;z)-e_2\big)\left(\begin{matrix}
1&0\\
-r_{-,\delta}(z)\mathrm{e}^{izx}&1
\end{matrix}\right),\]
\[F_\delta(x,z)=\big({\cal{P}}^+(r_{-,\delta}(z)\mathrm{e}^{izx})e_2, {\cal{P}}^-(\overline{r_{+,\delta}(z)}\mathrm{e}^{-izx})e_1\big).\]
\begin{lemma}\label{lem14}
For every $x_0\in\mathbb{R}^-$ and every $r_\pm(z)\in H^1(\mathbb{R})$, the unique solution of the integral equations in \eqref{mte54} and \eqref{mte55} admits the following estimate
\begin{equation}
\label{b50a}
\sup\limits_{x\in (-\infty,x_{0})}\| \langle x\rangle\eta^{(2)}_{-,\delta}(x;z)\|_{{L}^{2}_{z}(\mathbb{R})}\leq C\| r_{-,\delta}\|_{{H}^{1}(\mathbb{R})},
\end{equation}
\begin{equation}
\label{b51a}
\sup\limits_{x\in (-\infty x_{0})}\| \langle x\rangle\xi^{(1)}_{+,\delta}(x;z)\|_{{L}^{2}_{z}(\mathbb{R})}\leq C\| r_{+,\delta}\|_{{H}^{1}(\mathbb{R})},
\end{equation}
where $C$ is a positive constant depending on $\| r_{\pm,\delta}\|_{L^{\infty}(\mathbb{R})}$. In addition, for every $r_{\pm}\in {H}^{1}(\mathbb{R})\cap {{L}}^{2,1}(\mathbb{R})$, we have
\begin{equation}
\label{b52a}
\sup\limits_{x\in \mathbb{R}}\left\| \partial_{x}\eta^{(2)}_{-,\delta}(x;z)\right\|_{{L}^{2}_{z}(\mathbb{R})}\leq C\left(\| r_{+,\delta}\|_{{H}^{1}(\mathbb{R})\cap {L}^{2,1}(\mathbb{R})}+\| r_{-,\delta}\|_{{H}^{1}(\mathbb{R})\cap {L}^{2,1}(\mathbb{R})}\right),
\end{equation}
\begin{equation}
\label{b53a}
\sup\limits_{x\in \mathbb{R}}\| \partial_{x}\xi^{(1)}_{+,\delta}(x;z)\|_{{L}^{2}_{z}(\mathbb{R})}\leq C\left(\| r_{+,\delta}\|_{{H}^{1}(\mathbb{R})\cap L^{2,1}(\mathbb{R})}+\| r_{-,\delta}\|_{{H}^{1}(\mathbb{R})\cap {L}^{2,1}(\mathbb{R})}\right).
\end{equation}
\end{lemma}
{\bf Proof}~~ The proof is similar with that of Lemma \ref{lem11}, in view of \eqref{mte42}. \qquad$\square$

Using Proposition \ref{pro3} and Lemma \ref{lem14}, the statements of Lemma \ref{lem13} can be extended to the negative half-line.
\begin{lemma}\label{lem15}
If $r_{\pm}(z)\in {H}^{1}(\mathbb{R})\cap {L}^{2,1}(\mathbb{R})$ and admit the condition \eqref{mtc8}, then $v\in{H}^{2}(\mathbb{R}^-)\cap {H}^{1,1}(\mathbb{R}^-)$. In addition, we have
\begin{equation}
\label{b74bb}
\| v\|_{{H}^{2}(\mathbb{R}^{-})\cap {H}^{1,1}(\mathbb{R}^{-})}\leq C(\| r_{+,\delta}\|_{{H}^{1}(\mathbb{R})\cap {L}^{2,1}(\mathbb{R})}+\| r_{-,\delta}\|_{{H}^{1}(\mathbb{R})\cap {L}^{2,1}(\mathbb{R})},
\end{equation}
where $C$ is a positive constant depending on $\| r_{\pm,\delta}\|_{{H}^{1}(\mathbb{R})\cap {L}^{2,1}(\mathbb{R})}$.
\end{lemma}

\begin{lemma}\label{lem16}
The map
\begin{equation}
\label{b76a}
{H}^{1}(\mathbb{R})\cap {L}^{2,1}(\mathbb{R})\ni(r_{-},r_{+})\longmapsto v\in {H}^{2}(\mathbb{R}^-)\cap {H}^{1,1}(\mathbb{R}^-),
\end{equation}
is Lipschitz continuous. Moreover, let $r_{\pm}, \tilde{r}_{\pm}\in {H}^{1}(\mathbb{R})\cap {L}^{2,1}(\mathbb{R})$ admitting $\| r_{\pm}\|_{{H}^{1}\cap {L}^{2,1}(\mathbb{R})}$, $\|\tilde{r}_{\pm}\|_{{H}^{1}\cap {L}^{2,1}(\mathbb{R})}\leq\rho$, the associated potentials $v,\tilde{v}$ satisfy the bound
\begin{equation}
\label{b74cc}
\| v-\widetilde{v}\|_{{H}^{2}(\mathbb{R}^{-})\cap {H}^{1,1}(\mathbb{R}^{-})}\leq C(\rho)(\| r_{+}-\tilde{r}_{+}\|_{{H}^{1}(\mathbb{R})\cap {L}^{2,1}(\mathbb{R})}+\| r_{-}-\tilde{r}_{-}\|_{{H}^{1}(\mathbb{R})\cap {L}^{2,1}(\mathbb{R})}),
\end{equation}
for some constant $C(\rho)$.
\end{lemma}

\setcounter{equation}{0}
\section{Conservation laws and reconstruction for potential \texorpdfstring{$u$}{}}\label{sec6}
\subsection{Dressing method and conservation laws}
We firstly consider the time dependence of the scattering data. If $t\in[0,T]$ is fixed, then $v(x)=v(x,t), u(x)=u(x,t)$.
Now for every $t\in[0,T]$, we define fundamental solutions
\begin{equation}\label{mte1}
J_\pm(x,t;k)\mathrm{e}^{i\theta(x,t;k)\sigma_3}, \quad \theta=\frac{1}{2}(k^2x+k^{-2}t),
\end{equation}
to the massive Thirring spectral problem \eqref{mta2} and \eqref{mta3}, where the matrices
\begin{equation}\label{mte2}
J_\pm(x,t;k)=\big(\varphi_\pm(x,t;k),\phi_\pm(x,t;k)\big),
\end{equation}
admit the linear system
\begin{equation}\label{mte3}
\partial_xJ_\pm+\frac{i}{2}k^{2}[\sigma_{3},J_\pm]=-\frac{i}{2}\big(2kV-V^2\sigma_3\big)J_\pm,
\end{equation}
\begin{equation}\label{mte4}
\partial_tJ_\pm+\frac{i}{2}k^{-2}[\sigma_{3},J_\pm]=-\frac{i}{2}\big(2k^{-1}U-U^2\sigma_3\big)J_\pm.
\end{equation}
Here $\big(\varphi_\pm(x;k),\phi_\pm(x;k)\big)=\big(\varphi_\pm(x,t;k),\phi_\pm(x,t;k)\big)|_{t=0}$ are given in \eqref{mtb43}.
Thus, for every $t\in[0,T]$
\begin{equation}\label{mte5}
J_\pm(x,t;k)\to I, \quad x\to\pm\infty.
\end{equation}
It is noted that the Jost functions $\varphi_\pm(x,t;k), \phi_\pm(x,t;k)$ have the same analyticity and asymptotic behaviors as $\varphi_\pm(x;k),\phi_\pm(x;k)$. Furthermore, the associated Jost functions $m_\pm(x,t;z)$ and $n_\pm(x,t;z)$ can be defined via the same transformation \eqref{mtb43} with $v(x,t)$ instead of $v(x)$, and they inherit all the properties of the Jost functions $m_\pm(x;z)$ and $n_\pm(x;z)$.

Since the linear system \eqref{mte3} and \eqref{mte4} have zero trace, we obtain
\begin{equation}\label{mte6}
\begin{aligned}
a(k)=W[\varphi_-(x,t;k)\mathrm{e}^{-i\theta},\phi_+(x,t;k)\mathrm{e}^{i\theta}]=W[\varphi_-(0,0;k),\phi_+(0,0;k)],\\
b(k)=W[\varphi_+(x,t;k)\mathrm{e}^{-i\theta},\varphi_-(x,t;k)\mathrm{e}^{-i\theta}]=W[\varphi_+(0,0;k),\varphi_-(0,0;k)].
\end{aligned}
\end{equation}
In addition, the Riemann-Hilbert problems can be constructed similarly by the time-dependent scattering data
\begin{equation}\label{mte6a}
r_+(t;z)= -\frac{b(k)\mathrm{e}^{ik^{-2}t}}{2ka(k)}, \quad r_-(t;z)=\frac{2kb(k)\mathrm{e}^{ik^{-2}t}}{a(k)},
\end{equation}

\begin{proposition}\label{pro6}
If $r_\pm(z)\in H_z^1(\mathbb{R})\cap L_z^{2,1}(\mathbb{R})$, then for every $t\in[0,T]$, $r_\pm(t,z)\in {\mathcal{W}}(\mathbb{R})$, where
\begin{equation}\label{mte97}
{\cal{W}}(\mathbb{R}):=\{r(z): r(z)\in{H}^{1}(\mathbb{R})\cap {L}^{2,1}(\mathbb{R}), z^{-1}r(z)\in H^1(\mathbb{R})\}.
\end{equation}
\end{proposition}
{\bf Proof}.~~ Since $r_\pm(z)=r_\pm(t;z)\mathrm{e}^{-iz^{-1}t}$, then for every $t\in[0,T]$, \[\|r_\pm(z)\|_{L_z^{2,1}}=\|r_\pm(t;z)\|_{L_z^{2,1}}.\]
In addition, for every $t\in[0,T]$,
\begin{equation}\label{mte97a}
\|\partial_zr_\pm(t;z)\|_{L_z^2(\mathbb{R})}\leq\|\partial_zr_\pm(z)\|_{L_z^2(\mathbb{R})}+T\|z^{-2}r_\pm(z)\|_{L_z^2(\mathbb{R})}.
\end{equation}
The Proposition is prove in view of the Proposition \ref{pro3}. \qquad $\square$

Recall that the linear spectral problem \eqref{mta2} which we carried out the direct and inverse scattering tansform in section 2,3 and 4 only contain the potential $v$, without the potential $u$. To finish the potential reconstruction, we discuss the dressing procedure. According to the transformation \eqref{mtb43} and the asymptotic behaviors in Lemma \ref{lem2}, we let, for every $t\in[0,T]$ the Jost function $J_\pm(x,t;k)$ have the following expansion near $k=\infty$
\begin{equation}\label{mte7}
J_\pm(x,t;k)=J_\pm^{<0>}+k^{-1}J_\pm^{<1>}+k^{-2}J_\pm^{<2>}+k^{-3}J_\pm^{<3>}+O(k^{-4}),
\end{equation}
where $J_\pm^{<j>}=J_\pm^{<j>}(x,t), (j=0,1,2,\cdots)$  are independent of the spectral variable $k$.
We note that the Jost functions $J_\pm(x,t;k)$ satisfy the symmetry condition
\begin{equation}\label{mte8}
 \sigma_3J_\pm(x,t;-k)\sigma_3=J_\pm(x,t;k).
\end{equation}
Thus, we obtain that $J_\pm^{<2l>}$ are diagonal matrices and $J_\pm^{<2l+1>}$ are off-diagonal matrices.

Substituting \eqref{mte7} into \eqref{mte3}, we collect every powers of $k$ and find that
\begin{equation}\label{mte9}
 [\sigma_3,J_\pm^{<1>}]=-2VJ_\pm^{<0>},
\end{equation}
\begin{equation}\label{mte10}
\partial_xJ_\pm^{<2l>}-\frac{i}{2}V^2\sigma_3J_\pm^{<2l>}=-iVJ_\pm^{<2l+1>},
\end{equation}
\begin{equation}\label{mte11}
\partial_xJ_\pm^{<2l+1>}-\frac{i}{2}V^2\sigma_3J_\pm^{<2l+1>}+\frac{i}{2}[\sigma_3,J_\pm^{<2l+3>}]=-iVJ_\pm^{<2l+2>},
\end{equation}
where $l=0,1,2,\cdots$. Equation \eqref{mte9} implies that
\[J_\pm^{<1>}=V\sigma_3J_\pm^{<0>}.\]
Substituting it into \eqref{mte10} for $l=0$, we get
\[\partial_xJ_\pm^{<0>}=-\frac{i}{2}V^2\sigma_3J_\pm^{<0>}.\]
The boundary condition \eqref{mte5} implies that $J_\pm^{<0>}\to I, x\to\pm\infty$. Thus we have
\begin{equation}\label{mte12}
 J_\pm^{<0>}=\mathrm{e}^{i\nu_\pm(x,t)\sigma_3}, \quad J_\pm^{<1>}=V\sigma_3\mathrm{e}^{i\nu_\pm(x,t)\sigma_3},
\end{equation}
where $\nu_\pm$ is defined in \eqref{mtb14}. In addition, equation \eqref{mte11} for $l=0$ gives to
\[J_\pm^{<3>}=V\sigma_3J_\pm^{<2>}+\frac{1}{2}V^2J_\pm^{<1>}+i\sigma_3\partial_xJ_\pm^{<1>}.\]
Substituting it into \eqref{mte10} for $l=1$ and using the representations of $J_\pm^{<1>}$ and $J_\pm^{<0>}$ in \eqref{mte12}, we have
\[\partial_x(\mathrm{e}^{i\nu_\pm(x,t)\sigma_3}J_\pm^{<2>})=-VV_x,\]
which produces
\begin{equation}\label{mte13}
J_\pm^{<2>}=-\int_{\pm\infty}^xVV_y\mathrm{d}y\mathrm{e}^{i\nu_\pm(x,t)\sigma_3}.
\end{equation}
in terms of the condition  $J_\pm^{<l>}\to 0, x\to\pm\infty, {l=1,2,\cdots}$ derived from the boundary condition \eqref{mte5}.
Note that equations \eqref{mte10} and \eqref{mte11} are recurrence relations, from which we can obtain all the functions $J_\pm^{<j>}$. It is remarked that the obtained expansion \eqref{mte7} is equivalent to the series got from the transformation \eqref{mtb43} and the asymptotic behaviors in Lemma \ref{lem2}.

Substituting the obtained series \eqref{mte7} into \eqref{mte4}, we find that
\begin{equation}\label{mte14}
\partial_tJ_\pm^{<0>}=\frac{i}{2}U^2\sigma_3J_\pm^{<0>},
\end{equation}
\begin{equation}\label{mte15}
\partial_tJ_\pm^{<1>}-\frac{i}{2}U^2\sigma_3J_\pm^{<1>}=-iUJ_\pm^{<0>},
\end{equation}
\begin{equation}\label{mte16}
\begin{aligned}
\partial_tJ_\pm^{<2l>}-\frac{i}{2}U^2\sigma_3J_\pm^{<2l>}&=-iUJ_\pm^{<2l-1>},\\
\partial_tJ_\pm^{<2l+1>}-\frac{i}{2}U^2\sigma_3J_\pm^{<2l+1>}+i\sigma_3J_\pm^{<2l-1>}&=-iUJ_\pm^{<2l>},
\end{aligned}
\end{equation}
where $l=1,2,\cdots$. If $\partial_tv\in L_x^2(\mathbb{R})$ for everey $t\in[0,T]$, equation \eqref{mte14} and the definition $J_\pm^{<0>}$ in \eqref{mte12} imply that
\begin{equation}\label{mte17}
|u(x;t)|^2=-\int_{\pm\infty}^x\partial_t|v(y,t)|^2\mathrm{d}y,
\end{equation}
which gives the first conservation law
\begin{equation}\label{mte18}
 (|v(x,t)|^2)_t+(|u(x,t)|^2)_x=0.
\end{equation}
Substituting \eqref{mte12} into \eqref{mte15} for $l=1$, we obtain the massive Thirring equation
\begin{equation}\label{mte19}
iv_t(x,t)+u(x,t)-|u(x,t)|^2v(x,t)=0.
\end{equation}
We note that the second conservation law
\begin{equation}\label{mte20}
 i(v(x,t)\bar{v}_x(x,t))_t+(u(x,t)\bar{v}(x,t))_x=0,
\end{equation}
can be obtained by substituting \eqref{mte12} and \eqref{mte13} into \eqref{mte16} for $l=1$. Equations \eqref{mte19} and \eqref{mte20} imply the massive Thirring equation
\begin{equation}\label{mte21}
iu_{x}(x,t)+v(x,t)-\vert v(x,t)\vert^{2}u(x,t)=0.
\end{equation}
In fact, using the conservation law \eqref{mte20}, we find that
\[\begin{aligned}
&i\bar{v}(iv_t+u-|u|^2v)_x+iv(i\bar{v}_t-\bar{u}+|u|^2\bar{v})\\
&=(\bar{v}-|v|^2\bar{u})(iu_x-v+|v|^2u)+(v-|v|^2u)(i\bar{u}_x-\bar{v}+|v|^2\bar{u}).
\end{aligned}\]
It is interesting to note that more conservation laws can be obtained from the recurrence relation \eqref{mte16}.

\subsection{Reconstruction for the potential \texorpdfstring{$u$}{}}
In the following, we come to give the reconstruction for potential $u$. In general, the prior estimates for
the time-dependent Jost functions are still valid. So we need to consider the estimates for the derivative of the Jost functions with respect to the time variable.

\begin{lemma}\label{lem18}
For every $r_\pm(z)\in H^1(\mathbb{R})\cap L^{2,1}(\mathbb{R})$, 
we have
\begin{equation}
\label{b52b}
\sup\limits_{x\in \mathbb{R}^+}\left\|\langle x\rangle \partial_{t}\xi^{(2)}_{-}(x,t;z)\right\|_{{L}^{2}_{z}(\mathbb{R})}\leq C(\|z^{-1}{r_-}(t;z)\|_{H_z^1}+\|{r_-}(t;z)\|_{H_z^1}),
\end{equation}
\begin{equation}
\label{b53b}
\sup\limits_{x\in \mathbb{R}^+}\|\langle x\rangle \partial_{t}\eta^{(1)}_{+}(x,t;z)\|_{{L}^{2}_{z}(\mathbb{R})}\leq C(\|z^{-1}{r_+}(t;z)\|_{H_z^1}+\|{r_+}(t;z)\|_{H_z^1}).
\end{equation}
where the positive constant $C$ depends on $\|r_\pm(z)\|_{L^\infty}$, $\|k^{-1}r_+(z)\|_{L^\infty}$. In addition,
\begin{equation}
\label{b54a}
\sup\limits_{x\in \mathbb{R}}\left\| \partial_{xt}\xi^{(2)}_{-}(x,t;z)\right\|_{{L}^{2}_{z}(\mathbb{R})}\leq C\left(\| r_{+}\|_{{H}^{1}(\mathbb{R})\cap L^{2,1}}+\| r_{-}\|_{{H}^{1}(\mathbb{R})\cap {L}^{2,1}(\mathbb{R})}\right),
\end{equation}
\begin{equation}
\label{b55a}
\sup\limits_{x\in \mathbb{R}}\| \partial_{xt}\eta^{(1)}_{+}(x,t;z)\|_{{L}^{2}_{z}(\mathbb{R})}\leq C\left(\| r_{+}\|_{{H}^{1}(\mathbb{R})\cap L^{2,1}}+\| r_{-}\|_{{H}^{1}(\mathbb{R})\cap {L}^{2,1}(\mathbb{R})}\right).
\end{equation}
\end{lemma}
{\bf Proof}.~~By derivativing equation \eqref{b64} with respect to $t$, we have
\begin{equation}\label{mte70}
\tilde{G}(x,t;z)-{\mathcal{P}}^{-}(\tilde{G}(x,t;z)R(x,t;z))=\tilde{F}(x,t;z),
\end{equation}
where
\[\begin{aligned}
\tilde{G}(x,t;z)&=M_t(x,t;z)(I-R_+(x,t;z))\\
&=\left(\begin{matrix}
\xi_{-,t}^{(1)}&\eta_{+,t}^{(1)}-\overline{r_+(t;z)}\mathrm{e}^{-izx}\xi_{-,t}^{(1)}\\
\xi_{-,t}^{(2)}&\eta_{+,t}^{(2)}-\overline{r_+(t;z)}\mathrm{e}^{-izx}\xi_{-,t}^{(2)}
\end{matrix}\right),
\end{aligned}\]
and
\[\begin{aligned}
\tilde{F}(x,t;z)=&F_t(x,t;z)+{\cal{P}^+}(MR_{+,t})+{\cal{P}^-}(MR_{-,t})\\
=&i\big(e_{2}{\mathcal{P}}^{-}(z^{-1}r_{-}(t;z)\mathrm{e}^{izx}),e_{1}{\mathcal{P}}^{+}(-z^{-1}\overline{r}_{+}(t;z)\mathrm{e}^{-izx})\big)\\
&+i\left({\cal{P}^-}\big(z^{-1}r_-(t;z)e^{izx}(\eta_+-e_2)\big),{\cal{P}^+}\big(-z^{-1}\overline{r_+(t;z)}e^{-izx}(\xi_--e_1)\big)\right).
\end{aligned}\]
The first row vector of $\tilde{F}\tau_2$ gives
\[\begin{aligned}
(\tilde{F}\tau_2)_{[1]}=&i\big(0,-{\cal{P}^+}(z^{-1}\overline{r_+(t;z)}\mathrm{e}^{-izx})\big)\\
&+i\big((2k)^{-1}{\cal{P}^-}(z^{-1}r_-(t;z)\mathrm{e}^{izx}\eta_+^{(1)}),-{\cal{P}^+}(z^{-1}\overline{r_+(t;z)}\mathrm{e}^{-izx}(\xi_-^{(1)}-1))\big).
\end{aligned}\]
Thus, for every $x\in\mathbb{R}^+$, we have
\begin{equation}\label{mte72}
\|(2k)^{-1}\xi_{-,t}^{(1)}\|_{L_z^2}\leq\|(1-{\cal{P}}_s^-)^{-1}(\tilde{F}\tau_2)_{[1]}\|_{L_z^2},
\end{equation}
and
\begin{equation}\label{mte73}
\|\eta_{+,t}^{(1)}-\overline{r_+(t;z)}\mathrm{e}^{-izx}\xi_{-,t}^{(1)}\|_{L_z^2}
\leq\|(1-{\cal{P}}_s^-)^{-1}(\tilde{F}\tau_2)_{[1]}\|_{L_z^2}.
\end{equation}
Using Lemma \ref{lem9} and $\|{\cal{P}}^\pm f(z)\|_{L^2}\leq\|f(z)\|_{L^2}$, we have
\[\begin{aligned}
\|(2k)^{-1}\xi_{-,t}^{(1)}\|_{L_z^2}
&\leq C\bigg(\|{\cal{P}^+}(z^{-1}\overline{r_+}\mathrm{e}^{-izx})\|_{L_z^2}+2\|{\cal{P}^-}(k^{-1}r_+\mathrm{e}^{izx}\eta_+^{(1)})\|_{L_z^2}\\
&\quad +\|{\cal{P}^+}(z^{-1}\overline{r_+}\mathrm{e}^{-izx}(\xi_-^{(1)}-1))\|_{L_z^2}\bigg)\\
&\leq C\bigg(\|{\cal{P}^+}(z^{-1}\overline{r_+}\mathrm{e}^{-izx})\|_{L_z^2}+2\|(k^{-1}r_+\mathrm{e}^{izx}\eta_+^{(1)})\|_{L_z^2}\\
&\quad +\|(2k)z^{-1}\overline{r_+}\mathrm{e}^{-izx}(2k)^{-1}(\xi_-^{(1)}-1)\|_{L_z^2}\bigg)\\
&\leq C\bigg(\|{\cal{P}^+}(z^{-1}\overline{r_+}\mathrm{e}^{-izx})\|_{L_z^2}+2\|k^{-1}r_+\|_{L^\infty}\|\eta_+^{(1)}\|_{L_z^2}\\
&\quad +2\|k^{-1}\overline{r_+}\|_{L^\infty}\|(2k)^{-1}(\xi_-^{(1)}-1)\|_{L_z^2}\bigg),
\end{aligned}\]
then from Lemma \ref{lem10} and Proposition \ref{pro3}, as well as \eqref{b71} and \eqref{b73} we have
\begin{equation}\label{mte71}
\|(2k)^{-1}\xi_{-,t}^{(1)}\|_{L_z^2}\leq C_1(\|{\cal{P}^+}(z^{-1}\overline{r_+}\mathrm{e}^{-izx})\|_{L_z^2}+\|{\cal{P}^+}(\overline{r_+}\mathrm{e}^{-izx})\|_{L_z^2}).
\end{equation}

From \eqref{mte73} and Lemma \ref{lem9}, we similarly have
\begin{equation}\label{mte74}
\|\eta_{+,t}^{(1)}-\overline{r_+}\mathrm{e}^{-izx}\xi_{-,t}^{(1)}\|_{L_z^2}\leq C_1(\|{\cal{P}^+}(z^{-1}\overline{r_+}\mathrm{e}^{-izx})\|_{L_z^2}+\|{\cal{P}^+}(\overline{r_+}\mathrm{e}^{-izx})\|_{L_z^2}).
\end{equation}
Equations \eqref{mte71} and \eqref{mte74} imply that
\begin{equation}\label{mte75}
\begin{aligned}
\|\eta_{+,t}^{(1)}\|_{L_z^2}&\leq\|\eta_{+,t}^{(1)}-\overline{r_+}\xi_{-,t}^{(1)}\|_{L_z^2}
+\|(2k)\overline{r_+}\mathrm{e}^{-izx}\|_{L^\infty}\|(2k)^{-1}\xi_{-,t}^{(1)}\|_{L_z^2}\\
&\leq C(\|{\cal{P}^+}(z^{-1}\overline{r_+}\mathrm{e}^{-izx})\|_{L_z^2}+\|{\cal{P}^+}(\overline{r_+}\mathrm{e}^{-izx})\|_{L_z^2}).
\end{aligned}
\end{equation}
This prove the estimate \eqref{b53b} in view of the time-dependent bounds in Proposition \ref{pro5}.

The estimate \eqref{b52b} can be proved similarly by taking the second raw of $F\tau_1$.

For the estimate \eqref{b54a} and \eqref{b55a}, it is noted that
\[\partial_{xt}R_\pm(x,t;z)=-R_\pm(x,t;z), \quad \partial_{xt}F(x,t;z)=-F(x,t;z).\]
Hence, using Lemma \ref{lem11}, the estimate \eqref{b54a} and \eqref{b55a} can be proved similarly.   \qquad $\square$

\begin{corollary}\label{cor2}
Let $r_\pm(z)\in H^1(\mathbb{R})\cap L^{2,1}(\mathbb{R})$ and $z^{-1}r_\pm(z)\in L^2(\mathbb{R})$ satisfying the constraint \eqref{mtc10} and \eqref{mtc8}.
There is a positive constant $C$ that only depends on $\|r_\pm\|_{L^\infty}$ such that the unique solution to the
integral equations \eqref{mtd14} enjoys the estimate for every $x\in \mathbb{R}$
\begin{equation}\label{mtd75a}
\begin{aligned}
&\|\Phi_{\pm,t}(x,t;z)\|_{L_z^{2}(\mathbb{R})}\\
&\leq C(\| r_{+}\|_{{L}^{2}(\mathbb{R})}+\| r_{-}\|_{{L}^{2}(\mathbb{R})}+\| z^{-1}r_{+}\|_{{L}^{2}(\mathbb{R})}+\|z^{-1} r_{-}\|_{{L}^{2}(\mathbb{R})}).
\end{aligned}
\end{equation}
\end{corollary}

Now, we come to discuss the reconstruction for potential $u$. From \eqref{mte17} and \eqref{mte19}, we have
\begin{equation}\label{mte65}
u(x,t)=-iv_t(x,t)-v(x,t)\int_{\pm\infty}^x\partial_t|v(y,t)|^2\mathrm{d}y.
\end{equation}
We note that, for every $t\in[0,T]$, if $v,v_t\in H_x^2\cap L_x^{2,1}$, then $u\in H_x^2\cap L_x^{2,1}$, in view of
\[|u|\leq|v_t|+2|v|(\|v\|_{L_x^2}\|v_t\|_{L_x^2}).\]
In addtion, the associated reconstruction for $v(x,t)$ can be extended for every $x\in\mathbb{R}^+$
\begin{equation}\label{mte65a}
v(x,t)\mathrm{e}^{-2i\nu_+(x,t)}=\frac{1}{i\pi}\int_{\mathbb{R}}\overline{r_+(t;z)}\mathrm{e}^{-izx}\xi_-^{(1)}(x,t;z)\mathrm{d}z.
\end{equation}

Derivative equation \eqref{mte65a} with respect to $t$, using \eqref{mte17} and \eqref{mte19}, we have
\begin{equation}\label{mte67a}
\begin{aligned}
iu(x,t)\mathrm{e}^{-2i\nu_+(x,t)}&=\partial_t\big(v(x,t)\mathrm{e}^{-2i\nu_+(x,t)}\big)\\
&=-\frac{1}{\pi}\int_{\mathbb{R}}{z}^{-1}\overline{r_+(t;z)}\mathrm{e}^{-izx}\xi_-^{(1)}(x,t;z)\mathrm{d}z\\
&\quad+\frac{1}{i\pi}\int_{\mathbb{R}}\overline{r_+(t;z)}\mathrm{e}^{-izx}\xi_{-,t}^{(1)}(x,t;z)\mathrm{d}z.
\end{aligned}
\end{equation}
In addition, by virtue of the massive Thirring equation \eqref{mte21}, we find that
\begin{equation}\label{mte67c}
\begin{aligned}
\mathrm{e}^{-4i\nu_+(x,t)}\partial_x\big(u(x,t)\mathrm{e}^{2i\nu_+(x,t)}\big)&=iv(x,t)\mathrm{e}^{-2i\nu_+(x,t)}\\
&=\frac{1}{\pi}\int_{\mathbb{R}}\overline{r_+(t;z)}\mathrm{e}^{-izx}\xi_-^{(1)}(x,t;z)\mathrm{d}z.
\end{aligned}
\end{equation}
\begin{lemma}\label{lem19}
For every $t\in[0,T]$, if $r_{\pm}(t;z)\in {H}^{1}\cap {L}^{2,1}(\mathbb{R})$, $z^{-1}r_+(t;z)\in H^1(\mathbb{R})$ and admit the condition \eqref{mtc8}, then $u\in{H}^{2}(\mathbb{R}^+)\cap {H}^{1,1}(\mathbb{R}^+)$. In addition, we have
\begin{equation}\label{b74c1}
\begin{aligned}
&\| u(x,t)\|_{{H_x}^{3}(\mathbb{R}^{+})\cap {H_x}^{1,1}(\mathbb{R}^{+})}\\
&\leq C\big(\| r_{+}(t;\cdot)\|_{{H}^{1}(\mathbb{R})\cap {L}^{2,1}(\mathbb{R})}+\| r_{-}(t;\cdot)\|_{{H}^{1}(\mathbb{R})\cap {L}^{2,1}(\mathbb{R})}+\|z^{-1}r_+(t;\cdot)\|_{H_z^1}\big),
\end{aligned}
\end{equation}
where $C$ is a positive constant depending on $\| r_{\pm}(t;\cdot)\|_{{H}^{1}(\mathbb{R})\cap {L}^{2,1}(\mathbb{R})}$.
\end{lemma}
{\bf Proof}.~~ To give the estimate of $u$, we rewrite it as
\begin{equation}\label{mte67b}
\begin{aligned}
iu(x,t)\mathrm{e}^{-2i\nu_+}&=-\frac{1}{\pi}\int_{\mathbb{R}}{z}^{-1}\overline{r_+(t;z)}\mathrm{e}^{-izx}\xi_-^{(1)}(x,t,;z)\mathrm{d}z\\
&\quad+\frac{1}{i\pi}\int_{\mathbb{R}}\overline{r_+(t;z)}\mathrm{e}^{-izx}\xi_{-,t}^{(1)}(x,t,;z)\mathrm{d}z\\
&:=I_1(x,t)+I_2(x,t)+I_3(x,t),
\end{aligned}
\end{equation}
where
\[\begin{aligned}
I_1(x,t)&=-\frac{1}{\pi}\int_{\mathbb{R}}{z}^{-1}\overline{r_+(t;z)}\mathrm{e}^{-izx}\mathrm{d}z,\\
I_2(x,t)&=-\frac{1}{\pi}\int_{\mathbb{R}}{z}^{-1}\overline{r_+(t;z)}\mathrm{e}^{-izx}(\xi_-^{(1)}(x,t,;z)-1)\mathrm{d}z,\\
I_3(x,t)&=\frac{1}{i\pi}\int_{\mathbb{R}}\overline{r_+(t;z)}\mathrm{e}^{-izx}\xi_{-,t}^{(1)}(x,t,;z)\mathrm{d}z.
\end{aligned}\]
For the Fourier transformation $I_1(x)$ and $I_2(x)$, we similar have as in Lemma \ref{lem12}
\[\|I_1(x,t)\|_{L_x^{2,1}}=\frac{1}{\pi}\|{{z}^{-1}\overline{r_+(t;z)}}\|_{H_z^1},\]
For every $x\in\mathbb{R}^+$, using the weighted H\"older inequality,
\[\begin{aligned}
\sup\limits_{x\in (x_{0},+\infty)}|\langle x\rangle I_2(x,t)|
&\leq\frac{1}{\pi}\int_{\mathbb{R}}\langle x\rangle |r_-(t;z)\mathrm{e}^{izx}\eta_+^{(1)}{\cal{P}}^+({z}^{-1}\overline{r_+(t;z)}\mathrm{e}^{-izx})|\mathrm{d}z\\
&\leq\| r_{-}(t;z)\|_{{L}^{\infty}(\mathbb{R})}\sup\limits_{x\in (x_{0},+\infty)}\|\langle x\rangle{\mathcal{P}}^{+}({z}^{-1}\overline{r_{+}(t;z)}\mathrm{e}^{-izx})\|_{{L}^{2}_{z}(\mathbb{R})}\\
&\quad\times\sup\limits_{x\in (x_{0},+\infty)}\|\langle x\rangle\eta^{(1)}_{+}(x;z)\|_{{L}^{2}_{z}(\mathbb{R})}\\
&\leq C\| r_{+}(t;z)\|_{{H}_z^{1}(\mathbb{R})}\|{{z}^{-1}\overline{r_+(t;z)}}\|_{H_z^1},
\end{aligned}\]

To estimate the function $I_3(x)$, we derivative the first equation in\eqref{mtd18} with respect to $t$,
\[\xi_t(x,t;z)=i{\cal{P}}^-\left(z^{-1}r_-(t;z)\mathrm{e}^{izx}\eta_+(x;z)\right)+{\cal{P}}^-\left(r_-(t;z)\mathrm{e}^{izx}\eta_{+,t}(x,t;z)\right),\]
then
\[\begin{aligned}
I_3(x,t)=&\frac{1}{\pi}\int_{\mathbb{R}}\overline{r_+(t;z)}\mathrm{e}^{-izx}{\cal{P}}^-\left(z^{-1}r_-(t;z)\mathrm{e}^{izx}\eta_+^{(1)}(x;z)\right)\mathrm{d}z\\ &+\frac{1}{\pi}\int_{\mathbb{R}}\overline{r_+(t;z)}\mathrm{e}^{-izx}{\cal{P}}^-\left(r_-(t;z)\mathrm{e}^{izx}\eta_{+,t}^{(1)}(x,t;z)\right)\mathrm{d}z\\
=&-\frac{1}{\pi}\int_{\mathbb{R}}z^{-1}r_-(t;z)\mathrm{e}^{izx}\eta_+^{(1)}(x;z){\cal{P}}^+\left(\overline{r_+(t;z)}\mathrm{e}^{-izx}\right)\mathrm{d}z\\ &-\frac{1}{\pi}\int_{\mathbb{R}}r_-(t;z)\mathrm{e}^{izx}\eta_{+,t}^{(1)}(x,t;z){\cal{P}}^+\left(\overline{r_+(t;z)}\mathrm{e}^{-izx}\right)\mathrm{d}z.
\end{aligned}\]
For every $x_0\in\mathbb{R}^+$, using the weighted H\"older inequality,
\[\begin{aligned}
\sup\limits_{x\in (x_{0},+\infty)}&|\langle x\rangle I_3(x,t)|\\
&\leq\frac{4}{\pi}\|r_+\|_{L^\infty}\sup\limits_{x\in(x_0,\infty)}\|\langle x\rangle\eta_+^{(1)}\|_{L_z^2}\sup\limits_{x\in(x_0,\infty)}\|\langle x\rangle{\cal{P}}^+\left(\overline{r_+(t;z)}\mathrm{e}^{-izx}\right)\|_{L_z^2}\\
&\quad+\frac{1}{\pi}\|r_-\|_{L^\infty}\sup\limits_{x\in(x_0,\infty)}\|\langle x\rangle\eta_{+,t}^{(1)}\|_{L_z^2}\sup\limits_{x\in(x_0,\infty)}\|\langle x\rangle{\cal{P}}^+\left({\overline{r_+(t;z)}}\mathrm{e}^{-izx}\right)\|_{L_z^2}\\
&\leq C\big(\| r_{+}(t;z)\|_{H_z^{1}(\mathbb{R})}+\|{{z}^{-1}\overline{r_+(t;z)}}\|_{H_z^1}\big)\| r_{+}(t;z)\|_{H_z^{1}(\mathbb{R})}.
\end{aligned}\]

By combing the estimates for $I_j, j=1,2,3$ with the triangle inequality, we obtain
\begin{equation}\label{mte80}
\|u\|_{L_x^{2,1}(\mathbb{R}^+)}\leq C\bigg(\|{{z}^{-1}\overline{r_+}}\|_{H_z^1}+\| r_{+}\|_{H_z^{1}(\mathbb{R})}+\| r_{-}\|_{H_z^{1}(\mathbb{R})}\bigg).
\end{equation}

We differentiate the equation \eqref{mte67b} in $x$, that is
\begin{equation}\label{mte82}
\partial_x(iu(x,t)\mathrm{e}^{-2i\nu_+})=I_{1,x}+I_{2,x}+I_{3,x}.
\end{equation}
For $I_1$, we have
\[I_{1,x}(x,t)=\frac{i}{\pi}\int_{\mathbb{R}}\overline{r_+(t;z)}\mathrm{e}^{-izx}\mathrm{d}z,\]
which is the Fourier transformation of $\overline{r_+(t,z)}$ denoted by ${\cal{F}}[\overline{r_+}](t,x)$.
Thus for every $t\in[0,T]$,
\begin{equation}\label{mte83}
 \|\langle x\rangle I_{1,x}\|_{L_x^2(\mathbb{R}^+)}=\frac{1}{\pi}\|\langle x\rangle{\cal{F}}[\overline{r_+}](t,x)\|_{L_x^2(\mathbb{R})}\leq c\|r_+(t;z)\|_{H_z^1(\mathbb{R})}.
\end{equation}

For $I_{2,x}$, using \eqref{mtd18}, we find that
\begin{equation}\label{mte84}
\begin{aligned}
I_{2,x}(x,t)=&\frac{i}{\pi}\int_{\mathbb{R}}\overline{r_+(t;z)}\mathrm{e}^{-izx}(\xi_-^{(1)}(x,t;z)-1)\mathrm{d}z\\
&-\frac{1}{\pi}\int_{\mathbb{R}}{z}^{-1}\overline{r_+(t;z)}\mathrm{e}^{-izx}\partial_x\xi_-^{(1)}(x,t;z)\mathrm{d}z\\
=&-\frac{i}{\pi}\int_{\mathbb{R}}r_-(t;z)\mathrm{e}^{izx}\eta_+^{(1)}(x,t;z){\cal{P}}^+(\overline{r_+(t;z)}\mathrm{e}^{-izx})\mathrm{d}z\\
&+\frac{i}{\pi}\int_{\mathbb{R}}zr_-(t;z)\mathrm{e}^{izx}\eta_+^{(1)}(x,t;z){\cal{P}}^+(z^{-1}\overline{r_+(t;z)}\mathrm{e}^{-izx})\mathrm{d}z\\
&+\frac{1}{\pi}\int_{\mathbb{R}}r_-(t;z)\mathrm{e}^{izx}\partial_x\eta_+^{(1)}(x,t;z){\cal{P}}^+(z^{-1}\overline{r_+(t;z)}\mathrm{e}^{-izx})\mathrm{d}z,
\end{aligned}
\end{equation}
which gives for every $x_0\in\mathbb{R}^+$ and every $t\in[0,T]$
\[\begin{aligned}
|\langle x\rangle I_{2,x}|&\leq\frac{1}{\pi}\|r_-\|_{L^\infty}\sup\limits_{x\in(x_0,\infty)}\|\langle x\rangle\eta^{(1)}\|_{L_z^2(\mathbb{R})}\sup\limits_{x\in(x_0,\infty)}\|\langle x\rangle{\cal{P}}^+(\overline{r_+(t;z)}\mathrm{e}^{-izx})\|_{L_z^2(\mathbb{R})}\\
&+\frac{1}{\pi}\|r_-\|_{L^{2,1}}\sup\limits_{x\in(x_0,\infty)}\|\langle x\rangle\eta^{(1)}\|_{L_z^2(\mathbb{R})}\sup\limits_{x\in(x_0,\infty)}\|{\cal{P}}^+(z^{-1}\overline{r_+(t;z)}\mathrm{e}^{-izx})\|_{L_z^\infty(\mathbb{R})}\\
&+\frac{1}{\pi}\|r_-\|_{L^\infty}\sup\limits_{x\in(x_0,\infty)}\|\partial_x\eta^{(1)}\|_{L_z^2(\mathbb{R})}\sup\limits_{x\in(x_0,\infty)}\|\langle x\rangle{\cal{P}}^+(z^{-1}\overline{r_+(t;z)}\mathrm{e}^{-izx})\|_{L_z^2(\mathbb{R})}.
\end{aligned}\]
Using the bounds in Proposition \ref{pro5} and Lemma \ref{lem11}, for every $x_0\in\mathbb{R}^+$ and every $t\in[0,T]$, we have
\begin{equation}\label{mte85}
\begin{aligned}
\sup\limits_{x\in(x_0,\infty)}|\langle x\rangle I_{2,x}|\leq& C\| r_{-}\|_{{H}^{1}(\mathbb{R})\cap {L}^{2,1}(\mathbb{R})}\| r_{+}\|_{{H}^{1}\cap L^{2,1}(\mathbb{R})}\\
&\times(\| r_{+}\|_{{H}^{1}(\mathbb{R})\cap {L}^{2,1}(\mathbb{R})}+\|z^{-1}r_+\|_{H_z^1}\big).
\end{aligned}
\end{equation}

Similarly, for the derivative of $I_3(x)$ with respect to $x$, for every $x_0\in\mathbb{R}^+$ and every $t\in[0,T]$, we obtain
\[\begin{aligned}
|\langle x\rangle I_{3,x}|=&\frac{1}{\pi}\|r_-\|_{L^\infty}\sup\limits_{x\in(x_0,\infty)}\|\langle x\rangle\eta_+^{(1)}\|_{L_z^2}\sup\limits_{x\in(x_0,\infty)}\|\langle x\rangle{\cal{P}}^+\left(\overline{r_+(t;z)}\mathrm{e}^{-izx}\right)\|_{L_z^2}\\
&+\frac{4}{\pi}\|r_+\|_{L^\infty}\sup\limits_{x\in\mathbb{R}}\|\partial_x\eta_+^{(1)}(x;z)\|_{L_z^2}\sup\limits_{x\in(x_0,\infty)}\|\langle x\rangle{\cal{P}}^+\left(\overline{r_+(t;z)}\mathrm{e}^{-izx}\right)\|_{L_z^2}\\
&+\frac{4}{\pi}\|r_+\|_{L^\infty}\sup\limits_{x\in(x_0,\infty)}\|\langle x\rangle\eta_+^{(1)}\|_{L_z^2}\sup\limits_{x\in(x_0,\infty)}\|\langle x\rangle{\cal{P}}^+\left(z\overline{r_+(t;z)}\mathrm{e}^{-izx}\right)\|_{L_z^2}\\
 &+\frac{1}{\pi}\|zr_-\|_{L_z^2}\sup\limits_{x\in(x_0,\infty)}\|\langle x\rangle\eta_{+,t}^{(1)}\|_{L_z^2}\sup\limits_{x\in(x_0,\infty)}\|{\cal{P}}^+\left(\overline{r_+(t;z)}\mathrm{e}^{-izx}\right)\|_{L_z^\infty}\\
  &+\frac{1}{\pi}\|r_-\|_{L^\infty}\sup\limits_{x\in\mathbb{R}}\|\eta_{+,tx}^{(1)}\|_{L_z^2}\sup\limits_{x\in(x_0,\infty)}\|\langle x\rangle{\cal{P}}^+\left(\overline{r_+(t;z)}\mathrm{e}^{-izx}\right)\|_{L_z^2}\\
   &+\frac{1}{\pi}\|r_-\|_{L^\infty}\sup\limits_{x\in(x_0,\infty)}\|\langle x\rangle\eta_{+,t}^{(1)}\|_{L_z^2}\sup\limits_{x\in\mathbb{R}}\|{\cal{P}}^+\left(z\overline{r_+(t;z)}\mathrm{e}^{-izx}\right)\|_{L_z^2},
\end{aligned}\]
which implies the bound
\begin{equation}\label{mte86}
\begin{aligned}
\sup\limits_{x\in(x_0,\infty)}|\langle x\rangle I_{3,x}|\leq& C\| r_{-}\|_{{H}^{1}(\mathbb{R})\cap {L}^{2,1}(\mathbb{R})}\| r_{+}\|_{{H}^{1}\cap L^{2,1}(\mathbb{R})}\\
&\times(\| r_{+}\|_{{H}^{1}(\mathbb{R})\cap {L}^{2,1}(\mathbb{R})}+\|z^{-1}r_+\|_{H_z^1}\big).
\end{aligned}
\end{equation}

By combing the estimates for $I_{j,x}, j=1,2,3$ with the triangle inequality, we obtain estimate
\begin{equation}\label{mte81}
\|u\|_{H_x^{1,1}(\mathbb{R}^+)}\leq C\bigg(\|{{z}^{-1}\overline{r_+}}\|_{{H}^{1}(\mathbb{R})\cap {L}^{2,1}(\mathbb{R})}+\| r_{+}\|_{{H}^{1}(\mathbb{R})\cap {L}^{2,1}(\mathbb{R})}+\| r_{-}\|_{{H}^{1}(\mathbb{R})\cap {L}^{2,1}(\mathbb{R})}\bigg).
\end{equation}

The statement $u\in H_x^3(\mathbb{R}^+)$ can be proved by the statement $v\in H_x^2(\mathbb{R}^+)$ in Lemma \ref{lem12} and equation \eqref{mte67c} or
\[\partial_x\big(u(x,t)\mathrm{e}^{2i\nu_+(x,t)}\big)=iv(x,t)\mathrm{e}^{2i\nu_+(x,t)}\]
The Proof of the bound \eqref{b74c1} is complete. 
 \qquad $\square$

\begin{lemma}\label{lem20}
For every $t\in[0,T]$, if $r_{\pm}(t;z)\in {H}_z^{1}\cap {L}_z^{2,1}(\mathbb{R})$, $z^{-1}r_-(t;z)\in H_z^1(\mathbb{R})$ and admit the condition \eqref{mtc8}, then $u\in{H}_x^{2}(\mathbb{R}^-)\cap {H}_x^{1,1}(\mathbb{R}^-)$. In addition, we have
\begin{equation}\label{mte93}
\begin{aligned}
&\|u\|_{{H}_x^{3}(\mathbb{R}^{-})\cap {H}_x^{1,1}(\mathbb{R}^{-})}\\
&\leq C\big(\| r_{+}\|_{{H}^{1}(\mathbb{R})\cap {L}^{2,1}(\mathbb{R})}+\| r_{-}\|_{{H}^{1}(\mathbb{R})\cap {L}^{2,1}(\mathbb{R})}
+\|z^{-1}r_+\|_{H_z^1}\big),
\end{aligned}
\end{equation}
where $C$ is a positive constant depending on $\| r_{\pm}\|_{{H}^{1}(\mathbb{R})\cap {L}^{2,1}(\mathbb{R})}$.
\end{lemma}
{\bf Proof}.~~The bound can be proved similarly by using the integral equations \eqref{mte52}, \eqref{mte53} and the MT equation \eqref{mte19}. \qquad $\square$

\begin{lemma}\label{lem21}
For every $t\in[0,T]$, the map
\begin{equation}\label{mte92}
{\cal{W}}(\mathbb{R})\ni(r_+(t;z),r_-(t;z))\to u(x,t)\in H_x^3(\mathbb{R}^\pm)\cap H_x^{1,1}(R^\pm),
\end{equation}
is Lipshitz continuous.
\end{lemma}
{\bf Proof}.~~ Let $u, \tilde{u}\in {H}_x^{3}(\mathbb{R}^+)\cap {H}_x^{1,1}(\mathbb{R}^+)$, and the associated reflection coefficients are $r_\pm(z)$ and $\tilde{r}_\pm(z)$. For every $x\in\mathbb{R}^+$ and $t\in[0,T]$, repeating the procedure in Lemma \ref{lem13}, Corollary \ref{cor2} and Lemma \ref{lem12} for the reconstruction formula \eqref{mte67a}, we can get
\begin{equation}\label{mte96}
\begin{aligned}
&\| u-\tilde{u}\|_{{H}_x^{3}(\mathbb{R}^{+})\cap {H}_x^{1,1}(\mathbb{R}^{+})}\\
&\leq C\big(\| r_{+}-\tilde{r}_+\|_{{H}^{1}(\mathbb{R})\cap {L}^{2,1}(\mathbb{R})}+\| r_{-}-\tilde{r}_-\|_{{H}^{1}(\mathbb{R})\cap {L}^{2,1}(\mathbb{R})}+\|z^{-1}(r_+-\tilde{r}_+)\|_{H^1}\big).
\end{aligned}
\end{equation}
The bound for $x\in\mathbb{R}^-$ can be obtained similarly. The Proof is complete. \qquad $\square$

Combining the Lemma \ref{lem13}, Lemma \ref{lem16} and Lemma \ref{lem21}, we obtain the statement.
\begin{lemma}\label{lem22}
For every $t\in[0,T]$, the map
\begin{equation}\label{mte98}
{\cal{W}}(\mathbb{R})\ni\big(r_+(t;z),r_-(t;z)\big)\to \begin{array}{c}
v(x,t)\in H_x^2(\mathbb{R}^\pm)\cap H_x^{1,1}(R^\pm),\\
u(x,t)\in H_x^3(\mathbb{R}^\pm)\cap H_x^{1,1}(R^\pm)
\end{array},
\end{equation}
is Lipshitz continuous.
\end{lemma}

\setcounter{equation}{0}
\section{Global solutions to the MT model}\label{sec7}

Since $a(k)=a(z)$ is time independent, for $r_\pm(t;z)\in {\cal{W}}(\mathbb{R})$, the constraint \eqref{mtc8} and the relation \eqref{mtc10} are still valid for every $t\in[0,T]$.

\begin{proposition}\label{pro7}
For every $v_0\in H^2(\mathbb{R})\cap H^{1,1}(\mathbb{R})$ and $u_0\in H^3(\mathbb{R})\cap H^{1,1}(\mathbb{R})$ such that the linear spectral problem \eqref{mta2} admits no eigenvalues or resonances. For every $t\in[0,T]$, there exists a unique local solution
\[\begin{array}{c}
v(x,t)\in C\big(H_x^2(\mathbb{R})\cap H_x^{1,1}(\mathbb{R}), [0,T]\big),\\
u(x,t)\in C\big(H_x^3(\mathbb{R})\cap H_x^{1,1}(\mathbb{R}), [0,T]\big)
\end{array}
\]
to the Cauchy problem \eqref{mta1}. Moreover, the map
\begin{equation}\label{mtf1}
\begin{array}{c}
H^2(\mathbb{R})\cap H^{1,1}(\mathbb{R})\ni v_0\\
H^3(\mathbb{R})\cap H^{1,1}(\mathbb{R})\ni u_0
\end{array}
\mapsto \begin{array}{c}
v(x,t)\in C\big(H_x^2(\mathbb{R})\cap H_x^{1,1}(\mathbb{R}), [0,T]\big),\\
u(x,t)\in C\big(H_x^3(\mathbb{R})\cap H_x^{1,1}(\mathbb{R}), [0,T]\big)
\end{array}
\end{equation}
is Lipschitz continuous.
\end{proposition}
{\bf Proof}.~~The potentials $v(x,t)$ and $u(x,t)$ are recovered from the scattering data $r_\pm(t;z)$ in Section 5 and Section 6, and the potentials are controlled by $r_\pm(t;z)$ shown in Lemma \ref{lem21}. we note that $|u|\leq|v_t|+2|v|(\|v\|_{L_x^2}\|v_t\|_{L_x^2})$ for every $t\in[0,T]$. In addition, the Lemma \ref{lem8} shows that the scattering data $r_\pm(z)$ are controlled by the potentials in the direct scattering procedure. Combining the whole analysis in the previous sections, using Proposition \ref{pro6}, we can get, for every $t\in[0,T]$
\begin{equation}\label{mtf2}
\begin{aligned}
&\|v(x,t)\|_{H_x^2(\mathbb{R})\cap H_x^{1,1}(\mathbb{R})}+\|u(x,t)\|_{H_x^3(\mathbb{R})\cap H_x^{1,1}(\mathbb{R})}\\
&\leq c(\|r_+(t;z)\|_{H_z^1(\mathbb{R})\cap L_z^{2,1}(\mathbb{R})}+\|r_-(t;z)\|_{H_z^1(\mathbb{R})\cap L_z^{2,1}(\mathbb{R})})\\
&\leq c(T)(\|r_+(z)\|_{H_z^1(\mathbb{R})\cap L_z^{2,1}(\mathbb{R})}+\|r_-(z)\|_{H_z^1(\mathbb{R})\cap L_z^{2,1}(\mathbb{R})})\\
&\leq c(T)(\|v_0(x)\|_{H^2(\mathbb{R})\cap H^{1,1}(\mathbb{R})}+\|u_0(x)\|_{H^3(\mathbb{R})\cap H^{1,1}(\mathbb{R})}).
\end{aligned}
\end{equation}
where the positive constant $c(T)$ may grow at most polynomially in $T$ but is remains finite for every $T>0$.
According to discussions in the previous sections, we get that there exists a unique local solution
\[\begin{array}{c}
v(x,t)\in C\big(H_x^2(\mathbb{R})\cap H_x^{1,1}(\mathbb{R}), [0,T]\big),\\
u(x,t)\in C\big(H_x^3(\mathbb{R})\cap H_x^{1,1}(\mathbb{R}), [0,T]\big)
\end{array}
\]
to the Cauchy problem \eqref{mta1}.

Moreover, from the Lemma \ref{lem21}, we can also obtain, for every $x\in\mathbb{R}$
\[\begin{aligned}
&\|v(x,t+\Delta t)-v(x,t)\|_{H_x^2(\mathbb{R})\cap H_x^{1,1}(\mathbb{R})}+\|u(x,t+\Delta t)-u(x,t)\|_{H_x^3(\mathbb{R})\cap H_x^{1,1}(\mathbb{R})}\\
&\leq C_1\|r_\pm(z)(\mathrm{e}^{iz^{-1}(t+\Delta t)}-e^{iz^{-1}t})\|_{H_z^1(\mathbb{R})\cap L_z^{2,1}(\mathbb{R})}\\
&\leq C|\Delta t|\|r_\pm(z)\|_{H_z^1(\mathbb{R})\cap L_z^{2,1}(\mathbb{R})}.
\end{aligned}\]
Thus the Lipschitz continuity is proved.  \qquad $\square$

\vspace{1cm}
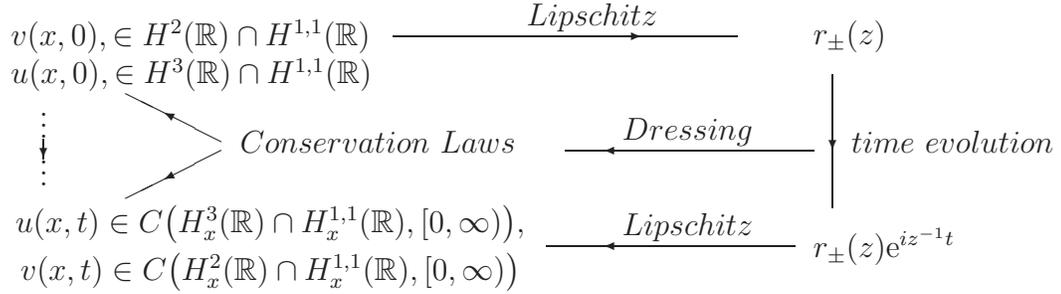
\begin{figure}[h]
\setlength{\unitlength}{0.1in}
\begin{picture}(20,10)
\put(3,11){\makebox(1,2)[l]{$v(x,0),\in H^2(\mathbb{R})\cap H^{1,1}(\mathbb{R})$}}
\put(23,12){\line(1,0){18}}\put(35,12){\vector(1,0){1}}
\put(30,12.5){\makebox(2,1)[l]{$Lipschitz$}}
\put(45,11){\makebox(1,2)[l]{$r_\pm(z)$}}
\put(46,3){\line(0,1){7}}\put(46,7){\vector(0,-1){1}}
\put(47,6){\makebox(2,1)[l]{$time~evolution$}}
\put(3,9){\makebox(1,2)[l]{$u(x,0),\in H^3(\mathbb{R})\cap H^{1,1}(\mathbb{R})$}}
\put(15,5.5){\makebox(1,2)[l]{$Conservation~Laws$}}
\put(35,6.5){\makebox(2,1)[l]{$Dressing$}}
\put(32,6){\line(1,0){13}}\put(35,6){\vector(-1,0){1}}
\put(14,6.5){\line(-2,1){5}}\put(12,7.5){\vector(-2,1){1}}
\put(14,6){\line(-2,-1){5}}\put(12,5.0){\vector(-2,-1){1}}
\put(4.5,6.7){\vdots}\put(4.5,4.2){\vdots}
\put(4.7,6.6){\vector(0,-1){1}}
\put(45,0){\makebox(1,2)[l]{$r_\pm(z)\mathrm{e}^{iz^{-1}t}$}}
\put(35,1.5){\makebox(2,1)[l]{$Lipschitz$}}
\put(31,1){\line(1,0){13}}\put(35,1){\vector(-1,0){1}}
\put(2,0){\makebox(1,2)[l]{
$\begin{array}{c}
u(x,t)\in C\big(H_x^3(\mathbb{R})\cap H_x^{1,1}(\mathbb{R}), [0,\infty)\big),\\
v(x,t)\in C\big(H_x^2(\mathbb{R})\cap H_x^{1,1}(\mathbb{R}), [0,\infty)\big)
\end{array}$}}
\end{picture}
\caption{The scheme for the proof of Theorem \ref{the1}.}
\end{figure}

\begin{theorem}\label{the1}
For every $v_0\in H^2(\mathbb{R})\cap H^{1,1}(\mathbb{R})$ and $u_0\in H^3(\mathbb{R})\cap H^{1,1}(\mathbb{R})$ such that the linear spectral problem \eqref{mta2} admits no eigenvalues or resonances. For every $t\in[0,\infty)$, there exists a unique global solution
\[\begin{array}{c}
v(x,t)\in C\big(H_x^2(\mathbb{R})\cap H_x^{1,1}(\mathbb{R}), (0,\infty)\big),\\
u(x,t)\in C\big(H_x^3(\mathbb{R})\cap H_x^{1,1}(\mathbb{R}), (0,\infty)\big)
\end{array}
\]
to the Cauchy problem \eqref{mta1}. Moreover, the map
\begin{equation}\label{mtf3}
\begin{array}{c}
H^2(\mathbb{R})\cap H^{1,1}(\mathbb{R})\ni v_0\\
H^3(\mathbb{R})\cap H^{1,1}(\mathbb{R})\ni u_0
\end{array}
\mapsto \begin{array}{c}
v(x,t)\in C\big(H_x^2(\mathbb{R})\cap H_x^{1,1}(\mathbb{R}), [0,\infty)\big),\\
u(x,t)\in C\big(H_x^3(\mathbb{R})\cap H_x^{1,1}(\mathbb{R}), [0,\infty)\big)
\end{array}
\end{equation}
is Lipschitz continuous.
\end{theorem}
{\bf Proof}.~~From the proof of the Proposition \ref{pro7}, we show that the local solution $v,u$ can not blow up in a finite time. Indeed, if there exists a maximal existence time $T_{m}>0$
such that
\[\lim\limits_{t\to T_{m}}\big(\|v(x,t)\|_{H_x^2\cap H_x^{1,1}}+\|u(x,t)\|_{H_x^3\cap H_x^{1,1}}\big)=\infty,\]
then the bound \eqref{mtf2} is contradiction as $t\to T$. Thus the local solution $v,u$ can be continued globally in time for every $T>0$. The Theorem \ref{the1} is proved. \qquad $\square$

\section*{Declarations}


\noindent{\bf Funding} The author declares that no funds were received during the preparation of this manuscript.

\noindent{\bf Data availability} No Data are used for the research described in the article.

\noindent{\bf Conflict of interest} The author declares that they have no conflicts of interest.


\end{document}